\documentclass[longauth]{aa}

\usepackage{graphicx}
\usepackage{stfloats}

\usepackage{txfonts}

\usepackage{amsmath}
\usepackage{amssymb}
\usepackage{xcolor}
\usepackage[pdfencoding=auto,psdextra]{hyperref}

\usepackage{euclid}

\renewcommand\linenumbers{}

\usepackage{cleveref}
\crefname{section}{Sect.}{Sects.}
\Crefname{section}{Section}{Sections}
\crefname{subsection}{Sect.}{Sects.}
\Crefname{subsection}{Section}{Sections}
\crefname{figure}{Fig.}{Figs.}
\Crefname{figure}{Figure}{Figures}
\crefname{equation}{Eq.}{Eqs.}
\Crefname{equation}{Equation}{Equations}
\crefname{table}{Table}{Tables}
\Crefname{table}{Table}{Tables}
\crefname{appendix}{Appendix}{Appendices}
\Crefname{appendix}{Appendix}{Appendices}

\hypersetup{
    colorlinks=true,
    linkcolor=black,
    urlcolor=black,
    citecolor=black
}

\renewcommand{\eprint}[2][]{{\if!#1!#2\else#1\arxivprefixesep\ignorespaces#2\fi}}

\newcommand{\github}[1]{\footnote{\url{#1}}}
\newcommand{\currentlenssample}{\reviewchange{15\,072}}
\DeclareRobustCommand{\reviewchange}[1]{{#1}}

\DeclareRobustCommand{\inspectionchange}[1]{{#1}}

\DeclareRobustCommand{\internalchange}[1]{#1}

\DeclareRobustCommand{\stylechange}[1]{#1}

\title{Euclid Data Release 1 (DR1)}
\subtitle{Euclid Lens Identification Pipeline and Scientific Exploration
(ECLIPSE) C: Lens modelling pipeline overview}
\titlerunning{ECLIPSE-C: Lens modelling}

\newcommand{\orcid}[1]{} 
\author{Euclid Collaboration: J.~W.~Nightingale\orcid{0000-0002-8987-7401}\thanks{\email{James.Nightingale@newcastle.ac.uk}}\inst{\ref{aff1}}
\and S.~H.~Vincken\orcid{0009-0005-7305-2359}\inst{\ref{aff2}}
\and L.~R.~Ecker\orcid{0009-0005-3508-2469}\inst{\ref{aff3},\ref{aff4},\ref{aff5}}
\and A.~Nersesian\orcid{0000-0001-6843-409X}\inst{\ref{aff6},\ref{aff7}}
\and D.~J.~Bowden\orcid{0009-0008-6114-1401}\inst{\ref{aff8}}
\and C.~M.~O'Riordan\orcid{0000-0003-2227-1998}\inst{\ref{aff5}}
\and N.~E.~P.~Lines\orcid{0009-0004-7751-1914}\inst{\ref{aff9}}
\and T.~E.~Collett\orcid{0000-0001-5564-3140}\inst{\ref{aff9}}
\and G.~Despali\orcid{0000-0001-6150-4112}\inst{\ref{aff10},\ref{aff11},\ref{aff12}}
\and S.~Schuldt\orcid{0000-0003-2497-6334}\inst{\ref{aff13},\ref{aff14},\ref{aff15}}
\and F.~Courbin\orcid{0000-0003-0758-6510}\inst{\ref{aff16},\ref{aff17},\ref{aff18}}
\and R.~B.~Metcalf\orcid{0000-0003-3167-2574}\inst{\ref{aff10},\ref{aff11}}
\and A.~Verma\orcid{0000-0002-0730-0781}\inst{\ref{aff8}}
\and A.~Angress\orcid{0009-0005-5250-9487}\inst{\ref{aff19},\ref{aff20}}
\and W.~J.~R.~Enzi\orcid{0009-0004-2992-3148}\inst{\ref{aff9},\ref{aff21}}
\and K.~Finner\orcid{0000-0002-4462-0709}\inst{\ref{aff22}}
\and P.~Fleury\orcid{0000-0001-9292-3651}\inst{\ref{aff23}}
\and M.~Fogliardi\orcid{0009-0006-4964-5311}\inst{\ref{aff24},\ref{aff11}}
\and A.~Galan\orcid{0000-0003-2547-9815}\inst{\ref{aff25}}
\and R.~Gavazzi\orcid{0000-0002-5540-6935}\inst{\ref{aff26},\ref{aff27}}
\and Q.~He\orcid{0000-0003-3672-9365}\inst{\ref{aff28}}
\and N.~B.~Hogg\orcid{0000-0001-9346-4477}\inst{\ref{aff29},\ref{aff30}}
\and P.~Holloway\orcid{0009-0002-8896-6100}\inst{\ref{aff9}}
\and D.~P.~Johnson\orcid{0000-0002-0311-2513}\inst{\ref{aff23}}
\and S.~K.~Li\orcid{0000-0002-4490-7304}\inst{\ref{aff31}}
\and T.~Li\orcid{0009-0005-5008-0381}\inst{\ref{aff9}}
\and N.~Massari\inst{\ref{aff32}}
\and A.~Melo\orcid{0000-0002-6449-3970}\inst{\ref{aff33},\ref{aff34},\ref{aff35}}
\and M.~Meneghetti\orcid{0000-0003-1225-7084}\inst{\ref{aff11},\ref{aff12}}
\and M.~Millon\orcid{0000-0001-7051-497X}\inst{\ref{aff36}}
\and L.~Moustakas\inst{\ref{aff37}}
\and C.~Olave\inst{\ref{aff38}}
\and H.~Qu\orcid{0009-0006-0299-0265}\inst{\ref{aff22}}
\and K.~Rojas\orcid{0000-0003-1391-6854}\inst{\ref{aff39}}
\and C.~Saulder\orcid{0000-0002-0408-5633}\inst{\ref{aff4},\ref{aff3}}
\and D.~Sluse\orcid{0000-0001-6116-2095}\inst{\ref{aff6}}
\and G.~Smadja\inst{\ref{aff40}}
\and H.~Srivastava\orcid{0009-0002-3254-0952}\inst{\ref{aff41}}
\and A.~Sonnenfeld\orcid{0000-0002-6061-5977}\inst{\ref{aff42},\ref{aff43}}
\and C.~J.~Storfer\orcid{0000-0002-0385-0014}\inst{\ref{aff44}}
\and T.~T.~Thai\orcid{0000-0002-8408-4816}\inst{\ref{aff45}}
\and C.~Tortora\orcid{0000-0001-7958-6531}\inst{\ref{aff46}}
\and N.~Wielders\orcid{0009-0001-8110-819X}\inst{\ref{aff1}}
\and Y.~Xie\inst{\ref{aff47}}
\and B.~Altieri\orcid{0000-0003-3936-0284}\inst{\ref{aff48}}
\and A.~Amara\inst{\ref{aff49}}
\and S.~Andreon\orcid{0000-0002-2041-8784}\inst{\ref{aff50}}
\and N.~Auricchio\orcid{0000-0003-4444-8651}\inst{\ref{aff11}}
\and C.~Baccigalupi\orcid{0000-0002-8211-1630}\inst{\ref{aff51},\ref{aff52},\ref{aff53},\ref{aff54}}
\and M.~Baldi\orcid{0000-0003-4145-1943}\inst{\ref{aff38},\ref{aff11},\ref{aff12}}
\and S.~Bardelli\orcid{0000-0002-8900-0298}\inst{\ref{aff11}}
\and P.~Battaglia\orcid{0000-0002-7337-5909}\inst{\ref{aff11}}
\and R.~Bender\orcid{0000-0001-7179-0626}\inst{\ref{aff4},\ref{aff3}}
\and A.~Biviano\orcid{0000-0002-0857-0732}\inst{\ref{aff52},\ref{aff51}}
\and M.~Bolzonella\orcid{0000-0003-3278-4607}\inst{\ref{aff11}}
\and E.~Branchini\orcid{0000-0002-0808-6908}\inst{\ref{aff55},\ref{aff56},\ref{aff50}}
\and M.~Brescia\orcid{0000-0001-9506-5680}\inst{\ref{aff57},\ref{aff46}}
\and S.~Camera\orcid{0000-0003-3399-3574}\inst{\ref{aff58},\ref{aff59},\ref{aff60}}
\and V.~Capobianco\orcid{0000-0002-3309-7692}\inst{\ref{aff60}}
\and C.~Carbone\orcid{0000-0003-0125-3563}\inst{\ref{aff15}}
\and V.~F.~Cardone\inst{\ref{aff61},\ref{aff62}}
\and J.~Carretero\orcid{0000-0002-3130-0204}\inst{\ref{aff63},\ref{aff64}}
\and M.~Castellano\orcid{0000-0001-9875-8263}\inst{\ref{aff61}}
\and G.~Castignani\orcid{0000-0001-6831-0687}\inst{\ref{aff11}}
\and S.~Cavuoti\orcid{0000-0002-3787-4196}\inst{\ref{aff46},\ref{aff65}}
\and K.~C.~Chambers\orcid{0000-0001-6965-7789}\inst{\ref{aff44}}
\and A.~Cimatti\inst{\ref{aff66}}
\and C.~Colodro-Conde\inst{\ref{aff67}}
\and G.~Congedo\orcid{0000-0003-2508-0046}\inst{\ref{aff68}}
\and C.~J.~Conselice\orcid{0000-0003-1949-7638}\inst{\ref{aff69}}
\and L.~Conversi\orcid{0000-0002-6710-8476}\inst{\ref{aff70},\ref{aff48}}
\and Y.~Copin\orcid{0000-0002-5317-7518}\inst{\ref{aff71}}
\and A.~Costille\inst{\ref{aff26}}
\and H.~M.~Courtois\orcid{0000-0003-0509-1776}\inst{\ref{aff72}}
\and M.~Cropper\orcid{0000-0003-4571-9468}\inst{\ref{aff73}}
\and H.~Degaudenzi\orcid{0000-0002-5887-6799}\inst{\ref{aff25}}
\and G.~De~Lucia\orcid{0000-0002-6220-9104}\inst{\ref{aff52}}
\and C.~Dolding\orcid{0009-0003-7199-6108}\inst{\ref{aff73}}
\and H.~Dole\orcid{0000-0002-9767-3839}\inst{\ref{aff74}}
\and F.~Dubath\orcid{0000-0002-6533-2810}\inst{\ref{aff25}}
\and X.~Dupac\inst{\ref{aff48}}
\and S.~Escoffier\orcid{0000-0002-2847-7498}\inst{\ref{aff75}}
\and M.~Farina\orcid{0000-0002-3089-7846}\inst{\ref{aff76}}
\and R.~Farinelli\inst{\ref{aff11}}
\and F.~Faustini\orcid{0000-0001-6274-5145}\inst{\ref{aff61}}
\and S.~Ferriol\inst{\ref{aff71}}
\and M.~Frailis\orcid{0000-0002-7400-2135}\inst{\ref{aff52}}
\and E.~Franceschi\orcid{0000-0002-0585-6591}\inst{\ref{aff11}}
\and M.~Fumana\orcid{0000-0001-6787-5950}\inst{\ref{aff15}}
\and L.~Gabarra\orcid{0000-0002-8486-8856}\inst{\ref{aff8}}
\and S.~Galeotta\orcid{0000-0002-3748-5115}\inst{\ref{aff52}}
\and K.~George\orcid{0000-0002-1734-8455}\inst{\ref{aff77}}
\and B.~Gillis\orcid{0000-0002-4478-1270}\inst{\ref{aff68}}
\and C.~Giocoli\orcid{0000-0002-9590-7961}\inst{\ref{aff11},\ref{aff12}}
\and P.~G\'omez-Alvarez\orcid{0000-0002-8594-5358}\inst{\ref{aff78},\ref{aff48}}
\and J.~Gracia-Carpio\orcid{0000-0003-4689-3134}\inst{\ref{aff4}}
\and A.~Grazian\orcid{0000-0002-5688-0663}\inst{\ref{aff79}}
\and F.~Grupp\inst{\ref{aff4},\ref{aff3}}
\and L.~Guzzo\orcid{0000-0001-8264-5192}\inst{\ref{aff80},\ref{aff50},\ref{aff81}}
\and W.~G.~Hartley\orcid{0000-0002-4563-5101}\inst{\ref{aff25}}
\and S.~V.~H.~Haugan\orcid{0000-0001-9648-7260}\inst{\ref{aff82}}
\and S.~Hemmati\orcid{0000-0003-2226-5395}\inst{\ref{aff22}}
\and H.~Hoekstra\orcid{0000-0002-0641-3231}\inst{\ref{aff83}}
\and M.~S.~Holliman\inst{\ref{aff68}}
\and W.~Holmes\orcid{0009-0007-8554-4646}\inst{\ref{aff84}}
\and I.~M.~Hook\orcid{0000-0002-2960-978X}\inst{\ref{aff85}}
\and F.~Hormuth\inst{\ref{aff86}}
\and A.~Hornstrup\orcid{0000-0002-3363-0936}\inst{\ref{aff87},\ref{aff88}}
\and M.~Huertas-Company\orcid{0000-0002-1416-8483}\inst{\ref{aff67},\ref{aff89},\ref{aff90}}
\and K.~Jahnke\orcid{0000-0003-3804-2137}\inst{\ref{aff91}}
\and M.~Jhabvala\inst{\ref{aff92}}
\and S.~Kermiche\orcid{0000-0002-0302-5735}\inst{\ref{aff75}}
\and A.~Kiessling\orcid{0000-0002-2590-1273}\inst{\ref{aff84}}
\and B.~Kubik\orcid{0009-0006-5823-4880}\inst{\ref{aff71}}
\and M.~K\"ummel\orcid{0000-0003-2791-2117}\inst{\ref{aff3}}
\and M.~Kunz\orcid{0000-0002-3052-7394}\inst{\ref{aff36}}
\and H.~Kurki-Suonio\orcid{0000-0002-4618-3063}\inst{\ref{aff14},\ref{aff93}}
\and A.~M.~C.~Le~Brun\orcid{0000-0002-0936-4594}\inst{\ref{aff94}}
\and D.~Le~Mignant\orcid{0000-0002-5339-5515}\inst{\ref{aff26}}
\and S.~Ligori\orcid{0000-0003-4172-4606}\inst{\ref{aff60}}
\and P.~B.~Lilje\orcid{0000-0003-4324-7794}\inst{\ref{aff82}}
\and V.~Lindholm\orcid{0000-0003-2317-5471}\inst{\ref{aff14},\ref{aff93}}
\and I.~Lloro\orcid{0000-0001-5966-1434}\inst{\ref{aff95}}
\and M.~Magliocchetti\orcid{0000-0001-9158-4838}\inst{\ref{aff76}}
\and G.~Mainetti\orcid{0000-0003-2384-2377}\inst{\ref{aff96}}
\and D.~Maino\inst{\ref{aff80},\ref{aff15},\ref{aff81}}
\and O.~Mansutti\orcid{0000-0001-5758-4658}\inst{\ref{aff52}}
\and S.~Marcin\inst{\ref{aff39}}
\and O.~Marggraf\orcid{0000-0001-7242-3852}\inst{\ref{aff97}}
\and M.~Martinelli\orcid{0000-0002-6943-7732}\inst{\ref{aff61},\ref{aff62}}
\and N.~Martinet\orcid{0000-0003-2786-7790}\inst{\ref{aff26}}
\and F.~Marulli\orcid{0000-0002-8850-0303}\inst{\ref{aff10},\ref{aff11},\ref{aff12}}
\and R.~J.~Massey\orcid{0000-0002-6085-3780}\inst{\ref{aff98}}
\and N.~Mauri\orcid{0000-0001-8196-1548}\inst{\ref{aff66},\ref{aff12}}
\and E.~Medinaceli\orcid{0000-0002-4040-7783}\inst{\ref{aff11}}
\and S.~Mei\orcid{0000-0002-2849-559X}\inst{\ref{aff99},\ref{aff100}}
\and M.~Melchior\inst{\ref{aff2}}
\and E.~Merlin\orcid{0000-0001-6870-8900}\inst{\ref{aff79}}
\and G.~Meylan\orcid{0000-0001-6503-0209}\inst{\ref{aff101}}
\and P.~Monaco\orcid{0000-0003-2083-7564}\inst{\ref{aff102},\ref{aff52},\ref{aff53},\ref{aff51}}
\and A.~Mora\orcid{0000-0002-1922-8529}\inst{\ref{aff103}}
\and M.~Moresco\orcid{0000-0002-7616-7136}\inst{\ref{aff10},\ref{aff11}}
\and C.~Moretti\orcid{0000-0003-3314-8936}\inst{\ref{aff52},\ref{aff51},\ref{aff53}}
\and L.~Moscardini\orcid{0000-0002-3473-6716}\inst{\ref{aff10},\ref{aff11},\ref{aff12}}
\and R.~Nakajima\orcid{0009-0009-1213-7040}\inst{\ref{aff97}}
\and C.~Neissner\orcid{0000-0001-8524-4968}\inst{\ref{aff104},\ref{aff64}}
\and R.~C.~Nichol\orcid{0000-0003-0939-6518}\inst{\ref{aff49}}
\and S.-M.~Niemi\orcid{0009-0005-0247-0086}\inst{\ref{aff105}}
\and C.~Padilla\orcid{0000-0001-7951-0166}\inst{\ref{aff104}}
\and S.~Paltani\orcid{0000-0002-8108-9179}\inst{\ref{aff25}}
\and F.~Pasian\orcid{0000-0002-4869-3227}\inst{\ref{aff52}}
\and W.~J.~Percival\orcid{0000-0002-0644-5727}\inst{\ref{aff106},\ref{aff107},\ref{aff108}}
\and V.~Pettorino\orcid{0000-0002-4203-9320}\inst{\ref{aff105}}
\and A.~Pezzotta\orcid{0000-0003-0726-2268}\inst{\ref{aff50}}
\and G.~Polenta\orcid{0000-0003-4067-9196}\inst{\ref{aff109}}
\and L.~A.~Popa\inst{\ref{aff110}}
\and L.~Pozzetti\orcid{0000-0001-7085-0412}\inst{\ref{aff11}}
\and F.~Raison\orcid{0000-0002-7819-6918}\inst{\ref{aff4}}
\and A.~Renzi\orcid{0000-0001-9856-1970}\inst{\ref{aff111},\ref{aff112},\ref{aff11}}
\and J.~Rhodes\orcid{0000-0002-4485-8549}\inst{\ref{aff84}}
\and G.~Riccio\inst{\ref{aff46}}
\and I.~Risso\orcid{0000-0003-2525-7761}\inst{\ref{aff55},\ref{aff56},\ref{aff50}}
\and E.~Romelli\orcid{0000-0003-3069-9222}\inst{\ref{aff52}}
\and M.~Roncarelli\orcid{0000-0001-9587-7822}\inst{\ref{aff11}}
\and B.~Rusholme\orcid{0000-0001-7648-4142}\inst{\ref{aff22}}
\and R.~Saglia\orcid{0000-0003-0378-7032}\inst{\ref{aff3},\ref{aff4}}
\and Z.~Sakr\orcid{0000-0002-4823-3757}\inst{\ref{aff113},\ref{aff114},\ref{aff115}}
\and D.~Sapone\orcid{0000-0001-7089-4503}\inst{\ref{aff116}}
\and B.~Sartoris\orcid{0000-0003-1337-5269}\inst{\ref{aff4},\ref{aff52}}
\and M.~Schirmer\orcid{0000-0003-2568-9994}\inst{\ref{aff91}}
\and P.~Schneider\orcid{0000-0001-8561-2679}\inst{\ref{aff97}}
\and T.~Schrabback\orcid{0000-0002-6987-7834}\inst{\ref{aff117}}
\and A.~Secroun\orcid{0000-0003-0505-3710}\inst{\ref{aff75}}
\and E.~Sihvola\orcid{0000-0003-1804-7715}\inst{\ref{aff118}}
\and C.~Sirignano\orcid{0000-0002-0995-7146}\inst{\ref{aff111},\ref{aff112}}
\and G.~Sirri\orcid{0000-0003-2626-2853}\inst{\ref{aff12}}
\and J.~Skottfelt\orcid{0000-0003-1310-8283}\inst{\ref{aff119}}
\and L.~Stanco\orcid{0000-0002-9706-5104}\inst{\ref{aff112}}
\and P.~Tallada-Cresp\'{i}\orcid{0000-0002-1336-8328}\inst{\ref{aff63},\ref{aff64}}
\and A.~N.~Taylor\inst{\ref{aff68}}
\and I.~Tereno\orcid{0000-0002-4537-6218}\inst{\ref{aff120},\ref{aff121}}
\and N.~Tessore\orcid{0000-0002-9696-7931}\inst{\ref{aff73}}
\and S.~Toft\orcid{0000-0003-3631-7176}\inst{\ref{aff122},\ref{aff123}}
\and R.~Toledo-Moreo\orcid{0000-0002-2997-4859}\inst{\ref{aff124},\ref{aff125}}
\and F.~Torradeflot\orcid{0000-0003-1160-1517}\inst{\ref{aff64},\ref{aff63}}
\and A.~Tsyganov\inst{\ref{aff126}}
\and I.~Tutusaus\orcid{0000-0002-3199-0399}\inst{\ref{aff127},\ref{aff128},\ref{aff114}}
\and J.~Valiviita\orcid{0000-0001-6225-3693}\inst{\ref{aff14},\ref{aff93}}
\and T.~Vassallo\orcid{0000-0001-6512-6358}\inst{\ref{aff52},\ref{aff77}}
\and A.~Veropalumbo\orcid{0000-0003-2387-1194}\inst{\ref{aff50},\ref{aff56},\ref{aff55}}
\and Y.~Wang\orcid{0000-0002-4749-2984}\inst{\ref{aff22}}
\and J.~Weller\orcid{0000-0002-8282-2010}\inst{\ref{aff3},\ref{aff4}}
\and A.~Zacchei\orcid{0000-0003-0396-1192}\inst{\ref{aff52},\ref{aff51}}
\and G.~Zamorani\orcid{0000-0002-2318-301X}\inst{\ref{aff11}}
\and E.~Zucca\orcid{0000-0002-5845-8132}\inst{\ref{aff11}}
\and M.~Ballardini\orcid{0000-0003-4481-3559}\inst{\ref{aff24},\ref{aff129},\ref{aff11}}
\and P.~Bergamini\orcid{0000-0003-1383-9414}\inst{\ref{aff11}}
\and S.~Borgani\orcid{0000-0001-6151-6439}\inst{\ref{aff102},\ref{aff51},\ref{aff52},\ref{aff53},\ref{aff130}}
\and E.~Bozzo\orcid{0000-0002-8201-1525}\inst{\ref{aff25}}
\and C.~Burigana\orcid{0000-0002-3005-5796}\inst{\ref{aff21},\ref{aff131}}
\and R.~Cabanac\orcid{0000-0001-6679-2600}\inst{\ref{aff114}}
\and M.~Calabrese\orcid{0000-0002-2637-2422}\inst{\ref{aff132},\ref{aff15}}
\and A.~Cappi\inst{\ref{aff133},\ref{aff11}}
\and F.~Caro\orcid{0009-0003-1053-0507}\inst{\ref{aff61}}
\and T.~Castro\orcid{0000-0002-6292-3228}\inst{\ref{aff134},\ref{aff52},\ref{aff51}}
\and D.~Di~Ferdinando\inst{\ref{aff12}}
\and J.~A.~Escartin~Vigo\inst{\ref{aff4}}
\and J.~Garc\'ia-Bellido\orcid{0000-0002-9370-8360}\inst{\ref{aff113}}
\and T.~Gasparetto\orcid{0000-0002-7913-4866}\inst{\ref{aff61}}
\and C.~M.~Gutierrez\orcid{0000-0001-7854-783X}\inst{\ref{aff67},\ref{aff135}}
\and A.~Loureiro\orcid{0000-0002-4371-0876}\inst{\ref{aff136},\ref{aff137}}
\and J.~Macias-Perez\orcid{0000-0002-5385-2763}\inst{\ref{aff138}}
\and R.~Maoli\orcid{0000-0002-6065-3025}\inst{\ref{aff139},\ref{aff61}}
\and A.~A.~Nucita\inst{\ref{aff140},\ref{aff141},\ref{aff142}}
\and M.~P\"ontinen\orcid{0000-0001-5442-2530}\inst{\ref{aff14}}
\and E.~Sarpa\orcid{0000-0002-1256-655X}\inst{\ref{aff52}}
\and V.~Scottez\orcid{0009-0008-3864-940X}\inst{\ref{aff143},\ref{aff144}}
\and M.~Sereno\orcid{0000-0003-0302-0325}\inst{\ref{aff11},\ref{aff12}}
\and M.~Tenti\orcid{0000-0002-4254-5901}\inst{\ref{aff12}}
\and M.~Tucci\inst{\ref{aff25}}
\and M.~Wiesmann\orcid{0009-0000-8199-5860}\inst{\ref{aff82}}
\and J.~A.~Acevedo~Barroso\orcid{0000-0002-9654-1711}\inst{\ref{aff84}}
\and Y.~Akrami\orcid{0000-0002-2407-7956}\inst{\ref{aff113},\ref{aff145}}
\and I.~T.~Andika\orcid{0000-0001-6102-9526}\inst{\ref{aff3}}
\and G.~Angora\orcid{0000-0002-0316-6562}\inst{\ref{aff46},\ref{aff24}}
\and S.~Anselmi\orcid{0000-0002-3579-9583}\inst{\ref{aff112},\ref{aff111},\ref{aff146}}
\and M.~Archidiacono\orcid{0000-0003-4952-9012}\inst{\ref{aff80},\ref{aff81}}
\and G.~Aric\`o\orcid{0000-0002-2802-2928}\inst{\ref{aff12}}
\and F.~Atrio-Barandela\orcid{0000-0002-2130-2513}\inst{\ref{aff147}}
\and M.~Baes\orcid{0000-0002-3930-2757}\inst{\ref{aff7}}
\and L.~Bazzanini\orcid{0000-0003-0727-0137}\inst{\ref{aff24},\ref{aff11}}
\and D.~Bertacca\orcid{0000-0002-2490-7139}\inst{\ref{aff111},\ref{aff79},\ref{aff112}}
\and M.~Bethermin\orcid{0000-0002-3915-2015}\inst{\ref{aff148}}
\and F.~Beutler\orcid{0000-0003-0467-5438}\inst{\ref{aff68}}
\and L.~Blot\orcid{0000-0002-9622-7167}\inst{\ref{aff149},\ref{aff94}}
\and M.~L.~Brown\orcid{0000-0002-0370-8077}\inst{\ref{aff69}}
\and S.~Bruton\orcid{0000-0002-6503-5218}\inst{\ref{aff150}}
\and A.~Calabro\orcid{0000-0003-2536-1614}\inst{\ref{aff61}}
\and B.~Camacho~Quevedo\orcid{0000-0002-8789-4232}\inst{\ref{aff51},\ref{aff54},\ref{aff52}}
\and C.~S.~Carvalho\inst{\ref{aff121}}
\and F.~Cogato\orcid{0000-0003-4632-6113}\inst{\ref{aff10},\ref{aff11}}
\and A.~R.~Cooray\orcid{0000-0002-3892-0190}\inst{\ref{aff151}}
\and P.~Corcho-Caballero\orcid{0000-0001-6327-7080}\inst{\ref{aff28}}
\and B.~Csizi\orcid{0000-0003-3227-6581}\inst{\ref{aff117}}
\and O.~Cucciati\orcid{0000-0002-9336-7551}\inst{\ref{aff11}}
\and T.~de~Boer\orcid{0000-0001-5486-2747}\inst{\ref{aff44}}
\and G.~Desprez\orcid{0000-0001-8325-1742}\inst{\ref{aff28}}
\and A.~D\'iaz-S\'anchez\orcid{0000-0003-0748-4768}\inst{\ref{aff152}}
\and S.~Di~Domizio\orcid{0000-0003-2863-5895}\inst{\ref{aff55},\ref{aff56}}
\and J.~M.~Diego\orcid{0000-0001-9065-3926}\inst{\ref{aff31}}
\and P.-A.~Duc\orcid{0000-0003-3343-6284}\inst{\ref{aff148}}
\and V.~Duret\orcid{0009-0009-0383-4960}\inst{\ref{aff75}}
\and M.~Y.~Elkhashab\orcid{0000-0001-9306-2603}\inst{\ref{aff52},\ref{aff53},\ref{aff102},\ref{aff51}}
\and A.~Enia\orcid{0000-0002-0200-2857}\inst{\ref{aff11}}
\and Y.~Fang\orcid{0000-0002-0334-6950}\inst{\ref{aff3}}
\and A.~Farina\orcid{0009-0000-3420-929X}\inst{\ref{aff50},\ref{aff56}}
\and A.~Finoguenov\orcid{0000-0002-4606-5403}\inst{\ref{aff14}}
\and A.~Fontana\orcid{0000-0003-3820-2823}\inst{\ref{aff61}}
\and A.~Franco\orcid{0000-0002-4761-366X}\inst{\ref{aff140},\ref{aff141},\ref{aff142}}
\and Y.~Fu\orcid{0000-0002-0759-0504}\inst{\ref{aff83},\ref{aff28}}
\and K.~Ganga\orcid{0000-0001-8159-8208}\inst{\ref{aff99}}
\and E.~Gaztanaga\orcid{0000-0001-9632-0815}\inst{\ref{aff127},\ref{aff128},\ref{aff9}}
\and Z.~Ghaffari\orcid{0000-0002-6467-8078}\inst{\ref{aff52},\ref{aff51}}
\and F.~Giacomini\orcid{0000-0002-3129-2814}\inst{\ref{aff12}}
\and F.~Gianotti\orcid{0000-0003-4666-119X}\inst{\ref{aff11}}
\and G.~Gozaliasl\orcid{0000-0002-0236-919X}\inst{\ref{aff153},\ref{aff14}}
\and A.~Gruppuso\orcid{0000-0001-9272-5292}\inst{\ref{aff11},\ref{aff12}}
\and M.~Guidi\orcid{0000-0001-9408-1101}\inst{\ref{aff38},\ref{aff11}}
\and A.~Hall\orcid{0000-0002-3139-8651}\inst{\ref{aff68}}
\and N.~A.~Hatch\orcid{0000-0001-5600-0534}\inst{\ref{aff154}}
\and T.~Heuberger\orcid{0009-0009-3474-8467}\inst{\ref{aff39}}
\and H.~Hildebrandt\orcid{0000-0002-9814-3338}\inst{\ref{aff155}}
\and J.~Hjorth\orcid{0000-0002-4571-2306}\inst{\ref{aff156}}
\and M.~Jauzac\orcid{0000-0003-1974-8732}\inst{\ref{aff157},\ref{aff98},\ref{aff158},\ref{aff159},\ref{aff114}}
\and J.~J.~E.~Kajava\orcid{0000-0002-3010-8333}\inst{\ref{aff160},\ref{aff13},\ref{aff161}}
\and Y.~Kang\orcid{0009-0000-8588-7250}\inst{\ref{aff117}}
\and V.~Kansal\orcid{0000-0002-4008-6078}\inst{\ref{aff162},\ref{aff163}}
\and D.~Karagiannis\orcid{0000-0002-4927-0816}\inst{\ref{aff24},\ref{aff164}}
\and J.~Kim\orcid{0000-0003-2776-2761}\inst{\ref{aff8}}
\and C.~C.~Kirkpatrick\inst{\ref{aff118}}
\and A.~Kov\'acs\orcid{0000-0002-5825-579X}\inst{\ref{aff165},\ref{aff166}}
\and I.~Kova{\v{c}}i{\'{c}}\orcid{0000-0001-6751-3263}\inst{\ref{aff7}}
\and K.~Koyama\orcid{0000-0001-6727-6915}\inst{\ref{aff9}}
\and S.~Kruk\orcid{0000-0001-8010-8879}\inst{\ref{aff48}}
\and M.~C.~Lam\orcid{0000-0002-9347-2298}\inst{\ref{aff68}}
\and M.~Lattanzi\orcid{0000-0003-1059-2532}\inst{\ref{aff129}}
\and G.~Leroy\orcid{0009-0004-2523-4425}\inst{\ref{aff157},\ref{aff98}}
\and G.~F.~Lesci\orcid{0000-0002-4607-2830}\inst{\ref{aff10},\ref{aff11}}
\and J.~Lesgourgues\orcid{0000-0001-7627-353X}\inst{\ref{aff167}}
\and L.~Linke\orcid{0000-0002-2622-8113}\inst{\ref{aff117}}
\and S.~J.~Liu\orcid{0000-0001-7680-2139}\inst{\ref{aff76}}
\and G.~Maggio\orcid{0000-0003-4020-4836}\inst{\ref{aff52}}
\and E.~A.~Magnier\orcid{0000-0002-7965-2815}\inst{\ref{aff44}}
\and F.~Mannucci\orcid{0000-0002-4803-2381}\inst{\ref{aff168}}
\and F.~R.~Marleau\orcid{0000-0002-1442-2947}\inst{\ref{aff117}}
\and C.~J.~A.~P.~Martins\orcid{0000-0002-4886-9261}\inst{\ref{aff169},\ref{aff170}}
\and M.~Migliaccio\inst{\ref{aff171},\ref{aff172}}
\and M.~Miluzio\inst{\ref{aff48},\ref{aff173}}
\and G.~Morgante\inst{\ref{aff11}}
\and K.~Naidoo\orcid{0000-0002-9182-1802}\inst{\ref{aff9},\ref{aff91}}
\and A.~Navarro-Alsina\orcid{0000-0002-3173-2592}\inst{\ref{aff97}}
\and S.~Nesseris\orcid{0000-0002-0567-0324}\inst{\ref{aff113}}
\and L.~Nicastro\orcid{0000-0001-8534-6788}\inst{\ref{aff11}}
\and M.~Oguri\orcid{0000-0003-3484-399X}\inst{\ref{aff174},\ref{aff175}}
\and F.~Oppizzi\orcid{0000-0003-3904-8370}\inst{\ref{aff56},\ref{aff112},\ref{aff111}}
\and F.~Pace\orcid{0000-0001-8039-0480}\inst{\ref{aff58},\ref{aff59},\ref{aff60},\ref{aff176}}
\and D.~Paoletti\orcid{0000-0003-4761-6147}\inst{\ref{aff11},\ref{aff131}}
\and F.~Passalacqua\orcid{0000-0002-8606-4093}\inst{\ref{aff112}}
\and K.~Paterson\orcid{0000-0001-8340-3486}\inst{\ref{aff91}}
\and L.~Patrizii\inst{\ref{aff12}}
\and C.~Pattison\orcid{0000-0003-3272-2617}\inst{\ref{aff9}}
\and A.~Pisani\orcid{0000-0002-6146-4437}\inst{\ref{aff75}}
\and D.~Potter\orcid{0000-0002-0757-5195}\inst{\ref{aff177}}
\and G.~W.~Pratt\inst{\ref{aff178}}
\and S.~Quai\orcid{0000-0002-0449-8163}\inst{\ref{aff10},\ref{aff11}}
\and M.~Radovich\orcid{0000-0002-3585-866X}\inst{\ref{aff79}}
\and G.~Rodighiero\orcid{0000-0002-9415-2296}\inst{\ref{aff111},\ref{aff79}}
\and W.~Roster\orcid{0000-0002-9149-6528}\inst{\ref{aff4}}
\and M.~Sahl\'en\orcid{0000-0003-0973-4804}\inst{\ref{aff179}}
\and D.~B.~Sanders\orcid{0000-0002-1233-9998}\inst{\ref{aff44}}
\and A.~Schneider\orcid{0000-0001-7055-8104}\inst{\ref{aff177}}
\and D.~Sciotti\orcid{0009-0008-4519-2620}\inst{\ref{aff61},\ref{aff62}}
\and E.~Sellentin\orcid{0009-0002-2655-3458}\inst{\ref{aff180},\ref{aff83}}
\and S.~Serjeant\orcid{0000-0002-0517-7943}\inst{\ref{aff181}}
\and L.~C.~Smith\orcid{0000-0002-3259-2771}\inst{\ref{aff29}}
\and I.~Szapudi\orcid{0000-0003-2274-0301}\inst{\ref{aff44}}
\and M.~Talia\orcid{0000-0003-4352-2063}\inst{\ref{aff10},\ref{aff11}}
\and K.~Tanidis\orcid{0000-0001-9843-5130}\inst{\ref{aff182}}
\and C.~Tao\orcid{0000-0001-7961-8177}\inst{\ref{aff75}}
\and G.~Testera\orcid{0000-0003-2970-766X}\inst{\ref{aff56}}
\and R.~Teyssier\orcid{0000-0001-7689-0933}\inst{\ref{aff183}}
\and S.~Tosi\orcid{0000-0002-7275-9193}\inst{\ref{aff55},\ref{aff50},\ref{aff56}}
\and A.~Troja\orcid{0000-0003-0239-4595}\inst{\ref{aff52}}
\and C.~Uhlemann\orcid{0000-0001-7831-1579}\inst{\ref{aff184},\ref{aff1}}
\and C.~Valieri\inst{\ref{aff12}}
\and D.~Vergani\orcid{0000-0003-0898-2216}\inst{\ref{aff11}}
\and G.~Vernardos\orcid{0000-0001-8554-7248}\inst{\ref{aff20},\ref{aff185}}
\and P.~Vielzeuf\orcid{0000-0003-2035-9339}\inst{\ref{aff75}}
\and E.~Vilenius\orcid{0000-0002-6184-7681}\inst{\ref{aff73}}
\and S.~Vinciguerra\orcid{0009-0005-4018-3184}\inst{\ref{aff26}}
\and M.~von~Wietersheim-Kramsta\orcid{0000-0003-4986-5091}\inst{\ref{aff98},\ref{aff157}}
\and N.~A.~Walton\orcid{0000-0003-3983-8778}\inst{\ref{aff29}}
\and L.~Wang\orcid{0000-0002-6736-9158}\inst{\ref{aff186},\ref{aff28}}
\and J.~Wilde\orcid{0000-0002-4460-7379}\inst{\ref{aff16}}
\and A.~H.~Wright\orcid{0000-0001-7363-7932}\inst{\ref{aff155}}
\and D.~Scott\orcid{0000-0002-6878-9840}\inst{\ref{aff41}}}
										   
\institute{School of Mathematics, Statistics and Physics, Newcastle University, Herschel Building, Newcastle-upon-Tyne, NE1 7RU, UK\label{aff1}
\and
University of Applied Sciences and Arts of Northwestern Switzerland, School of Engineering, 5210 Windisch, Switzerland\label{aff2}
\and
Universit\"ats-Sternwarte M\"unchen, Fakult\"at f\"ur Physik, Ludwig-Maximilians-Universit\"at M\"unchen, Scheinerstr.~1, 81679 M\"unchen, Germany\label{aff3}
\and
Max Planck Institute for Extraterrestrial Physics, Giessenbachstr. 1, 85748 Garching, Germany\label{aff4}
\and
Max-Planck-Institut f\"ur Astrophysik, Karl-Schwarzschild-Str.~1, 85748 Garching, Germany\label{aff5}
\and
STAR Institute, University of Li{\`e}ge, Quartier Agora, All\'ee du six Ao\^ut 19c, 4000 Li\`ege, Belgium\label{aff6}
\and
Universiteit Gent, Department of Physics and Astronomy, Proeftuinstraat 86 N3, 9000 Ghent, Belgium
\label{aff7}
\and
Department of Physics, University of Oxford, Keble Road, Oxford OX1 3RH, UK\label{aff8}
\and
Institute of Cosmology and Gravitation, University of Portsmouth, Portsmouth PO1 3FX, UK\label{aff9}
\and
Dipartimento di Fisica e Astronomia "Augusto Righi" - Alma Mater Studiorum Universit\`a di Bologna, via Piero Gobetti 93/2, 40129 Bologna, Italy\label{aff10}
\and
INAF-Osservatorio di Astrofisica e Scienza dello Spazio di Bologna, Via Piero Gobetti 93/3, 40129 Bologna, Italy\label{aff11}
\and
INFN-Sezione di Bologna, Viale Berti Pichat 6/2, 40127 Bologna, Italy\label{aff12}
\and
Finnish Centre for Astronomy with ESO (FINCA), Quantum, Vesilinnantie 5, University of Turku, 20014 Turku, Finland\label{aff13}
\and
Department of Physics, P.O. Box 64, University of Helsinki, 00014 Helsinki, Finland\label{aff14}
\and
INAF-IASF Milano, Via Alfonso Corti 12, 20133 Milano, Italy\label{aff15}
\and
Institut de Ci\`{e}ncies del Cosmos (ICCUB), Universitat de Barcelona (IEEC-UB), Mart\'{i} i Franqu\`{e}s 1, 08028 Barcelona, Spain\label{aff16}
\and
Instituci\'o Catalana de Recerca i Estudis Avan\c{c}ats (ICREA), Passeig de Llu\'{\i}s Companys 23, 08010 Barcelona, Spain\label{aff17}
\and
Institut de Ciencies de l'Espai (IEEC-CSIC), Campus UAB, Carrer de Can Magrans, s/n Cerdanyola del Vall\'es, 08193 Barcelona, Spain\label{aff18}
\and
CUNY Graduate Center, City University of New York, 365 5th Ave., New York, NY 10016, USA\label{aff19}
\and
Department of Physics and Astronomy, Lehman College of the CUNY, Bronx, NY 10468, USA\label{aff20}
\and
INAF, Istituto di Radioastronomia, Via Piero Gobetti 101, 40129 Bologna, Italy\label{aff21}
\and
Caltech/IPAC, 1200 E. California Blvd., Pasadena, CA 91125, USA\label{aff22}
\and
Laboratoire univers et particules de Montpellier, Universit\'e de Montpellier, CNRS, 34090 Montpellier, France\label{aff23}
\and
Dipartimento di Fisica e Scienze della Terra, Universit\`a degli Studi di Ferrara, Via Giuseppe Saragat 1, 44122 Ferrara, Italy\label{aff24}
\and
Department of Astronomy, University of Geneva, ch. d'Ecogia 16, 1290 Versoix, Switzerland\label{aff25}
\and
Aix-Marseille Universit\'e, CNRS, CNES, LAM, Marseille, France\label{aff26}
\and
Institut d'Astrophysique de Paris, UMR 7095, CNRS, and Sorbonne Universit\'e, 98 bis boulevard Arago, 75014 Paris, France\label{aff27}
\and
Kapteyn Astronomical Institute, University of Groningen, PO Box 800, 9700 AV Groningen, The Netherlands\label{aff28}
\and
Institute of Astronomy, University of Cambridge, Madingley Road, Cambridge CB3 0HA, UK\label{aff29}
\and
Kavli Institute for Cosmology Cambridge, Madingley Road, Cambridge, CB3 0HA, UK\label{aff30}
\and
Instituto de F\'isica de Cantabria, Edificio Juan Jord\'a, Avenida de los Castros, 39005 Santander, Spain\label{aff31}
\and
Laboratoire d'Astrophysique de Marseille, P\^{o}le de l'Etoile Site de Ch\^ateau-Gombert 38, rue Fr\'{e}d\'{e}ric Joliot-Curie, 13388 Marseille c\'{e}dex 13,  France\label{aff32}
\and
European Southern Observatory, Karl-Schwarzschild-Str.~2, 85748 Garching, Germany\label{aff33}
\and
Departamento de F\'isica, Universidad de Santiago de Chile, Av. Victor Jara 3493, Santiago, Chile\label{aff34}
\and
Center for Interdisciplinary Research in Astrophysics Space Science (CIRAS), Universidad de Santiago, Chile\label{aff35}
\and
Universit\'e de Gen\`eve, D\'epartement de Physique Th\'eorique and Centre for Astroparticle Physics, 24 quai Ernest-Ansermet, CH-1211 Gen\`eve 4, Switzerland\label{aff36}
\and
NASA Jet Propulsion Laboratory, California Institut of Technology - 4800 Oak Grove Drive - Pasadena, California 91109, USA\label{aff37}
\and
Dipartimento di Fisica e Astronomia, Universit\`a di Bologna, Via Gobetti 93/2, 40129 Bologna, Italy\label{aff38}
\and
University of Applied Sciences and Arts of Northwestern Switzerland, School of Computer Science, 5210 Windisch, Switzerland\label{aff39}
\and
Institut de Physique des 2 Infinis de Lyon, Domaine Scientifique de La Doua - 43, bd du 11 Novembre 1918, 69622 Villeurbanne c\'{e}dex, France\label{aff40}
\and
Department of Physics and Astronomy, University of British Columbia, Vancouver, BC V6T 1Z1, Canada\label{aff41}
\and
Department of Astronomy, School of Physics and Astronomy, Shanghai Jiao Tong University, Shanghai 200240, China\label{aff42}
\and
Shanghai Key Laboratory for Particle Physics and Cosmology, Shanghai 200240, China\label{aff43}
\and
Institute for Astronomy, University of Hawaii, 2680 Woodlawn Drive, Honolulu, HI 96822, USA\label{aff44}
\and
National Astronomical Observatory of Japan, 2-21-1 Osawa, Mitaka, Tokyo 181-8588, Japan\label{aff45}
\and
INAF-Osservatorio Astronomico di Capodimonte, Via Moiariello 16, 80131 Napoli, Italy\label{aff46}
\and
Max-Planck-Institut f\"{u}r Astrophysik, Karl-Schwarzschild-Str. 1 - 85741 Garching, Germany\label{aff47}
\and
ESAC/ESA, Camino Bajo del Castillo, s/n., Urb. Villafranca del Castillo, 28692 Villanueva de la Ca\~nada, Madrid, Spain\label{aff48}
\and
School of Mathematics and Physics, University of Surrey, Guildford, Surrey, GU2 7XH, UK\label{aff49}
\and
INAF-Osservatorio Astronomico di Brera, Via Brera 28, 20122 Milano, Italy\label{aff50}
\and
IFPU, Institute for Fundamental Physics of the Universe, via Beirut 2, 34151 Trieste, Italy\label{aff51}
\and
INAF-Osservatorio Astronomico di Trieste, Via G. B. Tiepolo 11, 34143 Trieste, Italy\label{aff52}
\and
INFN, Sezione di Trieste, Via Valerio 2, 34127 Trieste TS, Italy\label{aff53}
\and
SISSA, International School for Advanced Studies, Via Bonomea 265, 34136 Trieste TS, Italy\label{aff54}
\and
Dipartimento di Fisica, Universit\`a di Genova, Via Dodecaneso 33, 16146, Genova, Italy\label{aff55}
\and
INFN-Sezione di Genova, Via Dodecaneso 33, 16146, Genova, Italy\label{aff56}
\and
Department of Physics "E. Pancini", University Federico II, Via Cinthia 6, 80126, Napoli, Italy\label{aff57}
\and
Dipartimento di Fisica, Universit\`a degli Studi di Torino, Via P. Giuria 1, 10125 Torino, Italy\label{aff58}
\and
INFN-Sezione di Torino, Via P. Giuria 1, 10125 Torino, Italy\label{aff59}
\and
INAF-Osservatorio Astrofisico di Torino, Via Osservatorio 20, 10025 Pino Torinese (TO), Italy\label{aff60}
\and
INAF-Osservatorio Astronomico di Roma, Via Frascati 33, 00078 Monteporzio Catone, Italy\label{aff61}
\and
INFN-Sezione di Roma, Piazzale Aldo Moro, 2 - c/o Dipartimento di Fisica, Edificio G. Marconi, 00185 Roma, Italy\label{aff62}
\and
Centro de Investigaciones Energ\'eticas, Medioambientales y Tecnol\'ogicas (CIEMAT), Avenida Complutense 40, 28040 Madrid, Spain\label{aff63}
\and
Port d'Informaci\'{o} Cient\'{i}fica, Campus UAB, C. Albareda s/n, 08193 Bellaterra (Barcelona), Spain\label{aff64}
\and
INFN -- Sezione di Napoli, Via Cinthia 6, 80126, Napoli, Italy\label{aff65}
\and
Dipartimento di Fisica e Astronomia "Augusto Righi" - Alma Mater Studiorum Universit\`a di Bologna, Viale Berti Pichat 6/2, 40127 Bologna, Italy\label{aff66}
\and
Instituto de Astrof\'{\i}sica de Canarias, E-38205 La Laguna, Tenerife, Spain\label{aff67}
\and
Institute for Astronomy, University of Edinburgh, Royal Observatory, Blackford Hill, Edinburgh EH9 3HJ, UK\label{aff68}
\and
Jodrell Bank Centre for Astrophysics, Department of Physics and Astronomy, University of Manchester, Oxford Road, Manchester M13 9PL, UK\label{aff69}
\and
European Space Agency/ESRIN, Largo Galileo Galilei 1, 00044 Frascati, Roma, Italy\label{aff70}
\and
Universit\'e Lyon 1, CNRS/IN2P3, IP2I Lyon, Villeurbanne, France\label{aff71}
\and
Universit\'e Lyon 1, CNRS/IN2P3, IUF, IP2I Lyon, Villeurbanne, France \label{aff72}
\and
Mullard Space Science Laboratory, University College London, Holmbury St Mary, Dorking, Surrey RH5 6NT, UK\label{aff73}
\and
Universit\'e Paris-Saclay, CNRS, Institut d'astrophysique spatiale, 91405, Orsay, France\label{aff74}
\and
Aix-Marseille Universit\'e, CNRS/IN2P3, CPPM, Marseille, France\label{aff75}
\and
INAF-Istituto di Astrofisica e Planetologia Spaziali, via del Fosso del Cavaliere, 100, 00100 Roma, Italy\label{aff76}
\and
University Observatory, LMU Faculty of Physics, Scheinerstr.~1, 81679 Munich, Germany\label{aff77}
\and
FRACTAL S.L.N.E., calle Tulip\'an 2, Portal 13 1A, 28231, Las Rozas de Madrid, Spain\label{aff78}
\and
INAF-Osservatorio Astronomico di Padova, Via dell'Osservatorio 5, 35122 Padova, Italy\label{aff79}
\and
Dipartimento di Fisica "Aldo Pontremoli", Universit\`a degli Studi di Milano, Via Celoria 16, 20133 Milano, Italy\label{aff80}
\and
INFN-Sezione di Milano, Via Celoria 16, 20133 Milano, Italy\label{aff81}
\and
Institute of Theoretical Astrophysics, University of Oslo, P.O. Box 1029 Blindern, 0315 Oslo, Norway\label{aff82}
\and
Leiden Observatory, Leiden University, Einsteinweg 55, 2333 CC Leiden, The Netherlands\label{aff83}
\and
Jet Propulsion Laboratory, California Institute of Technology, 4800 Oak Grove Drive, Pasadena, CA, 91109, USA\label{aff84}
\and
Department of Physics, Lancaster University, Lancaster, LA1 4YB, UK\label{aff85}
\and
Felix Hormuth Engineering, Goethestr. 17, 69181 Leimen, Germany\label{aff86}
\and
Technical University of Denmark, Elektrovej 327, 2800 Kgs. Lyngby, Denmark\label{aff87}
\and
Cosmic Dawn Center (DAWN), Denmark\label{aff88}
\and
Universit\'e PSL, Observatoire de Paris, Sorbonne Universit\'e, CNRS, LERMA, 75014, Paris, France\label{aff89}
\and
Universit\'e Paris-Cit\'e, 5 Rue Thomas Mann, 75013, Paris, France\label{aff90}
\and
Max-Planck-Institut f\"ur Astronomie, K\"onigstuhl 17, 69117 Heidelberg, Germany\label{aff91}
\and
NASA Goddard Space Flight Center, Greenbelt, MD 20771, USA\label{aff92}
\and
Helsinki Institute of Physics, Gustaf H{\"a}llstr{\"o}min katu 2, University of Helsinki, 00014 Helsinki, Finland\label{aff93}
\and
Laboratoire d'etude de l'Univers et des phenomenes eXtremes, Observatoire de Paris, Universit\'e PSL, Sorbonne Universit\'e, CNRS, 92190 Meudon, France\label{aff94}
\and
SKAO, Jodrell Bank, Lower Withington, Macclesfield SK11 9FT, UK\label{aff95}
\and
Centre de Calcul de l'IN2P3/CNRS, 21 avenue Pierre de Coubertin 69627 Villeurbanne Cedex, France\label{aff96}
\and
Universit\"at Bonn, Argelander-Institut f\"ur Astronomie, Auf dem H\"ugel 71, 53121 Bonn, Germany\label{aff97}
\and
Department of Physics, Institute for Computational Cosmology, Durham University, South Road, Durham, DH1 3LE, UK\label{aff98}
\and
Universit\'e Paris Cit\'e, CNRS, Astroparticule et Cosmologie, 75013 Paris, France\label{aff99}
\and
CNRS-UCB International Research Laboratory, Centre Pierre Bin\'etruy, IRL2007, CPB-IN2P3, Berkeley, USA\label{aff100}
\and
Institute of Physics, Laboratory of Astrophysics, Ecole Polytechnique F\'ed\'erale de Lausanne (EPFL), Observatoire de Sauverny, 1290 Versoix, Switzerland\label{aff101}
\and
Dipartimento di Fisica - Sezione di Astronomia, Universit\`a di Trieste, Via Tiepolo 11, 34131 Trieste, Italy\label{aff102}
\and
Telespazio UK S.L. for European Space Agency (ESA), Camino bajo del Castillo, s/n, Urbanizacion Villafranca del Castillo, Villanueva de la Ca\~nada, 28692 Madrid, Spain\label{aff103}
\and
Institut de F\'{i}sica d'Altes Energies (IFAE), The Barcelona Institute of Science and Technology, Campus UAB, 08193 Bellaterra (Barcelona), Spain\label{aff104}
\and
European Space Agency/ESTEC, Keplerlaan 1, 2201 AZ Noordwijk, The Netherlands\label{aff105}
\and
Waterloo Centre for Astrophysics, University of Waterloo, Waterloo, Ontario N2L 3G1, Canada\label{aff106}
\and
Department of Physics and Astronomy, University of Waterloo, Waterloo, Ontario N2L 3G1, Canada\label{aff107}
\and
Perimeter Institute for Theoretical Physics, Waterloo, Ontario N2L 2Y5, Canada\label{aff108}
\and
Space Science Data Center, Italian Space Agency, via del Politecnico snc, 00133 Roma, Italy\label{aff109}
\and
Institute of Space Science, Str. Atomistilor, nr. 409 M\u{a}gurele, Ilfov, 077125, Romania\label{aff110}
\and
Dipartimento di Fisica e Astronomia "G. Galilei", Universit\`a di Padova, Via Marzolo 8, 35131 Padova, Italy\label{aff111}
\and
INFN-Padova, Via Marzolo 8, 35131 Padova, Italy\label{aff112}
\and
Instituto de F\'isica Te\'orica UAM-CSIC, Campus de Cantoblanco, 28049 Madrid, Spain\label{aff113}
\and
Institut de Recherche en Astrophysique et Plan\'etologie (IRAP), Universit\'e de Toulouse, CNRS, UPS, CNES, 14 Av. Edouard Belin, 31400 Toulouse, France\label{aff114}
\and
Universit\'e St Joseph; Faculty of Sciences, Beirut, Lebanon\label{aff115}
\and
Departamento de F\'isica, FCFM, Universidad de Chile, Blanco Encalada 2008, Santiago, Chile\label{aff116}
\and
Universit\"at Innsbruck, Institut f\"ur Astro- und Teilchenphysik, Technikerstr. 25/8, 6020 Innsbruck, Austria\label{aff117}
\and
Department of Physics and Helsinki Institute of Physics, Gustaf H\"allstr\"omin katu 2, University of Helsinki, 00014 Helsinki, Finland\label{aff118}
\and
Centre for Electronic Imaging, Open University, Walton Hall, Milton Keynes, MK7~6AA, UK\label{aff119}
\and
Departamento de F\'isica, Faculdade de Ci\^encias, Universidade de Lisboa, Edif\'icio C8, Campo Grande, PT1749-016 Lisboa, Portugal\label{aff120}
\and
Instituto de Astrof\'isica e Ci\^encias do Espa\c{c}o, Faculdade de Ci\^encias, Universidade de Lisboa, Tapada da Ajuda, 1349-018 Lisboa, Portugal\label{aff121}
\and
Cosmic Dawn Center (DAWN)\label{aff122}
\and
Niels Bohr Institute, University of Copenhagen, Jagtvej 128, 2200 Copenhagen, Denmark\label{aff123}
\and
Universidad Polit\'ecnica de Cartagena, Departamento de Electr\'onica y Tecnolog\'ia de Computadoras,  Plaza del Hospital 1, 30202 Cartagena, Spain\label{aff124}
\and
European University of Technology EUt+, European Union\label{aff125}
\and
Centre for Information Technology, University of Groningen, P.O. Box 11044, 9700 CA Groningen, The Netherlands\label{aff126}
\and
Institute of Space Sciences (ICE, CSIC), Campus UAB, Carrer de Can Magrans, s/n, 08193 Barcelona, Spain\label{aff127}
\and
Institut d'Estudis Espacials de Catalunya (IEEC),  Edifici RDIT, Campus UPC, 08860 Castelldefels, Barcelona, Spain\label{aff128}
\and
Istituto Nazionale di Fisica Nucleare, Sezione di Ferrara, Via Giuseppe Saragat 1, 44122 Ferrara, Italy\label{aff129}
\and
ICSC - Centro Nazionale di Ricerca in High Performance Computing, Big Data e Quantum Computing, Via Magnanelli 2, Bologna, Italy\label{aff130}
\and
INFN-Bologna, Via Irnerio 46, 40126 Bologna, Italy\label{aff131}
\and
Astronomical Observatory of the Autonomous Region of the Aosta Valley (OAVdA), Loc. Lignan 39, I-11020, Nus (Aosta Valley), Italy\label{aff132}
\and
Universit\'e C\^{o}te d'Azur, Observatoire de la C\^{o}te d'Azur, CNRS, Laboratoire Lagrange, Bd de l'Observatoire, CS 34229, 06304 Nice cedex 4, France\label{aff133}
\and
Department of Mathematical Physics, Institute of Physics, University of S\~ao Paulo, R. do Mat\~ao 1371, 05508-090, S\~ao Paulo, SP, Brazil\label{aff134}
\and
Universidad de La Laguna, Dpto. Astrof\'\i sica, E-38206 La Laguna, Tenerife, Spain\label{aff135}
\and
Oskar Klein Centre for Cosmoparticle Physics, Department of Physics, Stockholm University, Stockholm, SE-106 91, Sweden\label{aff136}
\and
Astrophysics Group, Blackett Laboratory, Imperial College London, London SW7 2AZ, UK\label{aff137}
\and
Univ. Grenoble Alpes, CNRS, Grenoble INP, LPSC-IN2P3, 53, Avenue des Martyrs, 38000 Grenoble, France\label{aff138}
\and
Dipartimento di Fisica, Sapienza Universit\`a di Roma, Piazzale Aldo Moro 2, 00185 Roma, Italy\label{aff139}
\and
Department of Mathematics and Physics E. De Giorgi, University of Salento, Via per Arnesano, CP-I93, 73100, Lecce, Italy\label{aff140}
\and
INFN, Sezione di Lecce, Via per Arnesano, CP-193, 73100, Lecce, Italy\label{aff141}
\and
INAF-Sezione di Lecce, c/o Dipartimento Matematica e Fisica, Via per Arnesano, 73100, Lecce, Italy\label{aff142}
\and
Institut d'Astrophysique de Paris, 98bis Boulevard Arago, 75014, Paris, France\label{aff143}
\and
ICL, Junia, Universit\'e Catholique de Lille, LITL, 59000 Lille, France\label{aff144}
\and
CERCA/ISO, Department of Physics, Case Western Reserve University, 10900 Euclid Avenue, Cleveland, OH 44106, USA\label{aff145}
\and
Laboratoire Univers et Th\'eorie, Observatoire de Paris, Universit\'e PSL, Universit\'e Paris Cit\'e, CNRS, 92190 Meudon, France\label{aff146}
\and
Departamento de F{\'\i}sica Fundamental. Universidad de Salamanca. Plaza de la Merced s/n. 37008 Salamanca, Spain\label{aff147}
\and
Universit\'e de Strasbourg, CNRS, Observatoire astronomique de Strasbourg, UMR 7550, 67000 Strasbourg, France\label{aff148}
\and
Center for Data-Driven Discovery, Kavli IPMU (WPI), UTIAS, The University of Tokyo, Kashiwa, Chiba 277-8583, Japan\label{aff149}
\and
California Institute of Technology, 1200 E California Blvd, Pasadena, CA 91125, USA\label{aff150}
\and
Department of Physics \& Astronomy, University of California Irvine, Irvine CA 92697, USA\label{aff151}
\and
Departamento F\'isica Aplicada, Universidad Polit\'ecnica de Cartagena, Campus Muralla del Mar, 30202 Cartagena, Murcia, Spain\label{aff152}
\and
Department of Computer Science, Aalto University, PO Box 15400, Espoo, FI-00 076, Finland\label{aff153}
\and
School of Physics and Astronomy, University of Nottingham, University Park, Nottingham NG7 2RD, UK\label{aff154}
\and
Ruhr University Bochum, Faculty of Physics and Astronomy, Astronomical Institute (AIRUB), German Centre for Cosmological Lensing (GCCL), 44780 Bochum, Germany\label{aff155}
\and
DARK, Niels Bohr Institute, University of Copenhagen, Jagtvej 155, 2200 Copenhagen, Denmark\label{aff156}
\and
Department of Physics, Centre for Extragalactic Astronomy, Durham University, South Road, Durham, DH1 3LE, UK\label{aff157}
\and
Astrophysics Research Centre, University of KwaZulu-Natal, Westville Campus, Durban 4041, South Africa\label{aff158}
\and
School of Mathematics, Statistics \& Computer Science, University of KwaZulu-Natal, Westville Campus, Durban 4041, South Africa\label{aff159}
\and
Department of Physics and Astronomy, Vesilinnantie 5, University of Turku, 20014 Turku, Finland\label{aff160}
\and
Serco for European Space Agency (ESA), Camino bajo del Castillo, s/n, Urbanizacion Villafranca del Castillo, Villanueva de la Ca\~nada, 28692 Madrid, Spain\label{aff161}
\and
ARC Centre of Excellence for Dark Matter Particle Physics, Melbourne, Australia\label{aff162}
\and
Centre for Astrophysics \& Supercomputing, Swinburne University of Technology,  Hawthorn, Victoria 3122, Australia\label{aff163}
\and
Department of Physics and Astronomy, University of the Western Cape, Bellville, Cape Town, 7535, South Africa\label{aff164}
\and
MTA-CSFK Lend\"ulet Large-Scale Structure Research Group, Konkoly-Thege Mikl\'os \'ut 15-17, H-1121 Budapest, Hungary\label{aff165}
\and
Konkoly Observatory, HUN-REN CSFK, MTA Centre of Excellence, Budapest, Konkoly Thege Mikl\'os {\'u}t 15-17. H-1121, Hungary\label{aff166}
\and
Institute for Theoretical Particle Physics and Cosmology (TTK), RWTH Aachen University, 52056 Aachen, Germany\label{aff167}
\and
INAF-Osservatorio Astrofisico di Arcetri, Largo E. Fermi 5, 50125, Firenze, Italy\label{aff168}
\and
Centro de Astrof\'{\i}sica da Universidade do Porto, Rua das Estrelas, 4150-762 Porto, Portugal\label{aff169}
\and
Instituto de Astrof\'isica e Ci\^encias do Espa\c{c}o, Universidade do Porto, CAUP, Rua das Estrelas, PT4150-762 Porto, Portugal\label{aff170}
\and
Dipartimento di Fisica, Universit\`a di Roma Tor Vergata, Via della Ricerca Scientifica 1, Roma, Italy\label{aff171}
\and
INFN, Sezione di Roma 2, Via della Ricerca Scientifica 1, Roma, Italy\label{aff172}
\and
HE Space for European Space Agency (ESA), Camino bajo del Castillo, s/n, Urbanizacion Villafranca del Castillo, Villanueva de la Ca\~nada, 28692 Madrid, Spain\label{aff173}
\and
Center for Frontier Science, Chiba University, 1-33 Yayoi-cho, Inage-ku, Chiba 263-8522, Japan\label{aff174}
\and
Department of Physics, Graduate School of Science, Chiba University, 1-33 Yayoi-Cho, Inage-Ku, Chiba 263-8522, Japan\label{aff175}
\and
Instituto de Astrof\'isica e Ci\^encias do Espa\c{c}o, Faculdade de Ci\^encias, Universidade de Lisboa, Campo Grande, 1749-016 Lisboa, Portugal\label{aff176}
\and
Department of Astrophysics, University of Zurich, Winterthurerstrasse 190, 8057 Zurich, Switzerland\label{aff177}
\and
Universit\'e Paris-Saclay, Universit\'e Paris Cit\'e, CEA, CNRS, AIM, 91191, Gif-sur-Yvette, France\label{aff178}
\and
Theoretical astrophysics, Department of Physics and Astronomy, Uppsala University, Box 524, 751 20 Uppsala, Sweden\label{aff179}
\and
Mathematical Institute, University of Leiden, Einsteinweg 55, 2333 CA Leiden, The Netherlands\label{aff180}
\and
School of Physical Sciences, The Open University, Milton Keynes, MK7 6AA, UK\label{aff181}
\and
Center for Astrophysics and Cosmology, University of Nova Gorica, Nova Gorica, Slovenia\label{aff182}
\and
Department of Astrophysical Sciences, Peyton Hall, Princeton University, Princeton, NJ 08544, USA\label{aff183}
\and
Fakult\"at f\"ur Physik, Universit\"at Bielefeld, Postfach 100131, 33501 Bielefeld, Germany\label{aff184}
\and
American Museum of Natural History, Department of Astrophysics, New York, NY 10024, USA\label{aff185}
\and
SRON Netherlands Institute for Space Research, Landleven 12, 9747 AD, Groningen, The Netherlands\label{aff186}}

\authorrunning{Euclid Collaboration: Nightingale et al.}

\date{}

\abstract{
\inspectionchange{Galaxy-scale strong lenses provide precise measurements of galaxy mass distributions,
magnified views of distant galaxies, and a competitive cosmological probe. Historically, such studies
have been limited to hundreds of lenses, with automated modelling methods proving difficult to scale.
We present automated lens modelling of 15\,072 strong-lens candidates from Euclid Data Release 1
(DR1), over an order of magnitude larger than any previously modelled sample. Using \texttt{PyAutoLens-JAX},
we perform three fits with multi-Gaussian expansion (MGE), pixelised, and S\'ersic representations
of the source, adopting an isothermal ellipsoid mass model with external shear. These fits provide
foreground lens-light subtraction, adaptive Delaunay source reconstructions, and deblended photometry.
More than 20 expert inspectors visually assess the resulting models, \internalchange{and each model is assigned the
classification on which at least $75\%$ of its inspectors agree. Of \stylechange{the 6073 grade A (secure)} lens
candidates, inspectors classify 4098 ($67.5\%$) as successfully modelled, 580 ($9.6\%$) as failures,
317 ($5.2\%$) as multi-galaxy lenses, 15 ($0.2\%$) as double-source-plane lenses\stylechange{,} and 7 ($0.1\%$) as
unsure if lens, while 1032 ($17.0\%$) reach no consensus and 24 ($0.4\%$) have incomplete models.} For the \stylechange{8999 grade B (probable)} candidates,
the modelling success rate falls to $43.1\%$, primarily reflecting an increase in cases without
consensus ($37.6\%$), rather than a comparable increase in agreed failures ($12.9\%$). We attribute
this greater disagreement to uncertainty over whether these candidates are genuine lenses. For 2990
candidates, inspectors also \stylechange{record} why models failed. Of the 210 failures, they attribute 39
($18.6\%$) to preprocessing and segmentation errors, 66 ($31.4\%$) to inaccurate MGE lens-light
subtraction\stylechange{,} and 12 ($5.7\%$) to source-reconstruction or mass-model failures, while 93
($44.3\%$) reach no consensus on the failure mode. Successfully modelled
grade A and B systems have a median effective Einstein radius of \ang{;;0.85} and a median total
\IE-band source magnification of $8.1$. We publicly release the successful lens models, their inferred
properties, and magnified-source reconstructions through the ECLIPSE catalogue, demonstrating
automated modelling at \Euclid sample sizes and enabling population-level strong-lensing science.}
}


\begin{document}

\maketitle

\section{Introduction}\label{sec:intro}

Galaxy-scale strong gravitational lensing occurs when a foreground galaxy
aligns closely along the line of sight with a more distant background source.
The mass of the foreground galaxy curves space-time, deflecting and
magnifying the light of the source into multiple images, arcs, or, in the most
aligned configurations, complete Einstein rings \reviewchange{\citep{Kochanek2004a,Treu2010}}. Each such system encodes an
unusually rich set of astrophysical measurements: the total mass and dark
matter distribution of the foreground galaxy \citep{Bolton2008,Vegetti2014,Oguri2014,Sonnenfeld2015}, and a magnified view of a
high-redshift source that would otherwise be inaccessible \reviewchange{\citep{Swinbank2015,Atek2015,Dye2017,Rizzo2020,Amvrosiadis2025}}.
\reviewchange{Strong lenses also provide cosmological constraints on the Hubble constant through time delays
\citep{Suyu2016,Wong2019,Birrer2020}, the matter density and dark energy through double-source-plane
systems \citep{Collett2014,Sahu2025DSPL,EuclidQ1DSPL}, and cosmic shear through line-of-sight
perturbations \citep{Birrer2017LOS,Birrer2018Shear,Hogg2026Shear,Fleury2026Shear}, for which \Euclid's \stylechange{upcoming} weak-lensing
catalogue further enables cross-correlations with galaxy shapes and tests \stylechange{for systematic errors in shear measurements} \citep{Hogg2023Shear,Fleury2026Shear}.}

Extracting that information requires strong lens modelling, a
simultaneous analysis of the light of the foreground lens galaxy, its mass
distribution, and the unlensed surface brightness of the background source.
Conceptually, a model of the light of the source is forward ray-traced via general
relativity, with its path distorted by a mass model of the foreground lens
galaxy. This produces an image of the lensed source, which is
added to a model image of the light of the foreground lens and compared to the data.
Lens modelling puts this process in an inference procedure, varying the model
parameters to determine the combination of lens light, mass, and source light
that best reproduces the observed strong lens system
\reviewchange{\citep[e.g.][]{Warren2003,Suyu2006,Vegetti2009,Nightingale2018,Birrer2018a,VernardosKoopmans2022,Galan2022}; see also the review by \citet{Shajib2024Review}.}

As strong lens samples have grown, scaling lens modelling has posed a significant challenge. Dedicated strong lens
surveys discovered hundreds of lenses \reviewchange{\citep{Bolton2008,Treu2011,Brownstein2012,Gavazzi2014,Shu2016,Tran2022}}, with modelling performed successfully but typically on a lens-by-lens basis
using procedures that relied heavily on manual human intervention. Automated and uniform modelling pipelines \stylechange{have since been}
introduced \citep{Nightingale2018,Nightingale2021} and demonstrated on real
data \citep{Shajib2018,Etherington2022,Schuldt2023HOLISMOKESX}, although these relied on manual preprocessing steps that \stylechange{have only recently been}
automated \citep{Shajib2025dolphin,Vincken2026}.
\reviewchange{Automated modelling has also been demonstrated for
lensed quasars \citep{Ertl2023TDCOSMO, Schmidt2023STRIDES}.} Despite the availability of samples containing hundreds
of strong lenses, automated modelling studies \stylechange{have} generally \stylechange{been} demonstrated on fewer than about 100 systems,
highlighting the difficulty of scaling lens modelling to survey-sized data sets. \stylechange{\citet{Nightingale2025COWLS} present lens} models
for \reviewchange{419} strong lens candidates discovered with the \textit{James Webb} Space Telescope
\reviewchange{as part of the COSMOS-Web Lens Survey} \stylechange{(COWLS)}.

The \Euclid mission \reviewchange{\citep{Scaramella-EP1,EuclidSkyOverview}} will provide high-resolution
\reviewchange{imaging from its visible instrument (VIS)}, with a pixel scale
of \ang{;;0.1} and point-spread function (PSF) full width at half maximum (FWHM) of approximately \ang{;;0.16} \reviewchange{\citep{EuclidSkyVIS,DR1-TP002}}, across $14\,000\,\deg^2$ of sky, alongside three
near-infrared bands \reviewchange{from its Near-Infrared Spectrometer and Photometer (NISP)}
that provide crucial colour information \citep{EuclidSkyNISP}.
This combination of image quality and survey scale is expected to yield in excess of $100\,000$ galaxy-scale strong
lenses \citep{Collett2015}, transitioning strong lensing into the era of large statistical samples.
\reviewchange{The precise number depends on the lens and source populations assumed and on how lenses
are identified \citep{Holloway2023, Ferrami2024}.} The VIS
imaging is sufficiently resolved to measure key lensing quantities directly from the data, including Einstein radii,
lens-galaxy mass distributions, and the morphologies of lensed source galaxies.

Euclid DR1 (\citeauthor{DR1cite}),
covering approximately $2000\,\deg^2$, already represents a major step towards this goal and is expected to
contain more strong lenses than all previous samples combined. Automated modelling of \Euclid lenses has
already been demonstrated on the Early Release Observations \citep{AcevedoBarroso24,ORiordan25} and, at larger
scale, by the \stylechange{Strong Lens Discovery Engine} (SLDE) collaboration on approximately 500 grade
A and B lenses discovered in Q1 \reviewchange{\citep{EuclidQ1Overview,EuclidQ1LensModel,EuclidQ1LensML,EuclidQ1DSPL,EuclidQ1LensEnsemble,EuclidQ1LensBright}}.
\reviewchange{Grade A denotes a secure lens with unambiguous lensing features, and grade B a
probable lens requiring additional evidence to confirm its lensing nature. In the ECLIPSE
catalogue and discovery pipeline \citep{DR1-SL01,DR1-SL02}, both grades are assigned solely
from visual inspection of the imaging, before lens modelling.}

This paper presents strong lens modelling results for \currentlenssample{} \reviewchange{grade A and B strong-lens candidates} discovered
across the Euclid DR1 footprint as part of the ECLIPSE survey \citep{DR1-SL01}, the first end-to-end
strong-lens discovery and characterisation effort at this scale. In doing so, we address a long-standing question
in strong lensing: can strong lens modelling be scaled beyond thousands of lenses? To quantify
how often modelling succeeds at this scale, the Euclid Strong Lensing Science Working Group (SL-SWG) performed a dedicated
visual inspection campaign involving over 20 expert lens modellers, with each inspected model
assessed by at least four inspectors. This enables us to quantify the success rate of automated lens modelling, provide a catalogue of validated
lens models for future scientific study,\footnote{\url{https://euclid-eclipse-lenses.org}} and systematically identify the dominant failure modes of automated lens modelling.

\reviewchange{For successfully modelled systems, we measure the principal lens and source properties to characterise the sample and provide a foundation for subsequent scientific studies.}
We first investigate the inferred Einstein radii which, when lens and source redshifts are known, can be converted
into Einstein masses, widely regarded as among the most precise galaxy mass measurements in astronomy
\reviewchange{\citep[see reviews by][]{Treu2010,Shajib2024Review}}, with
\reviewchange{systematic uncertainties of order $5\%$ reported in previous work \citep{Bolton2008}. Here we characterise the distribution of inferred Einstein radii, providing a basis for future population studies \citep{Sonnenfeld2021a,Sonnenfeld2021,Sonnenfeld2022}.} We also show the inferred source magnifications and make the source
reconstructions for every successfully modelled lens publicly available across the high-resolution VIS imaging and
all available NISP and \reviewchange{external (EXT) ground-based} wavebands. \reviewchange{Lens modelling of galaxy-scale strong lenses has enabled a broad range of
source-galaxy science \citep{Swinbank2015,Dye2017,Rizzo2020,Amvrosiadis2025}, and we aim to extend
this capability to large samples.}

This paper is part of a series describing the Euclid DR1 strong lens search and
catalogue. \citet[][hereafter ECLIPSE-A]{DR1-SL01} presents the catalogue of \currentlenssample{} grade A and B galaxy-scale
strong lenses discovered across the DR1 footprint, the largest homogeneous strong lens sample assembled to date.
The catalogue is the product of the discovery pipeline described by \citet[][hereafter ECLIPSE-B]{DR1-SL02}, which
combines machine-learning classification, citizen-science inspection, and expert visual validation to identify strong
lenses from approximately 68 million galaxies. Companion analyses of photometric redshifts and the physical properties of the
deflector galaxies are presented in \citet[][hereafter ECLIPSE-D]{DR1-SL04}.
\reviewchange{The automated segmentation and modelling inputs used here \stylechange{are} presented by
\citet{Vincken2026} through \stylechange{\texttt{AstroVink-TREX}}. This builds on the \stylechange{\texttt{AstroVink}} vision-transformer
lens-finding framework introduced for Euclid Q1 by \citet{Vincken2026AstroVink}.}
\reviewchange{Dedicated modelling of multi-galaxy lenses, in which more than one
foreground galaxy contributes to the lensing, is presented by \citet{Wielders2026MGL}.}

The remainder of this paper is organised as follows. \Cref{sec:data}
describes the Euclid DR1 imaging and strong-lens candidate sample.
\Cref{sec:preprocessing} sets out the automated preprocessing applied to each
lens prior to modelling. \Cref{sec:modeling} defines the lens model and the
\reviewchange{three-fit procedure}. \Cref{sec:visual} details the visual inspection
campaign, and \cref{sec:results} presents the catalogue together with the
success and failure statistics and the multi-wavelength results. We discuss
the implications in \cref{sec:discussion} and summarise in
\cref{sec:summary}.
\section{Data}\label{sec:data}

ECLIPSE-A describes how lens candidates come from a search of the \stylechange{approximately $2000\,\deg^2$} DR1 footprint. The primary data used for
lens modelling are the high-resolution images from the \Euclid VIS instrument \citep{EuclidSkyVIS}, observed
through the broad optical \IE filter. The VIS imaging provides the optimal data for lens modelling because of its high
spatial resolution and diffraction-limited image quality. The data were accessed via the European Space Agency (ESA) Science Archive Service
and ESA Datalabs.\footnote{\url{https://datalabs.esa.int/}} For each lens candidate, we extracted $100 \times 100$ pixel cutouts, corresponding
to \ang{;;10.0}\,$\times$\,\ang{;;10.0} at the VIS pixel scale of \ang{;;0.1}\,pixel$^{-1}$.
These cutouts \stylechange{included} the science image, root mean square (RMS) noise map, and PSF, all of which \stylechange{were} used in the lens modelling.

\reviewchange{We \stylechange{used} the effective PSFs supplied by the \Euclid MERge (MER) processing
pipeline, which propagates the input PSF models through resampling and stacking to provide
PSF stamps appropriate to each object's position and waveband \citep{Q1-TP004}.
For VIS, these combine the position-dependent PSF models of the contributing observations,
accounting for the mosaicing process. We \stylechange{did} not use the more detailed, colour-dependent
PSF treatment developed for the shear-measurement (SHE) processing function
\citep{Q1-TP002}. Using this PSF, we \stylechange{measured}
Einstein radii, lens-light properties, and source reconstructions, constrained by the
positions and appearance of the lensed images together with the foreground light distribution.
However, higher-fidelity PSFs may be required for more complex lens models, such as
power-law mass profiles that constrain the radial density slope. We recommend that readers
planning to fit more complex lens models to the public data first contact the SL-SWG for current guidance on PSF limitations and other
data-reduction issues. Improvements in these areas are expected following this work.}

\reviewchange{Following the VIS fit, lens modelling \stylechange{was} performed independently in each additional waveband using} the NISP near-infrared imaging
\citep{EuclidSkyNISP} and EXT ground-based data, for photometric-redshift estimation and spectral energy distribution (SED)
fitting. \reviewchange{The MER NISP and EXT images and their corresponding PSFs are sampled on the same
\ang{;;0.1}\,pixel$^{-1}$ grid as VIS; no additional PSF subsampling \stylechange{was} applied during lens modelling.}
\reviewchange{The multi-wavelength modelling approach and example NISP and EXT modelling results are presented in ECLIPSE-D.}
\section{Preprocessing}\label{sec:preprocessing}

Before lens modelling, each lens image \stylechange{was} preprocessed to define the inputs required by the
\stylechange{\Euclid strong-lens modelling pipeline}. These inputs are: (i) the circular mask, outside of which pixels are excluded
from the likelihood evaluation; (ii) \reviewchange{candidate conjugate points (image-plane positions at the brightest locations of the multiple
images, assumed to correspond to the same source-plane location)},
which are used during fitting to reject mass models that do not trace the images to consistent source-plane positions
\citep{Maresca2021}; and (iii) a binary mask flagging nearby galaxies whose emission overlaps
the fitting region and must therefore be removed. Historically, these tasks have been performed
manually by the scientist using one or more \reviewchange{graphical user interfaces (GUIs)}, as in \citet{Nightingale2025COWLS}.

\begin{figure}
    \centering
    \includegraphics[width=\columnwidth]{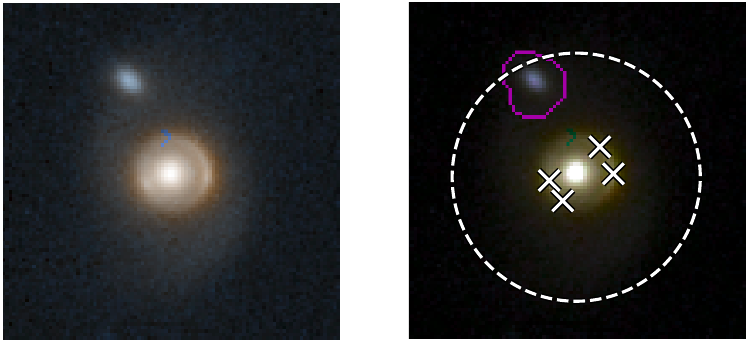}
    \caption{
        \reviewchange{Example results of automated preprocessing} for the lens
        Euclid-DR1\,J030313.592$-$580619.97. North is up and east is to the left, as in all
        images in this paper.
        \stylechange{\emph{Left}}: An RGB image using one colour scaling that contrasts the lens and source light
        with that of a nearby galaxy to the north-east. \stylechange{\emph{Right}}: An RGB image with different colour scaling that
        \reviewchange{shows the fitting mask, candidate conjugate points, and contaminating-galaxy contour used in lens modelling.} The white dashed
        circle shows the fitting mask, whose radius is computed from the furthest pixel in the
        segmented source emission, plus an additional buffer. \reviewchange{White crosses with black outlines} mark the automatically
        identified \reviewchange{candidate conjugate points}. The magenta contour marks a nearby contaminating galaxy whose
        light overlaps the fitting region and is therefore \reviewchange{effectively} excluded from the likelihood by zeroing
        its data values and inflating its RMS values.
    }
    \label{fig:segmentation-example}
\end{figure}

\reviewchange{To automate preprocessing for thousands of lenses, we \stylechange{used} the machine-learning framework
of \citet{Vincken2026}, which segments foreground lens light, lensed source emission, and nearby
galaxies (\cref{fig:segmentation-example}). The brightest pixel in the predicted lens-light map
\stylechange{defined} the fitting-mask centre, lens-light centre priors, and initial mass-model centre,
subsequently refined using the nearby VIS light. For the initial multi-Gaussian expansion (MGE) fit, the lens-light centre
\stylechange{had} a uniform prior spanning $\pm$\ang{;;0.1} in each coordinate around this refined centre.
The circular fitting mask \stylechange{extended} \ang{;;2.0}
beyond the segmented source emission, whose brightness peaks \stylechange{provided} candidate conjugate points.
Contaminating galaxies \stylechange{were} effectively excluded from the likelihood by zeroing their data values and
setting their RMS values to $10^8$ in the image's flux units, while retaining their pixels for ray tracing to avoid discontinuities in the
source-plane mesh \citep[\stylechange{section}~4.3.2]{Etherington2022}.}

\reviewchange{To constrain multiple images to originate from a common source, candidate conjugate
points \stylechange{were} ray-traced at each likelihood evaluation, penalising models whose maximum pairwise
source-plane separation \stylechange{exceeded} a specified threshold. Below this threshold the points \stylechange{did} not affect
the likelihood; above it, a penalty proportional to the excess separation \stylechange{was} subtracted from the
log-likelihood, with the threshold for the first fit set between \ang{;;0.3} and \ang{;;0.5} from a
pre-fit to the points (\cref{app:objective}). The candidate points taken from the segmentation
can occasionally miss images, for example when a counter-image lies close to the lens centre, and can
link peaks from different regions of the source plane. We therefore \stylechange{updated} them during the
\stylechange{\Euclid strong-lens modelling pipeline} (\cref{LensPipeline}).}

\section{Lens modelling}\label{sec:modeling}

\subsection{Overview}

We \stylechange{used} version \reviewchange{\texttt{2026.10.7.1}} of the lens modelling software
\texttt{PyAutoLens} \citep{Nightingale2021}.\footnote{\reviewchange{\url{https://github.com/PyAutoLabs/PyAutoLens}}}
\texttt{PyAutoLens} simultaneously fits the light and mass of the lens galaxy and the source galaxy. The
method assumes a model for the foreground light of the lens, which accounts for blurring by the
instrumental PSF and is subtracted from the observed image. A mass model ray-traces image pixels
from the image plane to the source plane, where either a parametric source model or a pixelised
source reconstruction on a Delaunay mesh is fitted. \reviewchange{Jupyter notebooks providing a
visual step-by-step guide to the \texttt{PyAutoLens} likelihood function used in this work are available online.}\footnote{\inspectionchange{\url{https://github.com/Jammy2211/autolens_likelihood_function}}}
\reviewchange{The \stylechange{\Euclid} strong-lens modelling pipeline used in this work, implemented in \texttt{PyAutoLens}, is publicly available.}\footnote{\reviewchange{\url{https://github.com/PyAutoLabs/euclid_strong_lens_modeling_pipeline}}}

\reviewchange{We \stylechange{fitted} the lens light, mass distribution, and background source simultaneously
using the semi-linear inversion framework \citep{Warren2003}. The model comprises 15, 10, and 12
non-linear free parameters in the MGE, pixelised, and S\'ersic fits (\cref{LensPipeline}),
describing the lens mass, external shear, light-profile geometry, and source regularisation, which \stylechange{were} sampled
using the nested sampler \texttt{nautilus} \citep{nautilus}. At each likelihood evaluation, the
foreground lens-light and background source intensities \stylechange{were} solved jointly as non-negative
linear parameters, so the potentially large number of Gaussian amplitudes and source-pixel
intensities need not be sampled non-linearly, substantially reducing the dimensionality of
the non-linear parameter space.}

\reviewchange{For accelerated inference, we \stylechange{used} \texttt{PyAutoLens-JAX}
(Nightingale et al., in preparation), which supports execution on both CPUs and
GPUs. We \stylechange{used} CPUs for this analysis because CPU hardware was more readily
available and inference time was not a limiting factor. For the full 15\,072-candidate
analysis, we also used \texttt{PyAutoLens-Assistant},\footnote{\reviewchange{\url{https://github.com/PyAutoLabs/autolens_assistant}}}
a natural-language interface that assists users in setting up and managing lens-modelling
workflows. Its development was informed by our experience modelling an initial subset of 2990 candidates from the \stylechange{Euclid} DR1 catalogue.}

\begin{figure*}
    \centering
    \includegraphics[width=\textwidth]{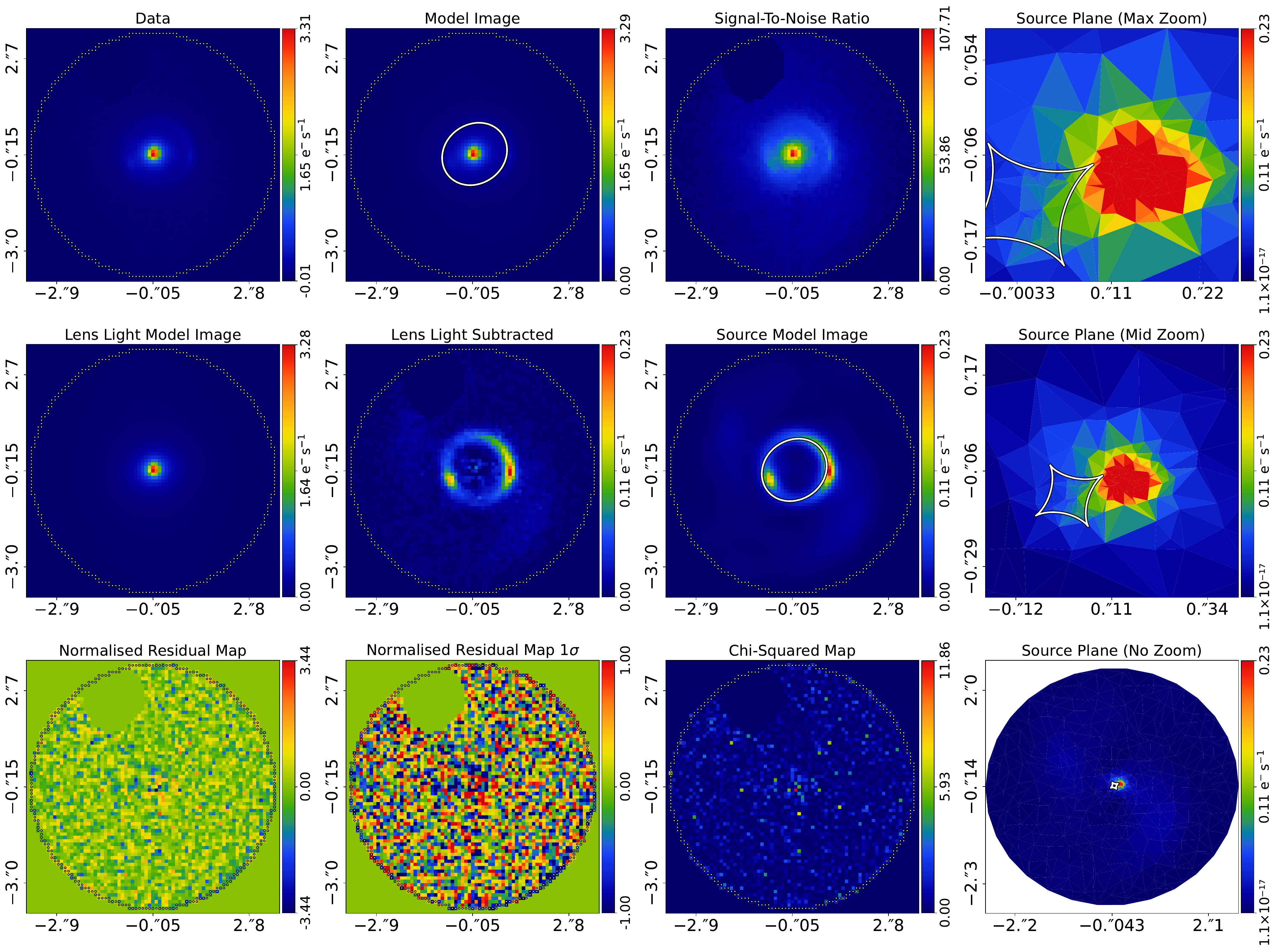}
    \caption{
        Example VIS \texttt{PyAutoLens} lens-model summary for the lens candidate
        J030313.592$-$580619.97, the same system shown in
        \cref{fig:segmentation-example}. The top row shows, from left to right, the VIS data, the
        full model image, the signal-to-noise \stylechange{ratio} map, and a maximally zoomed view of the reconstructed
        source plane. The middle row shows the foreground lens-light model, the lens-subtracted
        image, the lensed source model image, and a mid-zoom source-plane reconstruction. The bottom
        row shows the normalised residual map, the same residual map \reviewchange{plotted in the range from $-1\sigma$ to $+1\sigma$}, the
        $\chi^2$ map, and the full source-plane reconstruction. \reviewchange{The zeroed regions in the maps correspond}
        to the contaminating galaxy marked by the magenta
        contour in \cref{fig:segmentation-example}. \reviewchange{White dots with black edges mark the image-plane mask
        boundary, while white curves with black outlines show the critical curve and caustic structure where plotted.}
    }
    \label{fig:vis-pix-fit}
\end{figure*}

\Cref{fig:vis-pix-fit} shows an overview of a \texttt{PyAutoLens} lens model using the lens candidate
Euclid-DR1\,J030313.592$-$580619.97. The figure summarises the main
components of the \texttt{PyAutoLens} fit: the observed VIS image, the lens light and source model
images, the lens-subtracted lensed source, the fit residuals, and the reconstructed source plane
with different degrees of zooming. The mask used in this figure corresponds to the white circle in
\cref{fig:segmentation-example} and the extra galaxy marked by the magenta contour
in \cref{fig:segmentation-example} is visible in
\cref{fig:vis-pix-fit} as zeroed pixels in the VIS data panel and a corresponding region with zero
signal-to-noise \stylechange{ratio}. This masking removes the contaminating galaxy from the likelihood while retaining
the lensed source emission used to constrain the model. The white curves show the tangential critical curve in
the image plane and the corresponding caustic in the source plane.

\subsection{Lens light modelling}\label{LensLight}

\reviewchange{We \stylechange{modelled} the foreground lens light using an MGE
\citep{Cappellari2002}, comprising a sum of two-dimensional elliptical Gaussian profiles,
following \citet[][hereafter H24]{He2024}. The MGE provides a flexible and computationally
efficient representation of the lens galaxy emission, well suited to automated modelling of large
strong-lens samples. Appendix~\ref{app:likelihood} details the MGE equations, width convention,
and joint linear solution.}

\reviewchange{We \stylechange{modelled} the lens light using two sets of 20 Gaussians with logarithmically spaced
widths, sharing a common centre but with independent axis ratios and orientations, giving six
non-linear parameters. While H24 used up to six sets of 30 Gaussians for HST imaging, our tests on
\Euclid\ data found that two sets improved the fits over one, particularly for S0 galaxies with
distinct bulge and disc components. We therefore \stylechange{adopted} two sets as a compromise between flexibility
and computational efficiency, with visual inspection
(\cref{sec:visual}) used to identify inadequate lens-light models.}

\subsection{Mass profile}\label{MassProfiles}

For the lens mass model, we \stylechange{assumed} a singular isothermal ellipsoid (SIE) density profile of the form
\begin{equation}
\label{eqn:SPLEkap}
\kappa (\xi) = \frac{1}{1 + q^{\rm mass}} \bigg( \frac{\theta^{\rm mass}_\sfont{E}}{\xi} \bigg)\;,
\end{equation}
where $q^{\rm mass}$ is the mass-profile axis ratio,
$\xi = \sqrt{x^{\prime 2} + (y^\prime/q^{\rm mass})^2}$ is the elliptical radius in lens-centred
coordinates aligned with the mass distribution, and $\theta^{\rm mass}_\sfont{E}$ is the model
Einstein radius in arcseconds. Deflection angles \stylechange{were}
computed via an implementation of the method of \citet{Tessore2015} in \texttt{PyAutoLens}.

\reviewchange{We \stylechange{included} external shear, a standard approximation for tidal perturbations
from matter outside the main lens galaxy. It is described by a magnitude,
$\gamma^{\rm ext}$, and an orientation, $\phi^{\rm ext}$, measured anticlockwise from the
$+x$ axis of the modelling grid. In the fit we \stylechange{used} the two components
$(\gamma_1^{\rm ext},\gamma_2^{\rm ext})$, related to these quantities by
\begin{equation}
    \label{eq:shear}
    \gamma_1^{\rm ext}=\gamma^{\rm ext}\cos(2\phi^{\rm ext})\;,\qquad
    \gamma_2^{\rm ext}=\gamma^{\rm ext}\sin(2\phi^{\rm ext})\;.
\end{equation}
The deflection angles due to external shear \stylechange{were} computed analytically.
However, the fitted shear can also absorb missing complexity in the lens mass
distribution \citep{Etherington2023a}, so it need not represent a physical
measurement of the external tidal field.}

The catalogue reports the effective Einstein radius, denoted $\theta_\sfont{E}$ below, defined for
non-axisymmetric lenses as
\begin{equation}
    \label{eq:effective-einstein-radius}
    \theta_\sfont{E} = \sqrt{\frac{A_{\rm crit}}{\pi}}\;,
\end{equation}
where $A_{\rm crit}$ is the angular area enclosed by the tangential critical curve. Because this
area is measured from the full SIE plus external-shear model, the reported
$\theta_\sfont{E}$ includes the shear contribution to the critical-curve geometry and is not identical
to the SIE scale parameter $\theta^{\rm mass}_\sfont{E}$.

\subsection{Source model}\label{Source}

\reviewchange{We \stylechange{modelled} the background source using three representations: a multi-Gaussian
expansion \citep{Cappellari2002,He2024}, a pixelised Delaunay reconstruction, and an
elliptical S\'ersic profile. The pixelised reconstruction combines Hilbert-based adaptive mesh
placement \citep{He2025Jackpot}, barycentric interpolation \citep{Vegetti2009,VernardosKoopmans2022},
and adaptive split regularisation \citep{Nightingale2024,He2024}, whose strength \stylechange{was} inferred
alongside the lens mass parameters \citep{Warren2003,Suyu2006}. We \stylechange{evaluated} PSF-convolved models
using a Gaussian-noise likelihood, with spatially adaptive oversampling of the lens and source
light \citep{Nightingale2024}. Appendix~\ref{app:likelihood} provides the forward-model, joint
linear solution, adaptive mesh, regularisation, and likelihood equations.}

The source magnification,
$\mu$, is computed as the ratio of the lensed source flux in the image plane, $F_{\rm lensed}$, to
its intrinsic source-plane flux, $F_{\rm source}$,
\begin{equation}
    \label{eq:magnification}
    \mu = \frac{F_{\rm lensed}}{F_{\rm source}}\;.
\end{equation}
\reviewchange{The magnifications quoted in this paper use the S\'ersic source model to facilitate
comparison with previous literature and standard galaxy photometry (see ECLIPSE-D);
the catalogue also includes magnifications derived from the Delaunay source reconstruction.}

\subsection{Sequential lens model fits}\label{LensPipeline}

\reviewchange{The pipeline comprises three fits with complementary purposes. The first
\stylechange{used} MGE lens and source light to establish an initial mass model and foreground-light
subtraction. The second \stylechange{used} a flexible pixelised source to refine the mass model and
reconstruct the source morphology. The third \stylechange{held} the first-fit mass model fixed and
\stylechange{fitted} S\'ersic light profiles to obtain deblended photometry and structural measurements
of the lens and source. Appendix~\ref{app:likelihood} gives the likelihood equations,
and Table~\ref{tab:app-stages} summarises the free and fixed quantities in each stage.
The explicit priors and sampler settings are given in Appendix~\ref{app:priors}.
Each search \stylechange{stopped} on meeting the \texttt{nautilus} convergence criteria or reaching
its likelihood-evaluation limit; reaching that limit does not itself establish convergence.}

The first fit \stylechange{used} MGEs \citep{Cappellari2002, He2024} for both the
lens and source light, together with an SIE mass profile and external shear (see
\cref{MassProfiles}), and \stylechange{was} performed with the nested sampler
\texttt{nautilus} \citep{nautilus}.\footnote{\url{https://github.com/johannesulf/nautilus}} The
intensities of the MGE light profiles for the lens and source \stylechange{were} solved for linearly using the
semi-linear inversion. \reviewchange{For the source, we \stylechange{used} one set of 20 Gaussians, whose $\sigma$ values
\stylechange{spanned} a base-10 logarithmic range from \ang{;;0.0001} to the mask radius.}

\reviewchange{After inference with the simpler parametric MGE source model \stylechange{was} complete, the conjugate
points \stylechange{were} updated using its maximum-likelihood mass model, recovering missed images and replacing
peaks associated with different source regions with positions corresponding to a common source
location. The threshold used in the second fit \stylechange{was} then reset to three times the maximum source-plane
separation of the updated points under this mass model, with a minimum of \ang{;;0.2}
\citep[\stylechange{section}~4.3.2]{Etherington2022}.}

The second fit, also performed with \texttt{nautilus}, \stylechange{replaced} the parametric source model with a
Delaunay mesh source reconstruction, while retaining an MGE for the lens light and an SIE mass
profile with external shear. The non-linear parameters of the MGE lens light model \stylechange{were} fixed to the
values inferred in the first fit, but the Gaussian intensities \stylechange{were} re-solved linearly for every
model evaluated alongside the source reconstruction.
\reviewchange{The second fit \stylechange{released} the SIE centre within \ang{;;0.1} of the \stylechange{data set}
centre in each coordinate. Its SIE shape, Einstein-radius, and shear priors \stylechange{were} not
narrowed using the first-fit posterior; Table~\ref{tab:app-priors} specifies their
distributions, including the non-uniform priors on the SIE ellipticity components.}

The third fit \stylechange{was} performed to obtain deblended photometry of the lens and source galaxies. The
maximum-likelihood mass model inferred from the first fit \stylechange{was} fixed, and the MGE light profiles
of both the lens and source \stylechange{were} replaced with elliptical S\'ersic profiles. This fit \stylechange{was} performed
on the VIS imaging, where the structural parameters of the lens and source light profiles \stylechange{were} inferred.
Lens and source magnitudes \stylechange{were} computed from the total flux of the foreground S\'ersic lens model and the
lensed S\'ersic source model in the image plane, respectively. The source magnification reported in
the catalogue \stylechange{was} then computed using \cref{eq:magnification}, with the SIE mass model and external
shear fixed from the initial MGE fit.
In ECLIPSE-D, the same fixed mass model with S\'ersic lens and source \stylechange{was} then used to fit each available
NISP and EXT waveband independently for deblended photometry, photometric-redshift estimation, and SED fitting.

\section{Visual inspection}\label{sec:visual}

The goals of this visual inspection campaign \stylechange{were} (i) to produce a catalogue of galaxy-scale lenses whose models a majority of experts
agree are robust and accurate for subsequent scientific study; (ii) to quantify the overall success rate of the automated lens modelling
pipeline; and (iii) to understand and categorise the reasons why lens modelling fails, thereby guiding future automated lens modelling efforts.
\inspectionchange{More than 20 lens-modelling experts participated in the visual inspection campaign for the DR1 sample of \currentlenssample{} candidates.}

A preliminary round of visual inspection, described in Appendix~\ref{app:preliminary_visual_inspection}, was undertaken in which 12 expert
inspectors reviewed 200 lens models. The results of this exercise, together with follow-up discussion amongst the inspectors, were used to
refine the criteria defining a successful lens model, determine the categories used to describe modelling failures, and identify the diagnostic
images most informative for assessing model quality. This enabled the final inspection \reviewchange{GUI} used for \currentlenssample{} lenses to present the minimum
amount of information necessary to balance inspection speed with diagnostic power.

For the main visual inspection campaign, inspectors were shown two diagnostic images for each
candidate: (i) the RGB image with segmentation information, as shown in
\cref{fig:segmentation-example}, and (ii) the lens-model overview image shown in
\cref{fig:vis-pix-fit}. Together, these \stylechange{provided} the colour information needed to distinguish lens and
source emission, as well as the modelling diagnostics required to assess whether the data are
successfully fitted by a lens model. \reviewchange{\Cref{fig:vis-pix-fit} shows the Delaunay source
reconstruction shown to inspectors by default. Inspectors assigned classifications based solely on
the Delaunay modelling results. They could also switch in the GUI to a parametric MGE source model
for supplementary diagnostic information, but did not classify the MGE fits. S\'ersic fits were omitted because their close resemblance to the MGE fits would add redundant information and risk confusion.}

\reviewchange{The source-plane panels retain the zoom levels used during visual inspection
(\cref{fig:vis-pix-fit}), which are occasionally suboptimal (e.g. zoomed in too far to show the
full source). Each Delaunay triangle is displayed with a single colour to illustrate the mesh
resolution, although the underlying source reconstruction is continuous through piecewise-linear
barycentric interpolation (Appendix~\ref{app:mesh}).}

\subsection{Inspection classifications}\label{sec:inspection-classifications}

\reviewchange{The five primary classification options available to inspectors} \stylechange{were} ``Success'' (SC), ``Failure'' (FL), ``Unsure If Lens'' (UIL),
``Double Source Plane Lens'' (DSPL), and ``Multi Galaxy Lens'' (MGL). During preliminary inspection, it was
noted that the sample contained systems that violate the modelling assumptions in two common ways.
In a DSPL system, multiple sources at different redshifts are lensed by the same foreground
deflector; in an MGL system, more than one lens galaxy lies within the Einstein radius. The automated modelling pipeline assumes a single
MGE lens light model, a single SIE mass profile, and a single source plane; these systems are therefore not
expected to be modelled accurately. The visual inspection procedure therefore instructed reviewers
to identify DSPL and MGL systems as standalone categories in the initial classification stage.

Using the diagnostic images, inspectors assigned these classifications according to the following criteria:

\begin{enumerate}
    \item You have high confidence that the object is a genuine strong gravitational lens (otherwise classify it as UIL).
    \item The system does not contain sources at multiple redshifts (otherwise classify it as DSPL).
    \item The system contains only one lens galaxy within its Einstein radius (otherwise classify it as MGL).
    \item The lens model provides an accurate and reliable measurement of the Einstein radius suitable for publication in a peer-reviewed scientific article (otherwise classify it as FL).
    \item The critical-curve morphology and source reconstruction are physically plausible (otherwise classify it as FL).
    \item If all criteria above are satisfied, classify the system as SC.
\end{enumerate}

\inspectionchange{For a subset of 2990 lens candidates, inspectors were additionally asked to
identify the reason for unsuccessful lens models. When classifying a model as a failure (FL),
inspectors selected one of three categories: segmentation/masking (FL-Segmentation), lens-light
subtraction (FL-LensLight), or source reconstruction/mass modelling (FL-SourceMass). Inspectors
were instructed to consider these categories sequentially and select the first applicable failure,
reflecting how errors propagate through the modelling pipeline. For example, incorrect segmentation
can lead to poor lens-light subtraction, which can in turn cause the source reconstruction to fit
residual lens light rather than genuine source emission. The source reconstruction/mass modelling
category was reserved for cases where earlier stages were successful but the reconstructed source
or inferred mass model was physically implausible. This structured approach was introduced
following preliminary inspections, where less constrained classifications proved difficult to
interpret.}

\inspectionchange{The number of recorded votes \stylechange{varied} from four to seven per inspected object.
Detailed reasons for failure were not requested in the main campaign of all \currentlenssample{}
objects because of inspector time constraints.}
\reviewchange{Table~\ref{tab:classification-key} summarises the five primary classifications,
together with the outcome assigned when inspectors \stylechange{did} not reach consensus, defined in
\cref{sec:inspection-consensus-thresholds}.}

\begin{table*}
\reviewchange{
\caption{Key to the primary inspection classifications and the consensus outcome.}
\label{tab:classification-key}
\centering
\renewcommand{\arraystretch}{1.35}
\setlength{\tabcolsep}{6pt}
\begin{tabular}{lp{0.21\textwidth}p{0.62\textwidth}}
\hline\hline
Code & Classification & Description \\
\hline
SC & Success & Confident lens identification, a reliable Einstein radius, and physically plausible critical curves and source reconstruction; single-deflector, single-source-plane criteria satisfied. \\
FL & Failure & \stylechange{Confident} lens identification satisfying the single-deflector and single-source-plane criteria, but the model fails the reliability or physical-plausibility criteria. \\
UIL & Unsure If Lens & Insufficient confidence that the candidate is a genuine strong lens. \\
DSPL & Double Source Plane Lens & Multiple sources at different redshifts lensed by the same foreground deflector. \\
MGL & Multi Galaxy Lens & More than one lens galaxy within the Einstein radius. \\
\hline
NC & No Consensus & No unique primary classification satisfies the adopted consensus rule; assigned from the combined votes. \\
-- & \internalchange{Incomplete} & \internalchange{No completed pixelised model; not inspected.} \\
\hline
\end{tabular}
}
\end{table*}

\begin{figure*}
    \centering
    \includegraphics[width=0.96\textwidth]{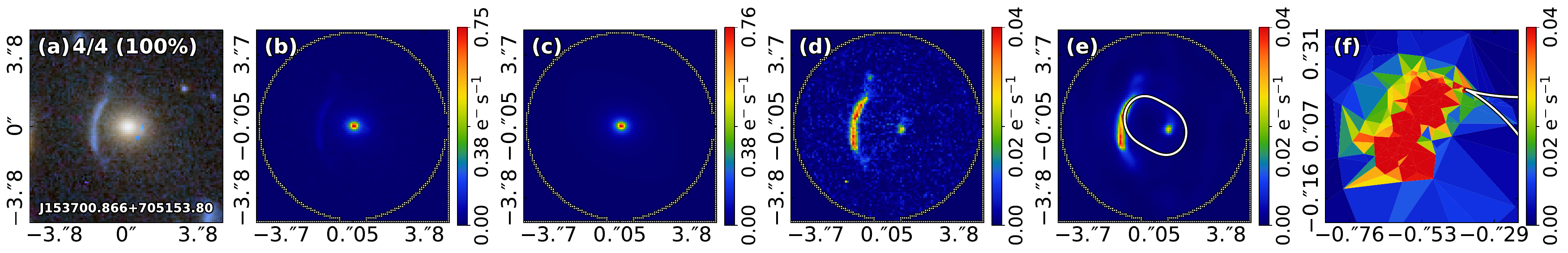}
    \includegraphics[width=0.96\textwidth]{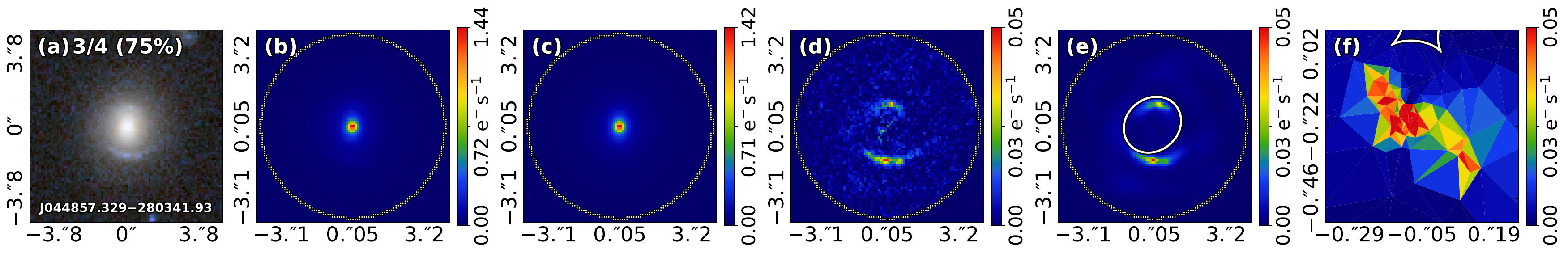}
    \includegraphics[width=0.96\textwidth]{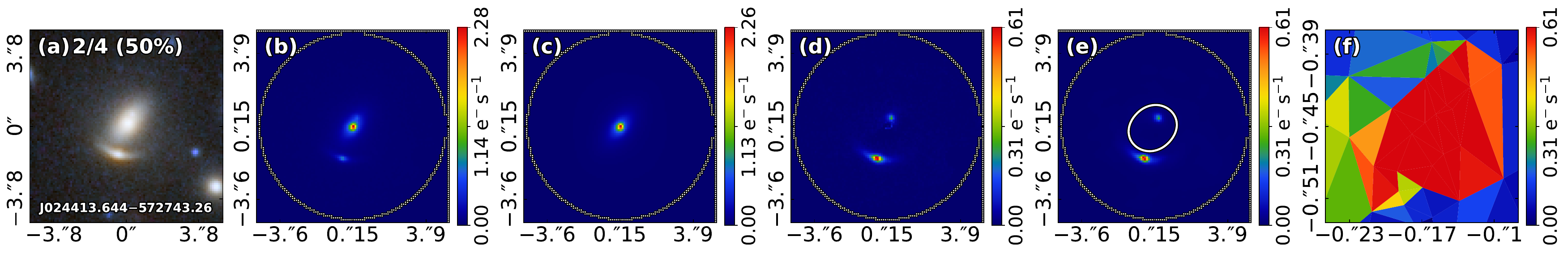}
    \caption{\inspectionchange{
        Representative grade A Success (SC) lens models spanning the observed SC-vote fractions.
        The \stylechange{systems in the} top, middle, and bottom rows \stylechange{have} $4/4$ ($100\%$), $3/4$ ($75\%$), and $2/4$ ($50\%$)
        SC votes, respectively; the bottom system is NC at the adopted $75\%$ threshold. Individual
        candidates were reviewed by subsets of the inspection team, so each RGB panel gives both the
        SC-vote tally and its fraction. Within each row, the panels show (a) the RGB image, (b) the VIS data,
        (c) the foreground lens-light model, (d) the lens-subtracted image, (e) the lensed source-model
        image, and (f) the source-plane reconstruction. Images are north up, east left; axes give
        offsets in arcseconds, with $y$ increasing to the north and $x$ to the west.
        White dots mark the fitting mask; critical curves and caustics
        have black outlines. The higher-agreement examples have plausible lens-light subtraction and
        source reconstructions, whereas the lower-agreement example illustrates the greater ambiguity
        introduced at lower consensus thresholds. Fourteen additional examples are shown in
        \cref{fig:success-consensus-additional-high}.
    }}
    \label{fig:success-consensus-quality}
\end{figure*}

\begin{figure*}
    \centering
    \includegraphics[width=\textwidth]{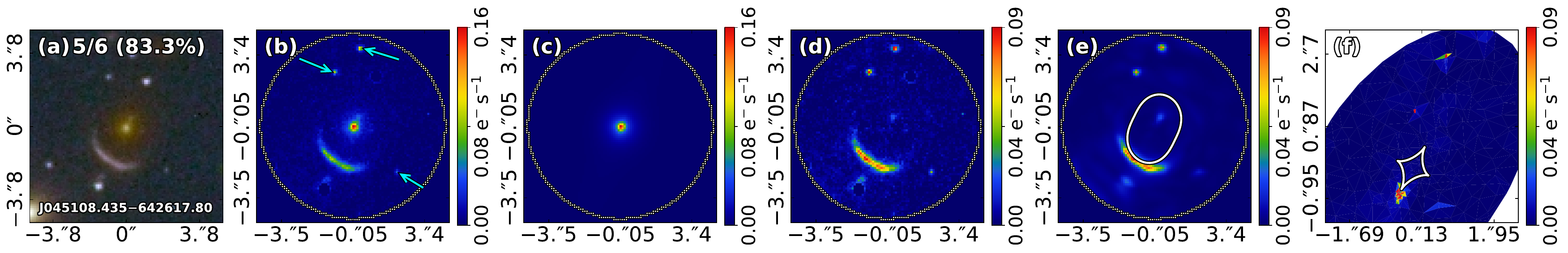}
    \includegraphics[width=\textwidth]{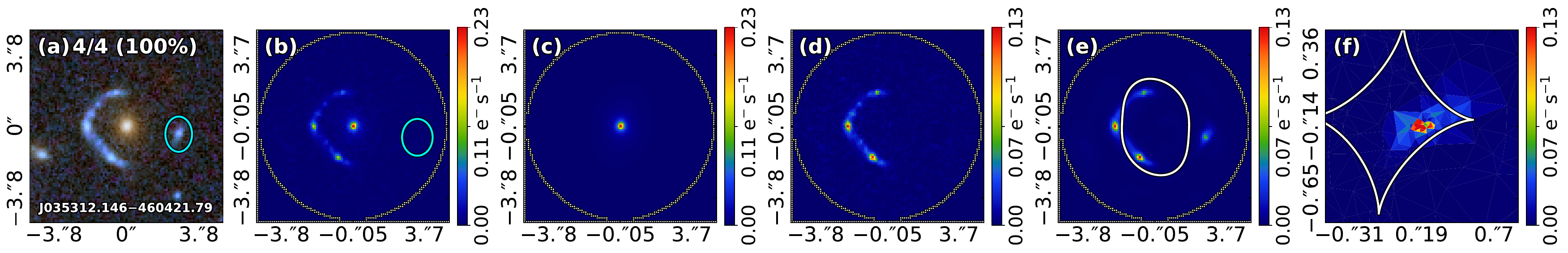}
    \caption{\inspectionchange{
        Grade A Success (SC) models above the $75\%$ SC consensus threshold despite residual
        segmentation issues. Panels follow the same layout as \cref{fig:success-consensus-quality}.
        In the top row, cyan arrows in panel (b) identify unmasked nearby objects whose light is
        mapped into the source reconstruction. In the bottom row, cyan ellipses in panels (a) and (b)
        highlight part of a genuine multiple image that is visible in the RGB image but incorrectly
        masked in the VIS data. The ellipses are visual guides rather than exact mask boundaries. The inferred
        Einstein radii remain reliable, but these issues limit the use of the models for detailed
        source reconstruction or more complex mass modelling.
    }}
    \label{fig:success-segmentation-caveats}
\end{figure*}

\begin{figure*}
    \centering
    \includegraphics[width=\textwidth]{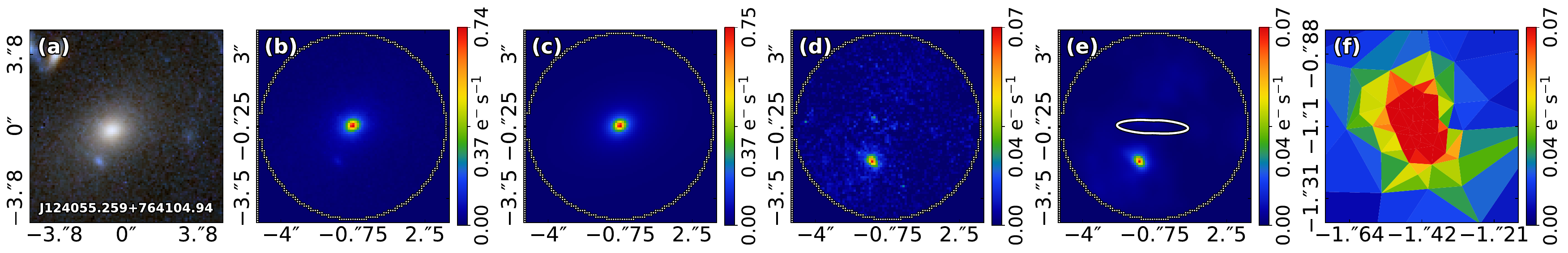}
    \includegraphics[width=\textwidth]{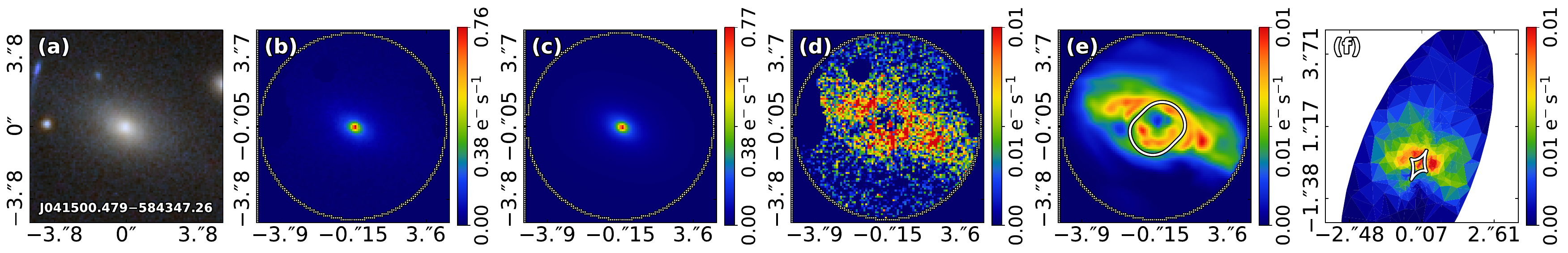}
    \caption{\inspectionchange{
        Two grade A Unsure If Lens (UIL) systems, each with $100\%$ UIL agreement in the
        inspection campaign. Panels follow the same layout as
        \cref{fig:success-consensus-quality}. From top to bottom, the examples illustrate a
        missing counter-image and faint residual emission with insufficient signal-to-noise
        ratio to establish a convincing lens.
    }}
    \label{fig:unsure-lens-examples}
\end{figure*}

\subsection{Consensus thresholds}\label{sec:inspection-consensus-thresholds}

\inspectionchange{All inspectors \stylechange{had} equal voting weight, with no individual veto. We \stylechange{assigned} a
consensus classification when at least a specified fraction of inspectors \stylechange{agreed}; otherwise, the
object \stylechange{was} labelled ``No Consensus'' (NC). This \stylechange{was} determined from the combined votes rather than being
an option available to individual inspectors. At our adopted $75\%$ threshold, three of four, four
of five, or five of six inspectors \stylechange{had to} agree. We \stylechange{applied} this threshold independently to the five
primary classifications (SC, FL, UIL, DSPL, and MGL) and the secondary failure categories
(FL-Segmentation, FL-LensLight, and FL-SourceMass) recorded for the 2990-candidate sample.
Primary classifications \stylechange{required} a unique consensus. At the $50\%$ threshold, ties between
classifications \stylechange{were} assigned NC. At the adopted $75\%$ threshold, no two primary classifications
can simultaneously meet the threshold. For secondary failure modes, ties \stylechange{were} instead resolved in the causal order
FL-Segmentation, FL-LensLight, then FL-SourceMass. No minimum number of votes \stylechange{was} imposed, while
objects without recorded votes \stylechange{received no classification but are included in the totals from which success rates are calculated}.}
\section{Results}\label{sec:results}

\Cref{tab:model-consensus-thresholds} reports the SC, UIL, DSPL, MGL, FL, and NC counts
\inspectionchange{for grades A and B separately} under thresholds of $100\%$, $75\%$, and $50\%$.
\internalchange{For 83 candidates, no completed pixelised model is
available; these were not inspected and are listed as incomplete.}
\begin{table*}
\centering
\caption{\inspectionchange{Primary inspection classifications at each consensus threshold for \internalchange{grade A ($N=6073$) and grade B ($N=8999$)}. Entries give counts and, in parentheses, percentages of all objects of the corresponding grade. \internalchange{Incomplete: candidates without a completed pixelised model, which were not inspected.} SC: Success; UIL: Unsure If Lens; DSPL: Double Source Plane Lens; MGL: Multi Galaxy Lens; FL: Failure; NC: No Consensus. Ties at $50\%$ are NC.}}
\label{tab:model-consensus-thresholds}
\begingroup\small
\renewcommand{\arraystretch}{1.2}\setlength{\tabcolsep}{2pt}
\begin{tabular*}{\textwidth}{@{\extracolsep{\fill}}ll@{\extracolsep{\fill}\hspace{4pt}}r@{\extracolsep{0pt}\,}l@{\extracolsep{\fill}\hspace{4pt}}r@{\extracolsep{0pt}\,}l@{\extracolsep{\fill}\hspace{4pt}}r@{\extracolsep{0pt}\,}l@{\extracolsep{\fill}\hspace{4pt}}r@{\extracolsep{0pt}\,}l@{\extracolsep{\fill}\hspace{4pt}}r@{\extracolsep{0pt}\,}l@{\extracolsep{\fill}\hspace{4pt}}r@{\extracolsep{0pt}\,}l@{\extracolsep{\fill}\hspace{4pt}}r@{\extracolsep{0pt}\,}l@{}}
\hline\hline
Grade & Threshold & \multicolumn{2}{c}{SC} & \multicolumn{2}{c}{UIL} & \multicolumn{2}{c}{DSPL} & \multicolumn{2}{c}{MGL} & \multicolumn{2}{c}{FL} & \multicolumn{2}{c}{NC} & \multicolumn{2}{c}{\internalchange{Incomplete}} \\
\hline
A & $100\%$ & 3011 & (49.6\%) & 3 & ($<$0.1\%) & 1 & ($<$0.1\%) & 172 & (2.8\%) & 272 & (4.5\%) & 2590 & (42.6\%) & 24 & (0.4\%) \\
 & $75\%$ & 4098 & (67.5\%) & 7 & (0.1\%) & 15 & (0.2\%) & 317 & (5.2\%) & 580 & (9.6\%) & 1032 & (17.0\%) & 24 & (0.4\%) \\
 & $50\%$ & 4411 & (72.6\%) & 24 & (0.4\%) & 25 & (0.4\%) & 394 & (6.5\%) & 788 & (13.0\%) & 407 & (6.7\%) & 24 & (0.4\%) \\
\hline
B & $100\%$ & 2163 & (24.0\%) & 16 & (0.2\%) & 1 & ($<$0.1\%) & 179 & (2.0\%) & 362 & (4.0\%) & 6219 & (69.1\%) & 59 & (0.7\%) \\
 & $75\%$ & 3882 & (43.1\%) & 123 & (1.4\%) & 4 & ($<$0.1\%) & 385 & (4.3\%) & 1161 & (12.9\%) & 3385 & (37.6\%) & 59 & (0.7\%) \\
 & $50\%$ & 4827 & (53.6\%) & 434 & (4.8\%) & 19 & (0.2\%) & 571 & (6.3\%) & 1896 & (21.1\%) & 1193 & (13.3\%) & 59 & (0.7\%) \\
\hline
\end{tabular*}
\endgroup
\end{table*}

The diagnostic figures in this section use a reduced set of panels for visual presentation in this
paper. These are not the full diagnostic images used during the visual inspection campaign
described in \cref{sec:visual}.

\subsection{\texorpdfstring{\inspectionchange{SC models}}{SC models}}\label{sec:grade-a-sc}

\inspectionchange{At the adopted $75\%$ threshold (\cref{sec:inspection-consensus-thresholds}),
4098 of 6073 grade A candidates ($67.5\%$) are classified as successfully modelled (SC), compared with
3882 of 8999 grade B candidates ($43.1\%$). The lower grade B rate does
not arise mainly from more agreed modelling failures ($9.6\%$ versus $12.9\%$), but from more cases
in which inspectors disagree ($17.0\%$ versus $37.6\%$). These no-consensus (NC) cases are included
in the total from which the success rate is calculated, but they are neither confirmed successes nor
confirmed failures. Instead, they form an ambiguous set of candidates for which inspectors' judgements
differ too much for any classification to be assigned. We discuss NC cases in \cref{sec:nc-grade-b}.
\Cref{tab:model-consensus-thresholds} presents results for other thresholds.}

\Cref{fig:success-consensus-quality} shows three examples with SC-vote fractions of $100\%$, $75\%$,
and $50\%$, respectively. The 14 additional lenses in
\cref{fig:success-consensus-additional-high} provide a qualitative visual illustration of how SC, defined as confidence in the inferred Einstein
radius, varies with inspector agreement. Models receiving at least $75\%$ SC votes generally appear robust,
whereas lower-agreement examples are more heterogeneous and more frequently exhibit features that could call the
Einstein radius measurement into question, for example a counter-image that is not modelled correctly or a
source reconstruction that fits light from a nearby galaxy missed by the segmentation process. However, by definition,
even the results with thresholds as low as $50\%$ are often of a quality for which many experts would consider the Einstein radius reliable.
This illustrates that there will always be some ambiguity over whether a lens model is definitely considered successful.

\begin{figure*}
    \centering
    \includegraphics[width=\textwidth]{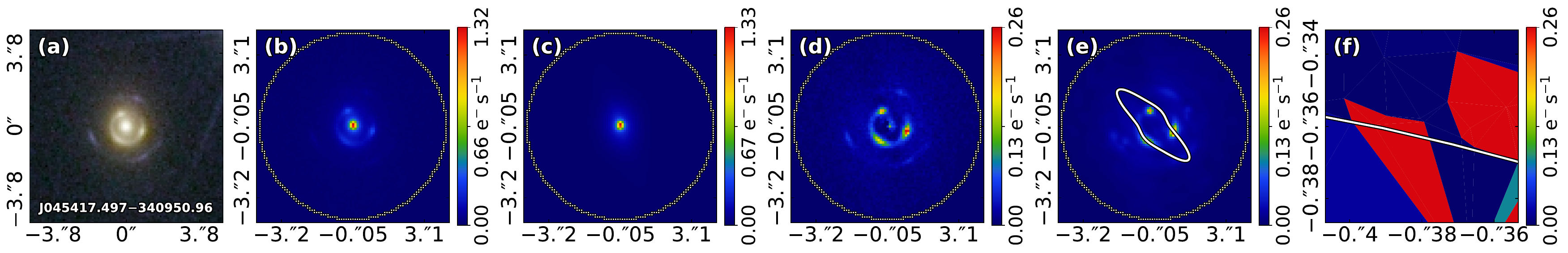}
    \includegraphics[width=\textwidth]{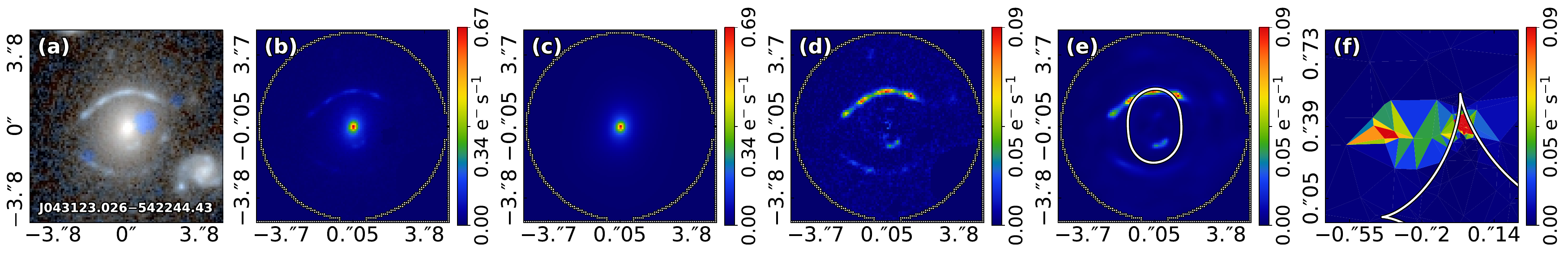}
    \caption{\inspectionchange{
        Two outcomes of the Double Source Plane Lens (DSPL) classification. Panels follow the same
        layout as \cref{fig:success-consensus-quality}. The top row shows a genuine DSPL that
        inspectors correctly identified. Whilst the RGB image already provides strong visual evidence
        for multiple lensed sources, the poor lens-model fit resulting from the violation of the
        single-source-plane assumption provides an additional diagnostic that the system is a DSPL.
        The bottom row shows a grade A candidate receiving $4/5$ ($80\%$) DSPL votes. The additional
        blue feature south of (below) the lens may have motivated the DSPL classification, but its nature is
        uncertain: it could be associated with the main lensed source, a second lensed source, or an
        unrelated line-of-sight object. Visual agreement alone therefore does not establish the
        presence of two lensed sources at distinct redshifts.
    }}
    \label{fig:dspl-results-examples}
\end{figure*}

A high SC-vote fraction does not imply that a model is free from limitations or suitable for
every scientific application. \inspectionchange{\Cref{fig:success-segmentation-caveats} shows two
grade A examples with residual segmentation issues that are above the $75\%$ consensus threshold
for SC.} In the top row, several nearby galaxies or
stars are not masked and their light is mapped into the source reconstruction. In the bottom row,
part of a genuine multiple image is incorrectly masked. In both cases the Einstein radius remains
reliably constrained, but the products are less suitable for detailed source reconstruction or more
complex mass modelling. These issues can be identified from the segmentation products and addressed
through revised masks and remodelling.

\subsection{\texorpdfstring{\inspectionchange{UIL systems}}{UIL systems}}\label{sec:grade-a-uil}

UIL systems are cases where the visual evidence is insufficient for inspectors to confidently
determine whether the object is a genuine strong lens, even after viewing the model diagnostics. At
\inspectionchange{the adopted $75\%$ threshold, \cref{tab:model-consensus-thresholds} reports \stylechange{seven} grade A
UIL systems ($0.1\%$ of the grade A sample) and 123 grade B UIL systems ($1.4\%$).}
\inspectionchange{\Cref{fig:unsure-lens-examples} shows two grade A systems with $100\%$
UIL agreement. In the top row, the lens model does not reveal the corresponding counter-image
to the north of the lens that would be expected for a convincing strong-lensing interpretation.
In the bottom row, faint emission remains after subtraction of the foreground galaxy light, but its
signal-to-noise ratio is too low to provide convincing evidence of lensing. These examples
illustrate that an acceptable model fit alone does not establish the lensing nature of a
candidate.}

\begin{figure*}
    \centering
    \includegraphics[width=\textwidth]{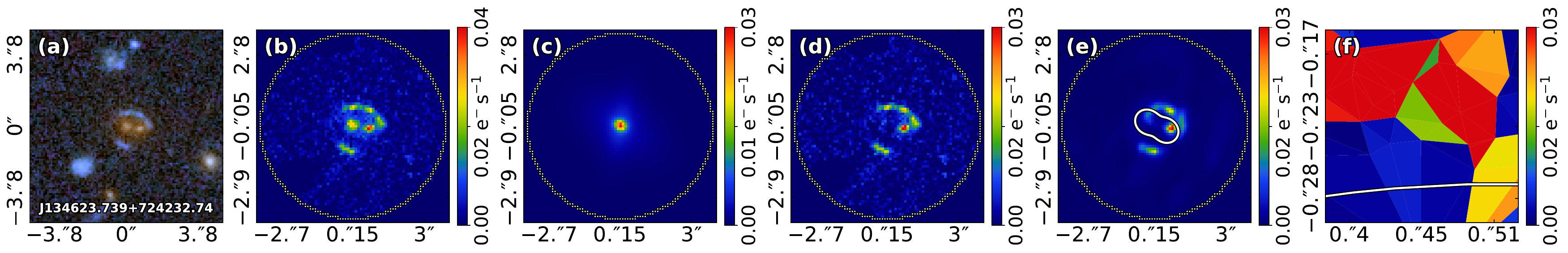}
    \includegraphics[width=\textwidth]{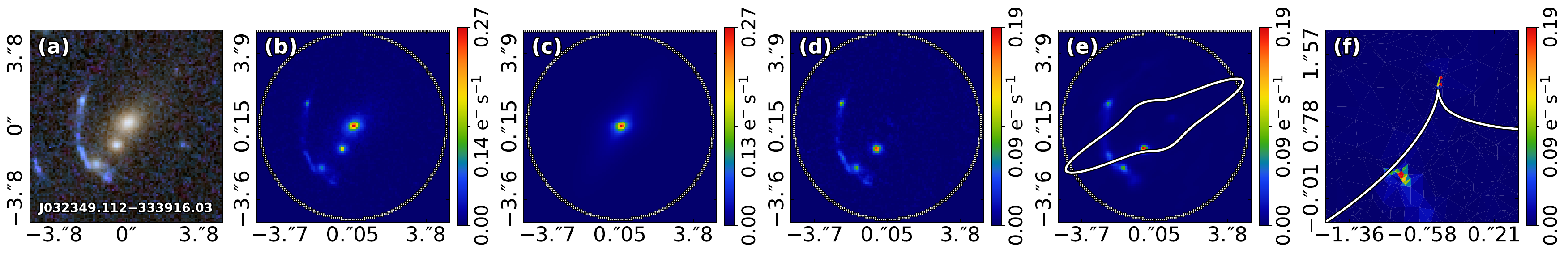}
    \includegraphics[width=\textwidth]{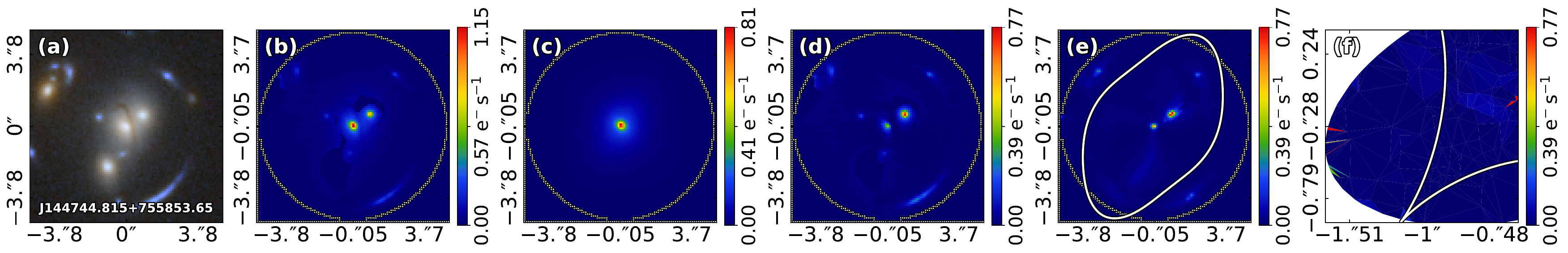}
    \caption{\inspectionchange{
        Examples of grade A multi-galaxy lens (MGL) systems, each with $100\%$ MGL agreement. Panels
        follow the same layout as \cref{fig:success-consensus-quality}. The top row shows a compact
        configuration of multiple foreground galaxies surrounded by arc-like emission, with an
        angular scale similar to that of single-galaxy lenses. The middle row shows multiple
        foreground galaxies with more extended arcs, whose lensed-source morphology would be
        difficult to reproduce with a single deflector. The bottom row shows a substantially larger
        MGL containing many foreground galaxies. Its scale and multiplicity are more characteristic
        of a group- or cluster-scale lens, conventionally associated in $\Lambda$CDM with a
        dark-matter halo mass above approximately $10^{14}\,\si{\solarmass}$.
    }}
    \label{fig:mgl-results-examples}
\end{figure*}

\subsection{\texorpdfstring{\inspectionchange{DSPL systems}}{DSPL systems}}

DSPL systems contain multiple background sources at different redshifts lensed by the same
foreground deflector. At the adopted $75\%$ threshold,
\inspectionchange{\cref{tab:model-consensus-thresholds} reports 15 grade A DSPL systems ($0.2\%$ of the grade A sample) and four grade B DSPL systems ($<0.1\%$).} An
important question is whether the inspectors can distinguish genuine DSPLs from conventional single-plane strong lenses.
The top row of \cref{fig:dspl-results-examples} shows a representative DSPL. The RGB image alone provides strong visual
evidence that the system contains multiple lensed sources and was likely sufficient for most inspectors to identify
it as a DSPL. However, because the lens model assumes a single source plane, it is unable to fit the system accurately,
leading to obvious modelling residuals. This demonstrates that violations of the single-source-plane assumption can
provide an additional diagnostic for identifying genuine DSPLs.

\begin{table}
    \centering
    \caption{
        \inspectionchange{Secondary FL classifications from the 2990-object campaign, without a grade A restriction, within its 210 systems receiving FL at the adopted}
        \inspectionchange{$75\%$ primary-classification threshold. These are not the failure-mode fractions of the current grade A or grade B samples.} Entries are percentages of this fixed FL sample;
        secondary agreement is measured among inspectors who voted FL on each system. The NC
        column gives the fraction for which
        inspectors did not reach the adopted agreement threshold on a secondary FL classification.
    }
    \label{tab:failure-mode-consensus-thresholds}
    \small
    \setlength{\tabcolsep}{5pt}
    \renewcommand{\arraystretch}{1.2}
    \begin{tabular}{lcccc}
        \hline\hline
        & \multicolumn{3}{c}{Secondary FL classification} & \\
        \cline{2-4}
        Threshold & Segmentation & LensLight & SourceMass & NC \\
        \hline
        $100\%$ & $11.4\%$ & $22.4\%$ & $\phantom{0}3.3\%$ & $62.9\%$ \\
        $75\%$  & $18.6\%$ & $31.4\%$ & $\phantom{0}5.7\%$ & $44.3\%$ \\
        $50\%$  & $29.0\%$ & $48.6\%$ & $15.7\%$ & $\phantom{0}5.7\%$ \\
        \hline
    \end{tabular}
\end{table}

\begin{figure*}
    \centering
    \includegraphics[width=\textwidth]{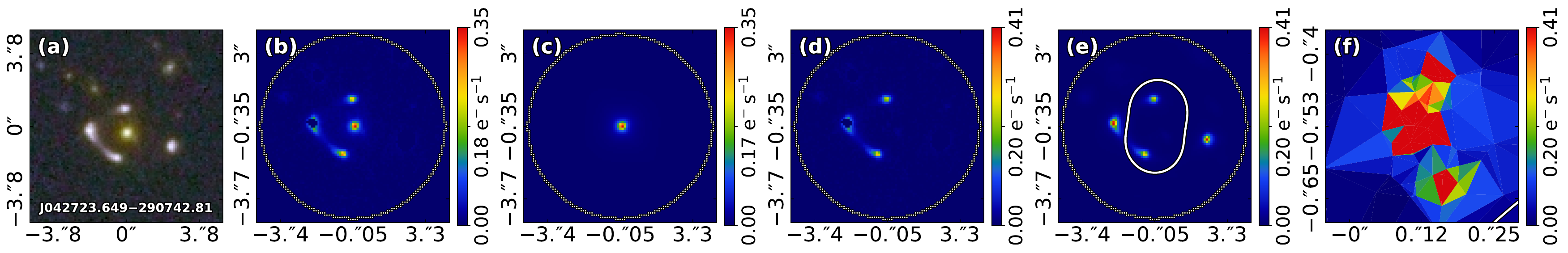}
    \includegraphics[width=\textwidth]{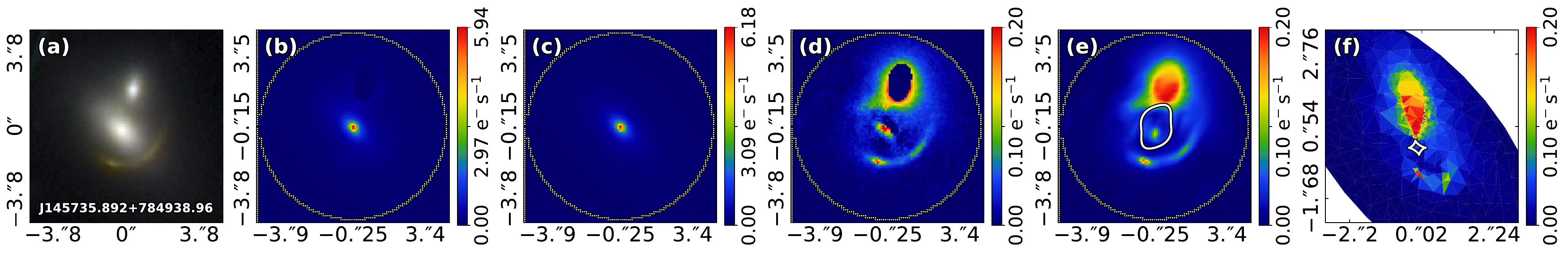}
    \includegraphics[width=\textwidth]{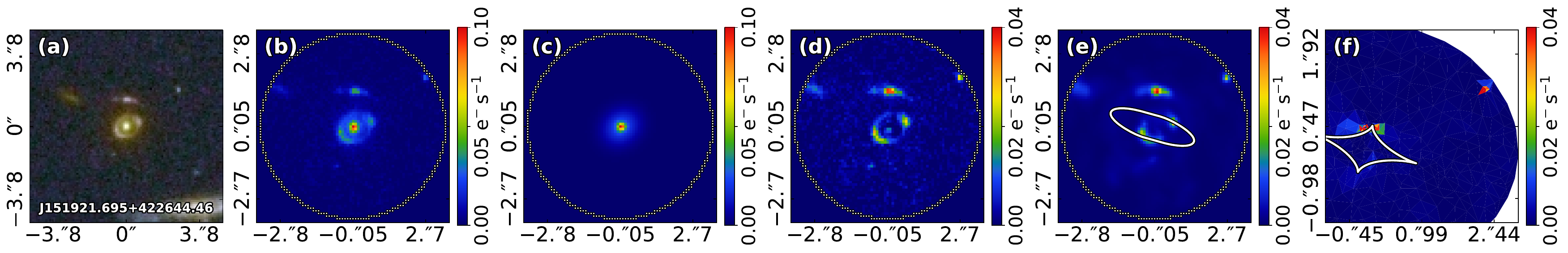}
    \caption{
        Examples of Failure (FL) systems assigned the secondary FL-Segmentation classification. Panels follow the same
        layout as \cref{fig:success-consensus-quality}. In the top row, the segmentation incorrectly
        classifies one of the lensed multiple images as a line-of-sight galaxy, masking source emission
        required for a successful lens model. In the middle row, a contaminating galaxy to the south is
        identified but not fully masked, leaving residual emission that affects the source reconstruction.
        In the bottom row, nearby galaxies are missed entirely, including a galaxy to the south and
        additional objects near the edge of the mask, causing the model to fit contaminating light rather
        than only the lens and source emission.
    }
    \label{fig:failed-segmentation-examples}
\end{figure*}

\inspectionchange{The bottom row of \cref{fig:dspl-results-examples} shows a further grade A
candidate receiving $4/5$ ($80\%$) DSPL votes.} A second question is whether inspectors incorrectly
identify single-plane strong lenses as DSPLs.
To quantify this effect, a team of DSPL experts reinspected the nine objects
\inspectionchange{in the 2990-candidate inspection} that achieved at least $75\%$ DSPL
consensus. They concluded that four were not convincing DSPL candidates, illustrating that visual
inspection alone can leave the interpretation of additional source-like features ambiguous. However, these four objects represent only $0.13\%$ of \inspectionchange{that}
sample and therefore have negligible impact on the overall classification statistics.

\subsection{\texorpdfstring{\inspectionchange{MGL systems}}{MGL systems}}

MGL systems contain more than one foreground lens galaxy within the Einstein radius. At the adopted
\inspectionchange{$75\%$ threshold, \cref{tab:model-consensus-thresholds} reports 317 grade A MGL systems
($5.2\%$ of the grade A sample) and 385 grade B MGL systems ($4.3\%$).} \Cref{fig:mgl-results-examples} shows that this category spans
\inspectionchange{compact configurations of multiple foreground galaxies to more extended arc structures.}
Interestingly, the models in the first two rows appear qualitatively reasonable: their tangential
critical curves suggest that the inferred Einstein radii may be approximately accurate, while their
source reconstructions are focused and form smooth, galaxy-like morphologies. Nevertheless, these
systems violate the single-deflector assumption of the current models and therefore require more
detailed modelling before their inferred parameters can be used for scientific analysis.
\reviewchange{Dedicated multi-deflector modelling of the \stylechange{\Euclid} MGL population is presented
by \citet{Wielders2026MGL}; we do not consider these systems further in this paper.}

\begin{figure*}
    \centering
    \includegraphics[width=\textwidth]{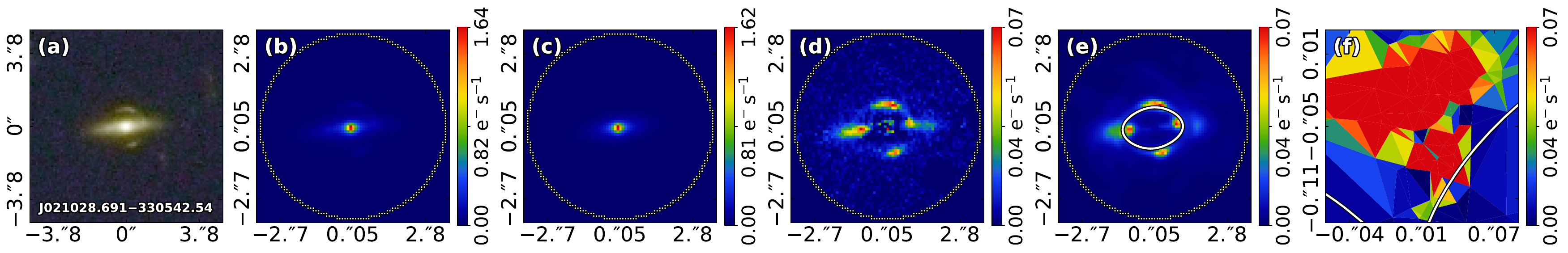}
    \includegraphics[width=\textwidth]{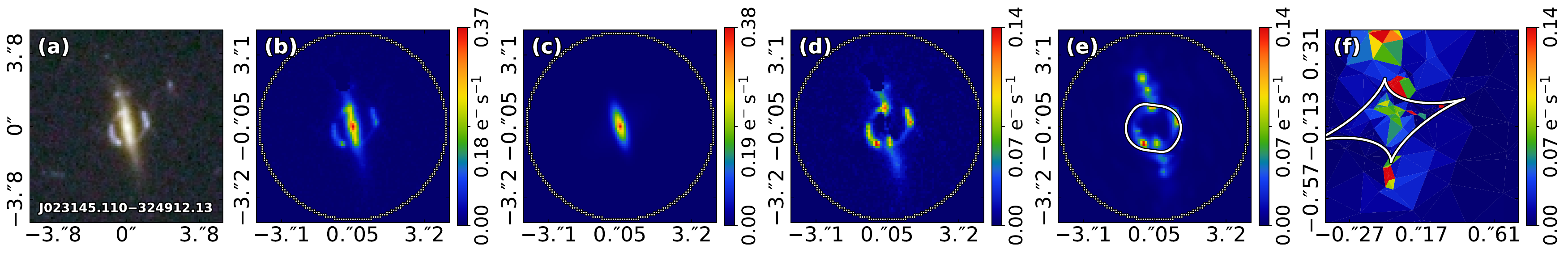}
    \caption{
        Examples of Failure (FL) systems assigned the secondary FL-LensLight classification. \reviewchange{Panels follow the same layout as \cref{fig:success-consensus-quality}.} \reviewchange{Residual foreground structure overlaps the lensed source emission in each system.} In the top row, an
        edge-on disc lens galaxy leaves residuals at both ends of the galaxy, which overlap the lensed
        source emission. The bottom row shows a similar failure, but with stronger ripple-like residuals,
        plausibly associated with jets from an active galactic nucleus.
    }
    \label{fig:failed-lens-light-examples}
\end{figure*}

\begin{figure*}
    \centering
    \includegraphics[width=\textwidth]{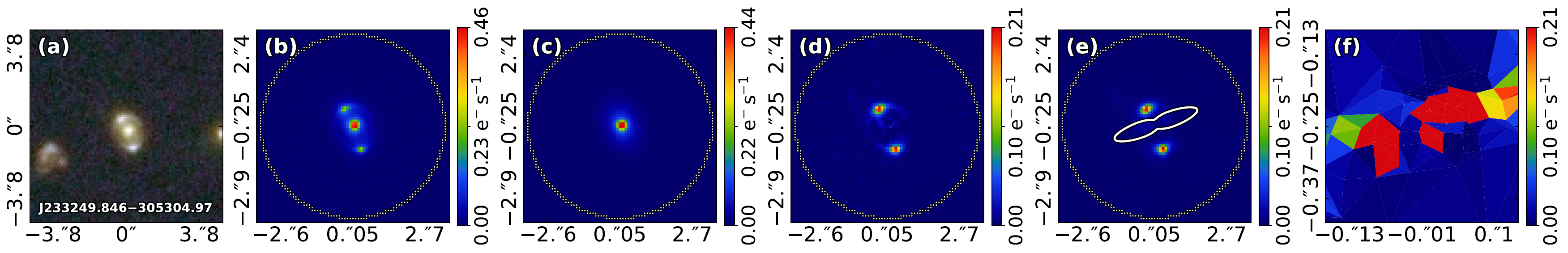}
    \includegraphics[width=\textwidth]{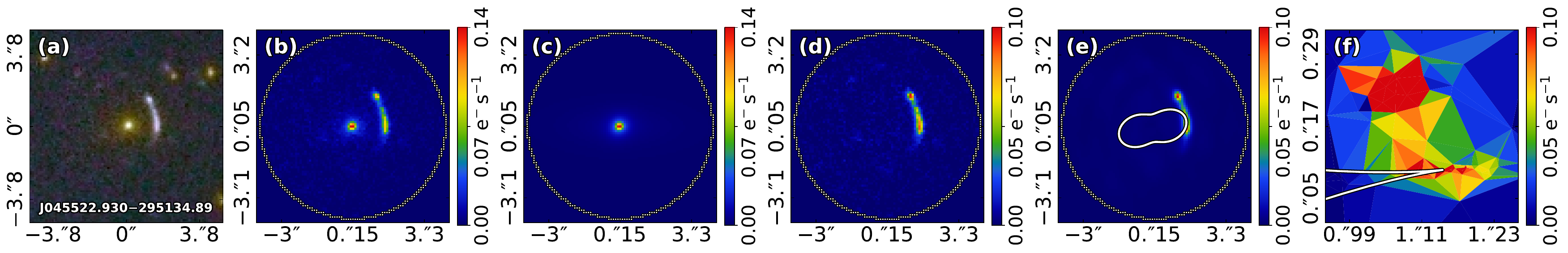}
    \caption{
        Examples of Failure (FL) systems assigned the secondary FL-SourceMass classification. \reviewchange{Panels follow the same layout as \cref{fig:success-consensus-quality}.} These systems
        pass the initial preprocessing and lens-light subtraction stages but yield an implausible
        reconstructed source, lensed source model, or inferred mass model. In the top row, the mass
        model converges to an unphysically elliptical solution and predicts two faint central images
        that are not observed. In the bottom row, the model reconstructs the multiple images in the
        arcs to the right but fails to reconstruct their counter-image to the left because the incorrect
        mass model places the source outside the tangential caustic.
    }
    \label{fig:failed-source-examples}
\end{figure*}

\subsection{\texorpdfstring{\inspectionchange{FL models}}{FL models}}\label{sec:grade-a-fl}

\inspectionchange{At the adopted $75\%$ threshold, 580 grade A systems receive the FL classification
($9.6\%$ of the grade A sample). Secondary failure classifications were recorded only for the
2990-candidate inspection, so the examples and failure-mode breakdown below are drawn from that
sample.} The following examples separate failures
using the secondary classifications adopted during visual inspection: FL-Segmentation, FL-LensLight,
and FL-SourceMass.
\Cref{tab:failure-mode-consensus-thresholds} gives the corresponding failure-mode consensus
breakdown for the subset of systems for which at least $75\%$ of inspectors selected FL in
the primary classification. The percentages in this table are therefore conditional on a system
being classified as FL.

FL-Segmentation failures make the likelihood inputs unreliable by excluding genuine lensed emission
or retaining contaminating light. FL-LensLight, the most common agreed secondary FL classification,
occurs when foreground lens light is not removed accurately enough for reliable source modelling.
FL-SourceMass failures arise when the preprocessing and lens-light subtraction are sufficiently
successful, but the inferred mass model or reconstructed source is not physically plausible.
\Cref{fig:failed-segmentation-examples,fig:failed-lens-light-examples,fig:failed-source-examples}
show representative examples.

\subsection{\texorpdfstring{\inspectionchange{No-consensus classifications}}{No-consensus classifications}}\label{sec:nc-grade-b}

\inspectionchange{A model is classified NC when no single classification reaches the $75\%$
threshold. NC classifications are considerably more common for grade B candidates ($37.6\%$) than
for grade A ($17.0\%$; \cref{tab:model-consensus-thresholds}), and grade B candidates also receive
more individual votes for both modelling failure (FL) and uncertain lens status (UIL).}

\begin{figure*}[t]
    \centering
    \includegraphics[width=\textwidth]{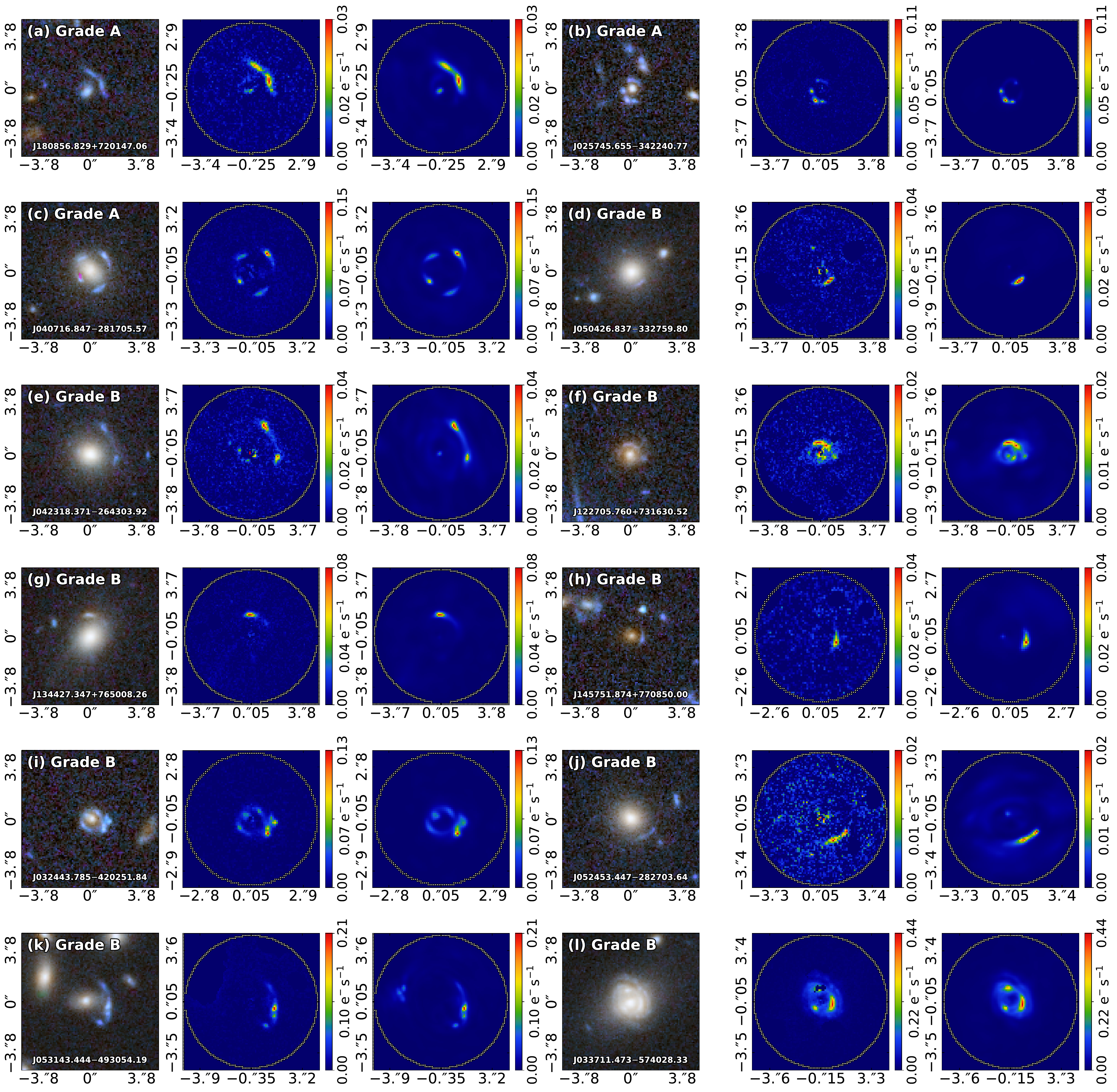}
    \caption{\inspectionchange{
        Selected systems with no consensus (NC) classification at the $75\%$ threshold.
        Reading left to right and top to bottom, the first three triplets show grade A
        candidates (a--c), and the remaining nine show grade B candidates (d--l). Each
        triplet displays the RGB image on the left, the VIS image after subtraction of
        the fitted foreground-galaxy light in the centre, and the lensed-source model
        image on the right, allowing a direct comparison of the residual emission and
        the model prediction.
        The examples illustrate the range of morphologies for which inspectors disagree,
        including uncertainty about whether the residual emission is genuinely lensed.
        They were selected for illustration and do not constitute a representative sample
        of either grade. The subtracted and model images share a colour scale within each
        triplet; this scale varies between systems.
    }}
    \label{fig:nc-grade-comparison}
\end{figure*}

\inspectionchange{\Cref{fig:nc-grade-comparison} illustrates this ambiguity through examples of
grade A and B candidates and their lens models. For many NC cases, the central difficulty is
determining whether the candidate is genuinely a strong lens, rather than simply whether the model
provides an adequate fit. A lens model may reproduce the observed emission without establishing its
lensing origin, leading inspectors to disagree over both the nature of the candidate and the quality
of the model. The lower grade B success rate therefore reflects greater uncertainty in lens
identification, although some reduction in modelling performance cannot be ruled out.}

\subsection{Scientific results}\label{sec:scientific-results}

\begin{figure}
    \centering
    \includegraphics[width=\columnwidth]{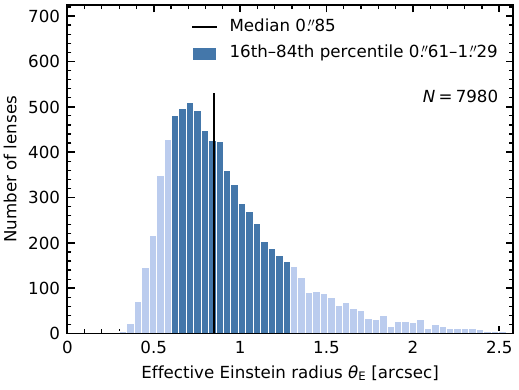}
    \caption{\inspectionchange{
        Distribution of inferred effective Einstein radii $\theta_\sfont{E}$ for the systems
        classified as SC models at the $75\%$ visual-inspection consensus
        threshold (\internalchange{$N = 7980$}; grades A and B). The vertical line marks the median and the darker bars
        the 16th--84th percentile range. The histogram is truncated at the 99.5th percentile.
    }}
    \label{fig:einstein-radius-distribution}
\end{figure}

\inspectionchange{\Cref{fig:einstein-radius-distribution} shows the distribution of Einstein radii for
the 7980 grade A and B systems classified as successfully modelled (SC) at the $75\%$ consensus
threshold. The distribution peaks near \ang{;;0.7}, with a median of \ang{;;0.85}, a 16th--84th
percentile range of \stylechange{\mbox{\ang{;;0.61}--\ang{;;1.29}}}, and a tail extending beyond \ang{;;2.0}, consistent
with strong-lens population forecasts. We consider these point estimates suitable for characterising
the sample but do not report individual parameter uncertainties, whose calibration against potential
\stylechange{systematic errors in the modelling} is deferred to future work. The quoted percentile ranges describe the spread
across systems rather than measurement uncertainties.}

\begin{figure}
\centering
\includegraphics[width=\columnwidth]{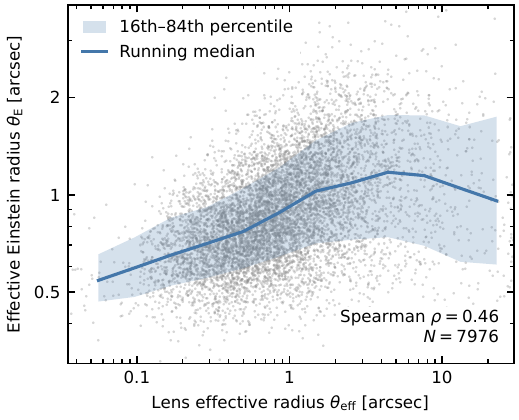}
\caption{\inspectionchange{
Comparison between the foreground lens-galaxy angular effective radius $\theta_{\rm eff}$ and the
inferred Einstein radius $\theta_\sfont{E}$ for SC models at the $75\%$ visual-inspection consensus threshold
(\internalchange{$N = 7976$}; grades A and B). \internalchange{Four of the 7980 SC systems lack a completed VIS S\'ersic fit and are not
shown here or in \cref{fig:source-magnification-distribution}.} The two quantities are positively correlated, indicating that larger lens galaxies
generally produce larger Einstein radii, but with substantial scatter due to variations in galaxy
mass structure and lens--source geometry. The solid line and shaded band show the running median and
16th--84th percentile range in log-spaced bins of the effective radius, and the annotation reports
the Spearman rank correlation coefficient.
}}
\label{fig:effective-radius-vs-einstein-radius}
\end{figure}

\inspectionchange{We \stylechange{obtained} lens-galaxy half-light radii from photometric fits that \stylechange{modelled} both the
lens and source light with S\'ersic profiles, although potential biases from fitting-mask truncation
remain unquantified. \Cref{fig:effective-radius-vs-einstein-radius} compares the lens-galaxy angular
effective radius $\theta_{\rm eff}$ with the Einstein radius for 7976 successfully modelled systems \internalchange{(four SC systems lack a completed
VIS S\'ersic fit)},
finding a positive correlation (Spearman coefficient of $0.46$). This is expected because more massive
galaxies are generally larger and produce larger Einstein radii. The substantial scatter reflects
variations in galaxy mass distributions, including dark matter, and differences in lens and source
redshifts.}

\begin{figure}
    \centering
    \includegraphics[width=\columnwidth]{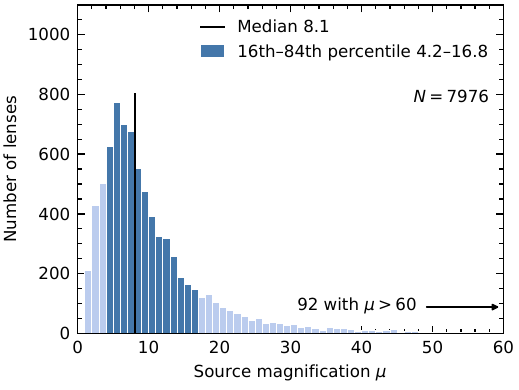}
    \caption{\inspectionchange{
        Distribution of the total source magnification $\mu$, the ratio of the lensed to the
        intrinsic source flux measured in the \IE\ band, for the SC models at the $75\%$
        visual-inspection consensus threshold (\internalchange{$N = 7976$}; grades A and B). \internalchange{Four of the 7980 SC
systems lack a completed VIS S\'ersic fit and are not shown here or in
\cref{fig:effective-radius-vs-einstein-radius}.} The
        vertical line marks the median and the darker bars the 16th--84th percentile range. The
        distribution is shown up to $\mu = 60$; the annotation gives the number of systems above
        this value.
    }}
    \label{fig:source-magnification-distribution}
\end{figure}

\inspectionchange{\Cref{fig:source-magnification-distribution} shows the distribution of total source
magnifications $\mu$, defined as the ratio of lensed to intrinsic source flux, measured in the \IE\
band for the same 7976 systems. The distribution peaks near \stylechange{\mbox{$\mu \simeq 4$--$6$}}, with a median of
$8.1$ and a 16th--84th percentile range of \stylechange{\mbox{$4.2$--$16.8$}}. A high-magnification tail extends beyond
$\mu \gtrsim 30$, with 92 systems exceeding $\mu = 60$ and 43 exceeding $\mu = 120$. Such extreme magnifications arise when
sources lie close to lens caustics, where magnification becomes very large but remains finite for
extended sources. These systems are particularly valuable for studying distant galaxies at enhanced
sensitivity and effective angular resolution.}

\inspectionchange{These magnifications assume an isothermal mass profile and may therefore be biased
for lenses whose radial density slopes differ significantly from isothermal, particularly in the
high-magnification tail. Although this can also bias inferred source sizes, simulations suggest that
the reconstructed source morphology can remain accurate \citep{Cao2022}. Future work will quantify
these \stylechange{systematic effects} using power-law mass models with free radial slopes. Researchers requiring accurate
magnifications or intrinsic source properties should account for this uncertainty, potentially by
remodelling individual lenses with more flexible mass profiles.}

\subsection{\texorpdfstring{\stylechange{ECLIPSE catalogue}}{ECLIPSE catalogue}}\label{sec:eclipse-catalogue-results}

\inspectionchange{All results presented in this work are publicly available through the ECLIPSE
catalogue (ECLIPSE-A; \cref{sec:data-availability}). Alongside inspection classifications and
inferred lens properties, including Einstein radii and source magnifications, we provide FITS images
of deblended lens and source light, and source reconstructions across the VIS, NISP, and EXT bands
(ECLIPSE-D), together with convergence, gravitational potential, and deflection-angle maps. Additional
products include COOLEST-format lens models \citep{Galan2023COOLEST} and Witt--Wynne diagnostics
\citep{Witt1996,WynneSchechter2018,SchechterWynne2019}, which approximate the fitted caustics using
elliptical potentials to rapidly predict image multiplicities, positions, and magnifications for
lensed-transient searches \citep{Schechter2026SN}.}\footnote{\inspectionchange{Our implementation
adapts \texttt{isit4or2or1} v1.0 \citep{SchechterLu2026Software}.}}
\inspectionchange{These approximations should be validated against the original lens models. The
catalogue enables population-level studies and provides the community with the modelling products
needed for independent analyses of individual strong lenses.}
\section{Discussion}\label{sec:discussion}

\subsection{Scaling up lens modelling}

\inspectionchange{\stylechange{Our results demonstrate that strong-lens modelling can be scaled to
survey-sized samples of more than ten thousand lenses, answering the question posed in the introduction.} At the $75\%$ consensus
threshold, 4098 of 6073 grade A candidates ($67.5\%$) are successfully modelled. The corresponding success rate for grade B candidates is $43.1\%$,
with the drop due primarily to the subjectivity of classifying lower-confidence candidates, which leads to more cases without consensus (\cref{sec:nc-grade-b}). Together,
the two grades yield 7980 successful models. The full campaign \stylechange{modelled}
\currentlenssample{} candidates, over an order of magnitude more than any previous modelled strong-lens sample. The
resulting ECLIPSE catalogue provides a foundation for population-level studies of galaxy formation,
evolution, and cosmology, although precise scientific inferences will require further validation of
measurement uncertainties and \stylechange{systematic errors in the modelling}.}

We now consider the breakthroughs that have made scaling lens modelling to \currentlenssample{} systems possible.
The first major breakthrough was not the lens modelling itself, but the machine-learning segmentation framework
used in this work and presented by \citet{Vincken2026}. Previous automated lens modelling studies all relied on a
human-intensive preprocessing stage where tasks such as masking the data and identifying contaminating galaxies were
performed manually. In the Euclid Q1 modelling effort, the lead author of this work performed these preprocessing
tasks manually for over 400 lenses, estimating that processing them all required approximately eight hours of repetitive
GUI interaction, with inevitable human fatigue and inconsistency. Automating these preprocessing steps was therefore
a critical requirement for scaling strong lens modelling to the sample sizes analysed in this work.
Automating these steps also makes the masking procedure, and hence the modelling workflow, reproducible.

The lens modelling software used in this work, \texttt{PyAutoLens}, has undergone many improvements over the past
decade that together enabled automated lens modelling at scale. The transition from S\'ersic lens light
models, as used by \citet{Etherington2022}, to the MGE formalism \citep{He2024} significantly reduced model failures
caused by residuals from inadequate lens light subtraction degrading other components of the fit. \reviewchange{In practical pipeline development, we observed fewer fits converging to local maxima
as we moved from \texttt{MultiNest} \citep{Feroz2009} to \texttt{Dynesty} \citep{Dynesty}
and finally \texttt{Nautilus}.} \reviewchange{\citet{Nightingale2025COWLS} also \stylechange{show} that using an MGE source parameterisation in
the first of the three sequential fits (\cref{LensPipeline}) significantly simplifies the parameter space
explored by the sampler: source size is represented through the linearly solved amplitudes of Gaussians
with prescribed widths rather than fitted directly, reducing degeneracies with the lens mass model.} In truth, automated strong lens modelling was not
enabled by a single methodological breakthrough, but instead by many gradual and incremental improvements developed over more than ten years.

\reviewchange{Machine-learning predictions offer a further route to improving model
initialisation. \citet{Pearson2021} \stylechange{demonstrate} the use of neural-network predictions and
uncertainties as priors for \texttt{PyAutoLens}, while \texttt{LeMoN}
\citep{Busillo2026} provides rapid estimates of lens mass and light parameters that can inform
subsequent modelling. Comparisons between \texttt{PyAutoLens} and \texttt{LeMoN} on subsets
of the DR1 sample indicate that combining the approaches could improve the success rate.
We did not incorporate machine-learning initialisation in the present analysis, prioritising
a simple pipeline design for this initial scaling to 15\,072 lenses.
Broader comparisons of modelling methods are underway within the SL-SWG to identify systematic biases, following \citet{Galan2024}.}

\inspectionchange{Computational cost was not a barrier to this study. Modelling 15\,072 strong-lens
candidates required fewer than 250\,000 CPU hours, relatively modest by modern high-performance
computing (HPC) standards, with individual fits typically completing in under five hours. We used
CPUs because they were readily available, while estimates from \texttt{PyAutoLens-Assistant}
suggested they would also be more energy efficient for this workload. \texttt{PyAutoLens-Assistant} supported
computational planning, monitoring, and aggregation of modelling results into catalogues shared
with the SL-SWG, motivating further development of its
natural-language interface for managing large modelling campaigns. \texttt{PyAutoLens} and other
lens-modelling packages, including \texttt{herculens} \citep{Galan2022}, now support GPU
acceleration, offering further potential speed-ups. These modest computational requirements
leave substantial scope for scaling to larger samples and more complex models.}

\inspectionchange{Human inspection presents a greater challenge to scaling. At approximately
10 seconds per lens, four inspections of all 15\,072 candidates require around 167 person-hours,
or about 8 hours each for a team of 20 inspectors. Such an approach becomes impractical for samples
exceeding 100\,000 lenses. Automated goodness-of-fit statistics alone are insufficient: models
can yield reliable Einstein radii despite imperfect lens-light subtraction \citep{Etherington2022},
while apparently good fits can contain unphysical source reconstructions. We therefore relied
on multiple expert inspectors to assess the residuals, lens-light subtraction, critical curves,
and source reconstructions. Developing reliable automated assessments is an important next step.}

\inspectionchange{Agentic AI offers a potential solution to this inspection bottleneck. By
assessing the imaging and model diagnostics together, an AI agent could identify reliable models,
diagnose failures, and recommend or perform targeted remodelling, such as correcting faulty
segmentation. Our visual-inspection campaign provides a \stylechange{data set} against which such automated
assessments can be developed and validated. Complementary machine-learning classifiers could
identify double-source-plane lenses, multi-galaxy lenses, and other systems requiring specialised
modelling. Together, these approaches could make the analysis of more than 100\,000 lenses
practical, with human inspectors focusing on ambiguous cases and validating automated decisions.}

\subsection{\texorpdfstring{\reviewchange{Fixing failures}}{Fixing failures}}\label{sec:fixing-failures}

\reviewchange{Within the 2990-object inspection subset, 39 of the 210 failures
were attributed to segmentation at the $75\%$ secondary-classification threshold
(\cref{tab:failure-mode-consensus-thresholds}). We anticipate that continued expansion of the
\stylechange{\texttt{AstroVink-TREX}} training \stylechange{data set} will reduce these failures. The initial labelled \stylechange{data set} comprised
377 Q1 systems \citep{Vincken2026}; during this study, it was expanded first with a few hundred
additional systems and then with 2990 DR1 lenses, producing the segmentation model
applied to all 15\,072 candidates. We observed improvements in segmentation performance as the
training \stylechange{data set} grew. The resulting models and visual-inspection assessments now
provide a much larger pool from which to select reliable training examples and identify difficult
cases for the next iteration. Further algorithmic improvements are also planned, including refinement
of the loss function and contaminant separation, and extending the architecture to use multi-band
information \citep{Vincken2026}.}

\reviewchange{At the same threshold, 66 systems were classified as lens-light failures. These \stylechange{are}
predominantly associated with dust or asymmetric structure, including radial light profiles that
rise in brightness away from the lens centre. Adding more concentric, symmetric Gaussian components,
each with a monotonically decreasing radial brightness profile, cannot reproduce these features,
so increasing the number of MGE sets is unlikely to resolve them. Such systems will likely
require a different approach using a non-parametric lens-light model and, where necessary, explicit
treatment of dust absorption. The matrix formalism incorporating lens light and dust extinction in
\citet[\stylechange{section}~4, \stylechange{equation}~13]{Suyu2009} provides a possible algebraic starting point for developing such a
method. Separately, some lenses could benefit from the central point-like component used by H24,
for example to represent a bright active galactic nucleus or other nuclear emission. Even without
this component in our models, these systems were typically judged to yield reliable Einstein-radius
measurements.}

\reviewchange{The 12 systems classified as source/mass failures suggest a further route to improvement
through more robust sampling. Where these failures reflect convergence to an incorrect local maximum,
increasing the number of live points in \texttt{Nautilus} could help locate the preferred solution.
Combining sampling with machine-learning initialisation, for example using \texttt{LeMoN}
\citep{Busillo2026} as discussed above, could also improve convergence and recover failed models.
Such approaches will also need to account for physical source--mass degeneracies, since source
structure and regularisation assumptions in pixelised reconstructions can affect the inferred lens
potential \citep{VernardosKoopmans2022}.}

\subsection{Science with large strong lens samples}

\inspectionchange{The ECLIPSE series models 15\,072 grade A and B strong-lens candidates,
providing the data and models needed for population-level studies and detailed investigations
of individual systems. For successfully modelled lenses, Einstein radii, galaxy sizes and
deblended photometry characterise the foreground galaxies, while magnifications and source
reconstructions provide insights into the background population. Einstein radii are relatively
robust to modelling choices \citep{Etherington2022}, while ECLIPSE-D evaluates the lens photometry
for photometric-redshift estimation through comparison with MER. Future work using more flexible
mass models will refine these measurements, particularly source magnifications and intrinsic
sizes, which depend on the assumed radial mass profile (\cref{sec:scientific-results}).}

The \Euclid strong-lens catalogue and lens models presented in this work provide a wealth of opportunities to
study galaxy evolution and cosmology with strong gravitational lenses. For the lens galaxies themselves, these
data establish strong lensing as a complementary tool to long-standing approaches such as stellar dynamics
\reviewchange{\citep[e.g.][]{Cappellari2011,Cappellari2013,Bundy2015}} and stellar-population modelling \reviewchange{\citep[e.g.][]{BruzualCharlot2003,Thomas2005,Conroy2013}}, providing
the statistically significant samples needed to use lensing-inferred
quantities as \reviewchange{measurements of galaxy structure and evolution across a wide range of galaxy masses and redshifts.} The S\'ersic
lens-light models released here will enable a census of lens-galaxy morphologies, including early-type, late-type,
and S0 systems \reviewchange{\citep[e.g.][]{Graham2001a,GrahamDriver2005}}, while providing structural measurements such as S\'ersic indices, effective radii, axis ratios, and luminosities.
\reviewchange{For both lenses and sources, these parameters provide familiar structural summaries
for comparison with studies of unlensed galaxies; even a morphologically complex source can be
characterised by a fitted effective radius and S\'ersic index.}
\reviewchange{Future bulge--disc decompositions could provide component sizes, luminosities and
stellar masses \citep[e.g.][]{Simard2011,Mendel2014}. Combined with lensing mass constraints, these would
connect bulge and disc growth to the total mass distribution and star-formation quenching
\citep[e.g.][]{Martig2009,Fang2013,Lang2014,Bluck2014,Bluck2022}.}

When combined with SIE lens mass models and redshift information, these observables provide direct measurements of
the projected total mass enclosed within the Einstein radius, \reviewchange{a measure of the enclosed mass of lens galaxies that is not available for unlensed galaxies.} This will enable extensions of classical galaxy scaling relations, such as the
fundamental plane \citep{DjorgovskiDavis1987,Dressler1987} and \stylechange{\citet{TullyFisher1977} relation}, in which lensing-inferred mass information is incorporated
alongside photometric and, where available, dynamical quantities. Unlike many traditional galaxy-evolution probes,
whose signal-to-noise \stylechange{ratio} or diagnostic power often decreases rapidly with redshift, the lensing constraint is set
primarily by the geometry and surface-brightness structure of the lensed images. Meaningful mass measurements can
therefore often be extracted for higher-redshift lens galaxies in much the same way as for lower-redshift systems.
The mass models released here thus provide a homogeneous view of the mass distributions of many thousands of massive
galaxies over the redshift range $z \simeq 0.0$--$2.0$, opening a new route to studying the connection between
galaxy structure, mass assembly, and evolution across cosmic time, complementary to local studies \citep[e.g.][]{Auger2009a} that provide the
essential $z \simeq 0$ baseline against which this evolution can be measured.

The strong-lens imaging modelled in this work contains substantially more information than the Einstein mass alone,
encoded in the extended surface-brightness distribution of the lensed source. Fully exploiting this information
will ultimately require more sophisticated lens models than those fitted here. Examples include power-law mass models \reviewchange{\citep[e.g.][]{Koopmans2009,Shajib2021,Etherington2023}},
models that decompose the lens mass distribution into stellar and dark matter components \reviewchange{\citep[e.g.][]{Sonnenfeld2013b,Nightingale2019,Li2026Jackpot,Li2026WFI2033}}, and
models that infer more detailed radial or angular structure in the mass distribution \citep[e.g.][]{Powell2022,Stacey2024}. Incorporating complementary probes
such as weak lensing or stellar dynamics pushes this further. \reviewchange{We therefore anticipate that future lens-modelling studies will build on our work to exploit this information.}

\reviewchange{The models presented here also enable studies of the lensed source galaxies themselves.} For
every successfully modelled system, this work provides source magnification estimates and pixel-grid source
reconstructions across the available imaging bands, enabling studies of intrinsic luminosities, morphologies, and
colours after correcting for the lensing distortion. There are already many examples of galaxy-scale strong lenses
being used to study background galaxies in detail \reviewchange{\citep[e.g.][]{Dye2014,Swinbank2015,Rizzo2020,Amvrosiadis2025}}, but such analyses have often been limited by sample size: researchers
have not generally been able to select, from large parent samples, the particular magnified sources best matched to
a given science case. The scale of the \Euclid sample changes this. On cluster scales, large magnifications
have enabled major progress in the study of galaxies at redshifts greater than 10 \reviewchange{\citep[e.g.][]{Coe2013,Atek2015,Lotz2017}}. \reviewchange{Galaxy-scale lenses offer comparatively simple mass models for source reconstruction,
while magnification improves the effective spatial resolution of source structure for a given observation
\citep[e.g.][]{Rizzo2020}. Their compact angular extent also allows an individual lens system, including its
multiple images, to fit within a single integral-field-unit field of view for suitably chosen targets
and instruments.} The catalogue therefore
provides a large, selectable sample of magnified galaxies for future studies of resolved galaxy structure,
stellar populations, and kinematics, with the richest constraints expected at redshifts $z \simeq 0.5$--$3.0$ and
smaller subsamples extending to $z \simeq 8.0$.

\reviewchange{One avenue for future work is to select lenses suited to specific science cases and obtain targeted follow-up observations. For dark-matter studies, such samples can constrain the abundance of low-mass haloes \citep[e.g.][]{Vegetti2018,Vegetti2024}, with sensitivity estimates guiding the selection of targets for deeper or higher-resolution imaging \citep[e.g.][]{ORiordan2023}. These analyses must also account for angular structure in the lens mass distribution, which can mimic substructure signals \inspectionchange{\citep[e.g.][]{ORiordanVegetti2024,Lange2025}}. For cosmography, the large parent sample, combined with detailed modelling and stellar-kinematic follow-up of selected lenses, could help constrain the diversity of galaxy mass profiles and the associated mass-sheet degeneracy \citep{SchneiderSluse2013,SchneiderSluse2014,Birrer2020}. Accounting for this degeneracy is essential both for time-delay cosmography and for cosmological inference from double-source-plane lenses \citep{Schneider2014MST}.}

\reviewchange{We anticipate high-quality photometric redshifts for most lens galaxies,
which are preferentially bright and massive (ECLIPSE-D). Spectroscopic follow-up will
provide complementary information: the 4MOST Strong Lensing Spectroscopic Legacy
Survey (4SLSLS) aims to obtain approximately 10\,000 source redshifts for \Euclid and
LSST lens candidates \citep{Collett2023FourMOST}. Spectroscopic completeness may therefore
be achievable for selected samples targeting cosmology or dark matter, but is unlikely
for the full population of more than 100\,000 lenses. Exploiting that population will
require improved photometric redshifts and methods that propagate redshift uncertainties
into lensing inference, including joint analyses that account for the source population
and lens selection function \citep[e.g.][]{Sonnenfeld2022,Sonnenfeld2022IV}.}
\section{Conclusions}\label{sec:summary}

\inspectionchange{We have presented automated lens modelling of 15\,072 strong-lens candidates
from Euclid Data Release 1 (DR1), over an order of magnitude more than any previous modelled
strong-lens sample. Using \texttt{PyAutoLens}, we \stylechange{combined} automated machine-learning
preprocessing, MGE lens-light subtraction, an isothermal ellipsoid mass
model with external shear, and parametric and pixelised source reconstructions. The resulting
models were visually assessed by more than 20 expert inspectors from the SL-SWG. Our main conclusions are:}

\begin{itemize}
\item \inspectionchange{\stylechange{\textit{Modelling success rates:}} At $75\%$ expert agreement, 4098 of
6073 \stylechange{grade A (secure)} candidates ($67.5\%$) and 3882 of 8999 \stylechange{grade B (probable)} candidates
($43.1\%$) are successfully modelled, yielding a total of 7980 successful models.}

\item \inspectionchange{\stylechange{\textit{Grade B uncertainty:}} The lower grade B success rate primarily
reflects greater disagreement between inspectors ($37.6\%$ without consensus, compared with
$17.0\%$ for grade A), rather than a comparable increase in agreed modelling failures
($12.9\%$ versus $9.6\%$). We attribute much of this disagreement to uncertainty over whether
grade B candidates are genuine lenses.}

\item \inspectionchange{\stylechange{\textit{Modelling failures:}} For a subset of 2990 candidates,
we \stylechange{analysed} 210 consensus modelling failures. The most common identified cause is inaccurate
lens-light subtraction ($31.4\%$), followed by segmentation errors ($18.6\%$) and
source-reconstruction or mass-model failures ($5.7\%$), with the remaining $44.3\%$ lacking
consensus on the failure mode.}

\item \inspectionchange{\stylechange{\textit{Lens properties:}} Successfully modelled systems have a median
Einstein radius of \ang{;;0.85}, with a 16th--84th percentile range of
\stylechange{\mbox{\ang{;;0.61}--\ang{;;1.29}}}. We consider these point estimates suitable for characterising the
population, while calibration of individual parameter uncertainties and \stylechange{systematic errors in the modelling}
is left to future work.}

\item \inspectionchange{\stylechange{\textit{Public catalogue:}} We release the successful lens models and
their inferred properties through the ECLIPSE catalogue,\footnote{\url{https://euclid-eclipse-lenses.org}} alongside masks, segmentation maps,
deblended multi-wavelength imaging, source reconstructions, and lens mass maps. These products
enable both population-level studies and detailed investigations of individual systems.}
\end{itemize}

\inspectionchange{Our results demonstrate that automated strong-lens modelling can scale to the
thousands of systems being discovered by \Euclid. The ECLIPSE catalogue provides a foundation
for studying galaxy mass distributions, the connection between baryons and dark matter, galaxy
evolution, and cosmology. The approach also provides a basis for automated modelling of strong
lenses from surveys with \stylechange{\textit{Roman}}, the China Space Station Telescope (CSST), and the Vera C. Rubin Observatory.
Extending this approach to future \Euclid releases, alongside
improvements in automated model assessment and uncertainty calibration, will establish strong
lensing as an increasingly powerful statistical probe of galaxy formation and cosmology.}

\section{Data availability}\label{sec:data-availability}

The lens catalogue, lens models, photometric redshifts, and stellar population properties are
available on Zenodo at \url{https://doi.org/10.5281/zenodo.19654904}. For every successfully
modelled lens, the release includes the inferred lens-model parameters and derived quantities,
together with the masks, segmentation maps, deblended multi-wavelength imaging, source
reconstructions, and lens mass maps described in this work. The catalogue can also be browsed
interactively at \url{https://euclid-eclipse-lenses.org}. The \Euclid DR1 data, including all the
catalogues and imaging used in this work, are available to download from
\url{https://eas.esac.esa.int/sas/} and are described at
\url{https://www.cosmos.esa.int/web/euclid/data-release-dr1}.

\section*{Software}\label{sec:software}

\begin{acknowledgements}
JWN is supported by an STFC/UKRI Ernest Rutherford Fellowship, Project Reference:
ST/X003086/1, and acknowledges travel support provided by Euclid:UK through STFC grant
APP91126. NL is supported through a graduate studentship from the UKRI STFC and the University
of Portsmouth, and acknowledges travel support provided by Euclid:UK through STFC grant
APP91126 and by STFC for UK participation in the Rubin LSST through grant ST/X001334/1.
This work has received funding from the European Research Council (ERC) under the European
Union's Horizon 2020 research and innovation programme (LensEra: grant agreement
No.~945536). TEC is funded by the Royal Society through a University Research Fellowship.
PH is funded by the Royal Society (RF/ERE/210033) and acknowledges travel support provided by
Euclid:UK through STFC grant APP91126. AN gratefully acknowledges the support of the Belgian
Federal Science Policy Office (BELSPO) for the provision of financial support in the
framework of the PRODEX Programme of the European Space Agency (ESA) under contract number
4000143347. CJS acknowledges funding from the NASA Euclid General Investigator Program Grant
No.~80NSSC25K7735. WJRE acknowledges the support by the Italian Ministry of University and
Research (grant FIS2023-01611, CUP C53C25000300001). AS acknowledges support from the National
Key R\&D Program of China (No.~2023YFA1607800, 2023YFA1607802), and from the Yangyang
Development Fund. AG acknowledges support by the SNSF through PostDoc Mobility Return grant
P5R5-2\_235360. QH was supported by the Dutch Research Council (NWO) through the grant Open
Competitie ENW M21-1. DJ acknowledges support by the First Rand Foundation, South Africa, and
the Centre National de la Recherche Scientifique of France. SKL acknowledges the generous
support from the Croucher Foundation through the Croucher Postdoctoral Fellowship. PB
acknowledges financial support from the Italian Space Agency (ASI) through contract
``Euclid - Phase E'', and INAF Grants ``The Big-Data era of cluster lensing'' and ``Probing
Dark Matter and Galaxy Formation in Galaxy Clusters through Strong Gravitational Lensing''.
JAAB's research was supported by an appointment to the NASA Postdoctoral Program at the NASA
Jet Propulsion Laboratory administered by Oak Ridge Associated Universities under contract
with NASA. Part of this research was carried out at the Jet Propulsion Laboratory, California
Institute of Technology, under a contract with the National Aeronautics and Space
Administration (80NM0018D0004). SHV is supported by the ELSA project. `ELSA: Euclid Legacy
Science Advanced analysis tools' (Grant Agreement no.~101135203) is funded by the European
Union. Views and opinions expressed are however those of the author(s) only and do not
necessarily reflect those of the European Union or Innovate UK. Neither the European Union nor
the granting authority can be held responsible for them. UK participation is funded through
the UK Horizon guarantee scheme under Innovate UK grant 10093177. Numerical computations were
carried out on the SCIAMA High Performance Compute (HPC) cluster, which is supported by the
ICG, SEPNet, and the University of Portsmouth. This work used IRIS computing resources funded
by the Science and Technology Facilities Council.
\AckDatalabs
\AckDRone
\AckEC
\end{acknowledgements}

\appendix
\section{\texorpdfstring{\reviewchange{Forward model, linear inversion, and likelihood}}{Forward model, linear inversion, and likelihood}}\label{app:likelihood}

\reviewchange{This appendix specifies the calculation underlying the three fits in
\cref{LensPipeline}. The joint treatment of MGE lens light and source light follows H24,
within the semi-linear framework of \citet{Warren2003} and \citet{Suyu2006}.
We describe mesh placement, interpolation, and regularisation separately, since these
choices differ between implementations. The related COWLS-I analysis
\citep{Nightingale2025COWLS} \stylechange{used} MGE light models and cross regularisation, but
\stylechange{interpolated} its pixelised source using Voronoi natural neighbours.}

\subsection{\texorpdfstring{\reviewchange{Image formation and pixel integration}}{Image formation and pixel integration}}\label{app:forward}

\reviewchange{Let $d_i$ and $n_i$ denote the measured intensity and noise standard deviation
of image pixel $i$ within the circular analysis mask. We \stylechange{assumed} independent Gaussian
noise, with covariance $C_{ij}=n_i^2\delta_{ij}$. The data and noise arrays include the
contaminant treatment described in \cref{sec:preprocessing}; pixels assigned very large
noise have negligible weight. For angular image coordinate $\boldsymbol\theta$, the
lens equation is
\begin{equation}
 \boldsymbol\beta(\boldsymbol\theta;\boldsymbol\eta)
 =\boldsymbol\theta-\boldsymbol\alpha(\boldsymbol\theta;\boldsymbol\eta)\;,
 \label{eq:app-lens-equation}
\end{equation}
where $\boldsymbol\eta$ contains the non-linear mass and light parameters, and
$\boldsymbol\alpha$ includes the SIE and external shear. If $L$ and $S$ are the lens
and source surface-brightness distributions, the model image is
\begin{equation}
 m_i=\left\{\mathcal B\mathcal P
 [L(\boldsymbol\theta)+S(\boldsymbol\beta(\boldsymbol\theta))]\right\}_i\;.
 \label{eq:app-forward}
\end{equation}
Here $\mathcal P$ averages the sampled brightness within each image pixel and
$\mathcal B$ applies the discrete, normalised PSF convolution, including light from
outside the mask that is blurred into fitted pixels. Both operators preserve the
intensity convention of the input image.}

\reviewchange{We \stylechange{evaluated} a pixel using a uniform $n_{{\rm sub},i}\times n_{{\rm sub},i}$
subgrid, averaging the $n_{{\rm sub},i}^2$ contributions before PSF convolution.
Image-plane subgridding also \stylechange{entered} the source mapping in \citet{Nightingale2024},
who \stylechange{used} $4\times4$ sampling. Here the sampling \stylechange{varied} spatially. In the MGE fit,
$n_{{\rm sub},i}=4,4,2$ at image-plane radii respectively below \ang{;;0.1},
between \ang{;;0.1} and \ang{;;0.3}, and above \ang{;;0.3} from the \stylechange{data set} centre.
The same light-profile sampling \stylechange{was} retained for the lens MGE in the pixelised fit;
the source mapping instead \stylechange{used} $4\times4$ sampling where the adaptation map
defined below \stylechange{exceeded} a signal-to-noise ratio of three, and $2\times2$ elsewhere.}

\reviewchange{For the S\'ersic fit, the first-fit mass model \stylechange{defined} the source-plane
coordinates of the image pixels. Using the same two radial boundaries, sampling
\stylechange{was} $16,4,2$ subpixels per dimension around both the inferred source centre and the lens
\stylechange{data set} centre, and each image pixel \stylechange{adopted} the larger of the two values. Thus the source refinement \stylechange{followed} the locations of its lensed
images, rather than a single image-plane centre. These sampling maps \stylechange{were} set before
the corresponding fit.}

\subsection{\texorpdfstring{\reviewchange{MGE basis and joint linear solution}}{MGE basis and joint linear solution}}\label{app:mge}

\reviewchange{Following the notation of H24, the MGE surface brightness is
\begin{equation}
 I_{\rm MGE}(x,y)=\sum_{k=1}^{N_{\rm g}^{\rm tot}}G_k(x,y)\;,\qquad
 G_k(x,y)=I_k\exp\!\left[-\frac{R_k^2(x,y)}{2\sigma_k^2}\right]\;,
 \label{eqn: MGE}
\end{equation}
where $I_k\geq0$ is the central intensity and $\sigma_k$ is the major-axis Gaussian width.
The elliptical radius and rotated coordinates are
\begin{align}
 R_k^2&=x_k'^2+(y_k'/q_k)^2\;,\\
 x_k'&=(x-x_k^{\rm c})\cos\phi_k+(y-y_k^{\rm c})\sin\phi_k\;,\\
 y_k'&=-(x-x_k^{\rm c})\sin\phi_k+(y-y_k^{\rm c})\cos\phi_k\;.
\end{align}
Here $q_k$ is the axis ratio and $\phi_k$ the orientation of the major axis, measured
anticlockwise from the $+x$ axis of the modelling grid. Components within
one Gaussian set share their centre, axis ratio, and orientation. The software
parameterises each width as $q_k\sigma_k$; these prescribed values \stylechange{were} logarithmically
spaced from \ang{;;0.0001} to the mask radius for both galaxies.
There \stylechange{were} two sets of 20 lens Gaussians and one set of 20 source Gaussians. The lens
sets \stylechange{shared} a centre but \stylechange{had} independent axis ratios and orientations, giving
six non-linear parameters.}

\reviewchange{At fixed $\boldsymbol\eta$, let $A_{ik}$ be the pixel-integrated,
PSF-convolved unit-intensity lens profile and $f_{ij}$ the corresponding lensed
source mapping, following H24. The source basis consists of Gaussians in the first
fit and interpolated mesh functions in the second. With lens intensities
$\boldsymbol I$ and source intensities $\boldsymbol S$, the model is
\begin{equation}
 \begin{gathered}
 m_i=\sum_j f_{ij}S_j+\sum_k A_{ik}I_k\;,\\
 X_{ik}=\frac{A_{ik}}{n_i}\;,\quad Z_{ij}=\frac{f_{ij}}{n_i}\;,\quad
 Y_i=\frac{d_i}{n_i}\;.
 \end{gathered}
 \label{eq:app-design}
\end{equation}
Writing the source regularisation matrix as $\stylechange{\tens{H}}=\stylechange{\tens{\Gamma}}^{\rm T}\stylechange{\tens{\Gamma}}$ gives the
joint quadratic objective in H24's source-first ordering:
\begin{equation}
 Q=\frac12\left\|
 \begin{pmatrix}\stylechange{\tens{Z}}&\stylechange{\tens{X}}\\\stylechange{\tens{\Gamma}}&0\end{pmatrix}
 \begin{pmatrix}\boldsymbol S\\\boldsymbol I\end{pmatrix}
 -\begin{pmatrix}\boldsymbol Y\\0\end{pmatrix}
 \right\|^2\;,\qquad \stylechange{S_j\geq0\,,\ I_k\geq0}\;.
 \label{eq:app-nnls}
\end{equation}
Thus $Q=\tfrac12\chi^2+\tfrac12\boldsymbol S^{\rm T}\stylechange{\tens{H}}\boldsymbol S$, where
$\chi^2$ is the full residual sum defined in Eq.~\eqref{eq:app-image-likelihood}.
\stylechange{The constraints require every source and lens intensity to be non-negative.} Without these constraints, the solution obeys
\begin{equation}
 \begin{pmatrix}\stylechange{\tens{Z}}^{\rm T}\stylechange{\tens{Z}}+\stylechange{\tens{H}}&\stylechange{\tens{Z}}^{\rm T}\stylechange{\tens{X}}\\\stylechange{\tens{X}}^{\rm T}\stylechange{\tens{Z}}&\stylechange{\tens{X}}^{\rm T}\stylechange{\tens{X}}\end{pmatrix}
 \begin{pmatrix}\widehat{\boldsymbol S}\\\widehat{\boldsymbol I}\end{pmatrix}
 =\begin{pmatrix}\stylechange{\tens{Z}}^{\rm T}\boldsymbol Y\\\stylechange{\tens{X}}^{\rm T}\boldsymbol Y\end{pmatrix}\;.
\end{equation}
All three fits \stylechange{used} the \texttt{JAX} implementation, which enforces
non-negativity using the primal--dual interior-point solver adapted from
\texttt{JaxNNLS} \citep{Krawczyk2025JaxNNLS},\footnote{\url{https://github.com/CKrawczyk/JaxNNLS}.}
which specialises the \texttt{qpax} method of \citet{TracyManchester2024} to
non-negative least squares. H24 instead used the modified fast non-negative
least-squares algorithm of \citet{Bro1997}. The off-diagonal blocks couple the lens and source intensities.
The lens MGE amplitudes carry no regularisation, corresponding to a zero
lens-regularisation block in H24's notation, and $\stylechange{\tens{H}}=0$ for the MGE source. For
numerical stability, a ridge of $10^{-3}$ \stylechange{was} added to the diagonal of the curvature
matrix for each such unregularised amplitude; it \stylechange{entered} the linear solve but not
$\chi^2$. For adaptive source regularisation, $\stylechange{\tens{H}}$ is the spatially weighted matrix
defined below.
For the S\'ersic stage, each galaxy contributes a unit-intensity S\'ersic basis,
with shape parameters sampled non-linearly and intensity solved linearly.}

\subsection{\texorpdfstring{\reviewchange{Hilbert placement and Delaunay interpolation}}{Hilbert placement and Delaunay interpolation}}\label{app:mesh}

\reviewchange{The first MGE fit \stylechange{supplied} an adaptation map $u_i$: the data after
subtracting the fitted foreground light, divided by the noise map, with values
floored at one per cent of the map maximum. This map \stylechange{was} held fixed during the
pixelised fit. Image-plane mesh placement \stylechange{followed} a Hilbert space-filling curve,
as also used by \citet{He2025Jackpot}; the weight form below is that of the
\texttt{PyAutoLens} implementation, with the power and floor chosen for this pipeline. The adaptation map \stylechange{was} interpolated
along the curve, and the relative sampling weights are
\begin{equation}
 w_i^{\rm mesh}=\max\!\left[\left(\frac{u_i}{\max_{i'} u_{i'}}\right)^{3.5},\,0.01\right]\;.
 \label{eq:app-hilbert}
\end{equation}
The weights \stylechange{were} normalised and accumulated into a one-dimensional cumulative
distribution. Interpolating its inverse at uniformly
spaced cumulative weights places more mesh points in high-signal regions while
retaining coverage of faint regions. The image-plane vertices \stylechange{were} fixed for the
fit; their source-plane positions \stylechange{were} ray-traced for each trial mass model.
We \stylechange{placed} 500 points using the Hilbert sampling and \stylechange{added} 30 points on a circle just outside the image-plane mask, at
radius $r_{\rm mask}+\Delta\theta/2$, where $\Delta\theta$ is the image pixel scale;
the designated boundary amplitudes \stylechange{were} constrained to zero. Before interpolation,
traced sub-pixels and vertices outside an ellipse fitted to the traced mask border \stylechange{were} moved
radially onto that ellipse, limiting
the influence of rays that would otherwise map to remote source-plane positions.}

\reviewchange{The traced vertices $\boldsymbol v_j$ define a Delaunay triangulation.
For a ray landing at $\boldsymbol\beta$ in a triangle with vertices
$\boldsymbol v_1,\boldsymbol v_2,\boldsymbol v_3$, the barycentric weights satisfy
\begin{equation}
 \boldsymbol\beta=\sum_{j=1}^3 b_j\boldsymbol v_j\;,\qquad
 \sum_{j=1}^3 b_j=1\;,\qquad
 S(\boldsymbol\beta)=\sum_{j=1}^3 b_j S_j\;.
 \label{eq:app-barycentric}
\end{equation}
This is the three-vertex linear interpolation of \citet{Vegetti2009}, explicitly
described as barycentric interpolation by \citet{VernardosKoopmans2022}.
For image pixel $i$, its unconvolved mapping matrix is
\begin{equation}
 M_{ij}=\frac{1}{n_{{\rm sub},i}^2}
 \sum_{s=1}^{n_{{\rm sub},i}^2}b_j(\boldsymbol\beta_{is})\;,\qquad \stylechange{\tens{f}}=\mathcal B\stylechange{\tens{M}}\;,
 \label{eq:app-mapping}
\end{equation}
where $\boldsymbol\beta_{is}$ is the traced position of subpixel $s$, with zero weights
for vertices outside the enclosing triangle. Queries outside
the triangulation use the nearest vertex. Intensities therefore reside at vertices
and the model is continuous and piecewise linear within the mesh. Assigning one
colour to each triangle in the source-plane plots is a display convention and
does not describe the interpolation used to fit the image. H24 and COWLS-I
\citep{Nightingale2025COWLS} instead \stylechange{used} Voronoi natural-neighbour interpolation;
Hilbert placement does not require either choice of interpolant.}

\subsection{\texorpdfstring{\reviewchange{Adaptive split regularisation}}{Adaptive split regularisation}}\label{app:regularisation}

\reviewchange{We \stylechange{used} an adaptive cross, or split, regularisation following the approach of
\citet{Nightingale2024} and H24, also used in COWLS-I \citep{Nightingale2025COWLS},
here evaluated with the Delaunay interpolant. Regularising each vertex against its mesh
neighbours makes the number of regularisation terms change whenever a small change in the
mass model re-wires the traced mesh, producing discontinuities in the likelihood surface;
the cross keeps exactly four terms per vertex, a motivation set out in appendix~A.2
(`Likelihood surface noise') of H24. The implementation differs from
\citet{Nightingale2024} in its cross geometry: they \stylechange{placed} cross points on the diagonals
at a distance $\sqrt{\mathcal A_j}$, whereas we \stylechange{used} the Cartesian axes and half that
distance. For vertex $j$, define its associated area $\mathcal A_j$ as one-third of
the summed areas of triangles sharing that vertex. Four cross points \stylechange{were} placed at
\begin{equation}
 \boldsymbol v_j\pm h_j\hat{\boldsymbol x}\;,\quad
 \boldsymbol v_j\pm h_j\hat{\boldsymbol y}\;,\qquad
 h_j=\tfrac12\sqrt{\mathcal A_j}\;,
 \label{eq:app-cross}
\end{equation}
indexed by $a=1,\dots,4$. Let $\omega_{jal}$ be the interpolation weight of source
vertex $l$ at cross point $(j,a)$, and $D_{(j,a)l}=\delta_{jl}-\omega_{jal}$. Then
\begin{equation}
 \boldsymbol S^{\rm T}\stylechange{\tens{H}}\boldsymbol S
 =\sum_j\sum_{a=1}^4\lambda_j
 \left[S_j-\sum_l \omega_{jal}\,S_l\right]^2
 +\epsilon\sum_j S_j^2\;,\qquad \epsilon=10^{-8}\;.
 \label{eq:app-split}
\end{equation}
Equivalently, $\stylechange{\tens{H}}=\stylechange{\tens{D}}^{\rm T}\stylechange{\tens{\Lambda}}\stylechange{\tens{D}}+\epsilon\stylechange{\tens{I}}$, where
$\Lambda_{(j,a),(j,a)}=\lambda_j$. The small diagonal term stabilises the matrix.}

\reviewchange{The local regularisation strength \stylechange{was} determined from the fixed adaptation
map projected through the current mapping. \citet{Nightingale2024} instead \stylechange{derived} this
signal from a previous model of the lensed source. If $N_j$ is the number of
subpixel-to-vertex mappings contributing to vertex $j$, define
\begin{equation}
 t_j=\frac{1}{N_j}\sum_{(i,s)\in\mathcal M_j}
 b_j(\boldsymbol\beta_{is})u_i\;,\qquad
 \nu_j=\left(\frac{t_j}{\max_l t_l}\right)^p\;.
 \label{eq:app-signal}
\end{equation}
$\mathcal M_j$ is the set of subpixel mappings that include vertex $j$, and $N_j=|\mathcal M_j|$. Unmapped vertices have $t_j=0$. The exponent $p$ controls how rapidly regularisation
changes with signal. In the \texttt{AdaptSplit} coefficient convention of the
\texttt{PyAutoLens} implementation,
\begin{equation}
 \lambda_j=\left[c_{\rm in}\nu_j+c_{\rm out}(1-\nu_j)\right]^4\;.
 \label{eq:app-reg-weight}
\end{equation}
The fourth power follows from squaring the interpolated coefficient to form a
regularisation weight and squaring that weight when constructing $\stylechange{\tens{H}}$; in
\citet{Nightingale2024} the interpolated weight \stylechange{was} linear and \stylechange{entered} $\stylechange{\tens{H}}$ squared.
The two positive coefficients and $p$ \stylechange{were} inferred non-linearly. They allow bright
and faint source regions to have different smoothing strengths without changing
the linearity of the intensity solution at each evaluation.}

\subsection{\texorpdfstring{\reviewchange{Statistical objective and fit sequence}}{Statistical objective and fit sequence}}\label{app:objective}

\reviewchange{For the fitted model image, define
\begin{equation}
 \chi^2=\sum_i\frac{(d_i-m_i)^2}{n_i^2}\;,\qquad
 \ln\mathcal L_{\rm img}=-\frac12\left[\chi^2+\sum_i\ln(2\pi n_i^2)\right]\;.
 \label{eq:app-image-likelihood}
\end{equation}
The MGE and S\'ersic stages \stylechange{used} this Gaussian image likelihood evaluated at the
non-negative linear solution. For the pixelised stage, the statistical objective
additionally accounts for regularisation and source complexity
\citep{Suyu2006,Nightingale2024,He2024}:
\begin{align}
 -2\ln\mathcal E={}&\chi^2+\widehat{\boldsymbol S}^{\rm T}\stylechange{\tens{H}}\widehat{\boldsymbol S}
 +\ln\det(\stylechange{\tens{Z}}^{\rm T}\stylechange{\tens{Z}}+\stylechange{\tens{H}})\nonumber\\
 &-\ln\det\stylechange{\tens{H}}+\sum_i\ln(2\pi n_i^2)\;.
 \label{eq:app-evidence}
\end{align}
Here the determinant matrices are the source blocks used by the implementation;
the unregularised lens-light amplitudes are excluded from these determinants,
although they enter the joint solution and $\chi^2$. The reconstruction imposes
the designated zero-amplitude boundary constraints; the determinant calculation
retains the source block. This specifies the objective evaluated by the code,
not an exact marginalisation over all non-negative lens and source intensities:
no additional integration over the positivity-truncated domain \stylechange{was} performed.}

\reviewchange{The candidate conjugate points introduced during preprocessing also constrain
the fit. If $\Delta_{\rm pos}$ is their maximum pairwise separation after ray tracing
to the source plane and $\delta_{\rm pos}$ is the allowed threshold, the positional
penalty subtracted from the log objective is
\begin{equation}
 P_{\rm pos}=10^8\max(0,\Delta_{\rm pos}-\delta_{\rm pos})\;,
 \label{eq:app-positions}
\end{equation}
with separations expressed in arcseconds. For the first fit, a pre-fit check \stylechange{removed}
candidate points within \ang{;;0.15} of the lens-light centre and, where needed, a single
outlier, then \stylechange{fitted} an SIE with external shear (centre fixed at the lens-light centre) to
find the smallest achievable maximum source-plane separation $s$. The threshold \stylechange{was}
$\min[\max(2s,\ang{;;0.3}),\ang{;;0.5}]$; candidates with fewer than two retained points
\stylechange{were} fitted without the positional constraint. This discourages configurations in which
putative multiple images map to unrelated source locations. In the pixelised fit,
the conjugate points \stylechange{were} updated from the first-fit model, and the threshold \stylechange{was} the larger of
three times their first-fit source-plane separation and \ang{;;0.2}. This is an
auxiliary constraint, rather than an additional independent photometric measurement.}

\reviewchange{The non-linear sampler \stylechange{explored} the mass and free light-shape parameters,
and the regularisation parameters when present. With their adopted prior density
$\pi(\boldsymbol\eta)$, the sampled density is proportional to
$\mathcal E(\boldsymbol\eta)\exp[-P_{\rm pos}]\pi(\boldsymbol\eta)$ for the pixelised
fit, replacing $\mathcal E$ by $\mathcal L_{\rm img}$ for the analytic fits and
setting $P_{\rm pos}=0$ where no positional constraint \stylechange{was} supplied.
The quantity $\mathcal E$ is the per-model inversion objective; it is distinct from
the evidence obtained by the outer sampler integrating over non-linear parameters.
Table~\ref{tab:app-stages} summarises which quantities change between stages.
As in the multiband analysis of COWLS-I \citep{Nightingale2025COWLS}, subsequent
photometric fits can hold the reference-band mass model fixed. Our NISP and EXT
fits \stylechange{were} performed independently using the fixed VIS mass model, with their
photometric results presented in ECLIPSE-D.}

\begin{table*}
\reviewchange{
\caption{Quantities fitted and fixed in the three VIS modelling stages.}
\label{tab:app-stages}
\centering
\renewcommand{\arraystretch}{1.3}
\setlength{\tabcolsep}{6pt}
\begin{tabular}{p{0.09\textwidth}p{0.25\textwidth}p{0.27\textwidth}p{0.29\textwidth}}
\hline\hline
Stage & Non-linear parameters & Fixed quantities & Linear solution \\
\hline
MGE & Lens and source MGE geometry; SIE and shear & Gaussian widths & Lens and source Gaussian intensities; no regularisation \\
Pixelised & SIE and shear; adaptive regularisation & First-fit lens MGE geometry; adaptation map and image-plane mesh & Lens Gaussian intensities and source vertex intensities; split regularisation on source \\
S\'ersic & Lens and source S\'ersic shapes & First-fit SIE and shear & Lens and source intensities; no regularisation \\
\hline
\end{tabular}
}
\end{table*}

\subsection{\texorpdfstring{\reviewchange{Priors and stopping criteria}}{Priors and stopping criteria}}\label{app:priors}

\reviewchange{Table~\ref{tab:app-priors} lists the priors for the three VIS fits.
We denote a uniform distribution by $\mathcal U(a,b)$, a distribution uniform in
$\ln x$ by $\mathcal U_{\ln}(a,b)$, and a Gaussian with mean $m$ and standard
deviation $s$ by $\mathcal N(m,s^2)$. The notation $\mathcal N_{[a,b]}$ denotes
that Gaussian truncated to $[a,b]$. Angular coordinates, radii, and their prior
widths are in arcseconds. Each coordinate or ellipticity component receives the
listed prior. The \stylechange{data set} centre $\boldsymbol c_0$ anchors the initial lens-centre
priors. For the MGE and SIE shapes, the code \stylechange{sampled} ellipticity components
$e_0=[(1-q)/(1+q)]\sin(2\phi)$ and
$e_1=[(1-q)/(1+q)]\cos(2\phi)$, rather than $q$ and $\phi$ directly.
The two lens MGE sets \stylechange{shared} a centre and \stylechange{had} independent shape priors, with
$e_1^{(1)}>e_1^{(2)}$ imposed to select one labelling of the exchangeable sets.
The linear light intensities have non-negativity constraints rather than sampled
prior distributions; source-pixel smoothing is specified in
\cref{app:regularisation}.}

\reviewchange{The second fit \stylechange{fixed} the lens MGE geometry but \stylechange{re-solved} its intensities.
The SIE parameters \stylechange{were} assigned fresh priors, and the shear \stylechange{retained} its original
prior distributions; neither \stylechange{was} narrowed using the first-fit posterior.
For the S\'ersic fit, centre-prior means \stylechange{were} taken from the first-fit posterior
medians and the configured standard deviation \stylechange{was} \ang{;;0.05}. The SIE and shear
\stylechange{were} fixed to the first-fit maximum-likelihood model. NISP and EXT photometry \stylechange{used}
that same fixed VIS mass model, as described in \cref{LensPipeline} and ECLIPSE-D.}

\begin{table*}
\reviewchange{
\caption{Non-linear priors for the VIS modelling stages. Fixed quantities and linear
amplitudes are distinguished from sampled parameters.}
\label{tab:app-priors}
\centering
\renewcommand{\arraystretch}{1.25}
\setlength{\tabcolsep}{6pt}
\begin{tabular}{p{0.14\textwidth}p{0.36\textwidth}p{0.38\textwidth}}
\hline\hline
Stage & Parameter & Prior or fixed value \\
\hline
MGE & Lens MGE centre, $c_k$ & $\mathcal U(c_{0,k}-0.1,c_{0,k}+0.1)$ \\
 & Lens MGE ellipticity, each set & $\mathcal N_{[-0.5,0.5]}(0,0.3^2)$ \\
 & Source MGE centre, $c_k$ & $\mathcal N(c_{0,k},0.3^2)$ \\
 & Source MGE ellipticity & $\mathcal N_{[-0.7,0.7]}(0,0.3^2)$ \\
 & SIE centre & Fixed to $\boldsymbol c_0$ \\
MGE/pixelised & SIE ellipticity & $\mathcal N_{[-1,1]}(0,0.3^2)$ \\
 & SIE scale $\theta^{\rm mass}_{\rm E}$ & $\mathcal U(0,8)$ \\
 & Shear components $\gamma_1^{\rm ext},\gamma_2^{\rm ext}$ & $\mathcal U(-0.3,0.3)$ \\
Pixelised & SIE centre, $c_k$ & $\mathcal U(c_{0,k}-0.1,c_{0,k}+0.1)$ \\
 & Lens MGE geometry & Fixed to first-fit maximum-likelihood values \\
 & $c_{\rm in},c_{\rm out}$ & $\mathcal U_{\ln}(10^{-6},10^6)$ \\
 & Signal exponent $p$ & $\mathcal U(0,1)$ \\
S\'ersic & Lens/source centre, $c_k$ & $\mathcal N(\widetilde c_k^{\rm MGE},0.05^2)$ \\
 & Lens/source ellipticity & $\mathcal N_{[-1,1]}(0,0.3^2)$ \\
 & Lens/source effective radius & $\mathcal U(0,30)$ \\
 & Lens S\'ersic index & $\mathcal U(0.5,10)$ \\
 & Source S\'ersic index & $\mathcal U(0.8,5)$ \\
 & SIE and shear & Fixed to first-fit maximum-likelihood values \\
\hline
\end{tabular}
}
\end{table*}

\reviewchange{We \stylechange{used} the \texttt{nautilus} convergence criteria \citep{nautilus},
with exploration ending when the estimated fraction of evidence in the live set
\stylechange{fell} below $f_{\rm live}=0.01$. Subsequent sampling \stylechange{required} an effective sample
size $N_{\rm eff}\geq500$ and at least one sample in each shell.
These are convergence targets, rather than a requirement that the maximum likelihood
or estimated evidence cease increasing. Table~\ref{tab:app-sampler} gives the
live-point counts and maximum likelihood-evaluation budgets. A search \stylechange{could} also stop
on reaching its evaluation budget before satisfying the convergence targets.}

\begin{table}
\reviewchange{
\caption{\texttt{Nautilus} settings for the three VIS searches.}
\label{tab:app-sampler}
\centering
\renewcommand{\arraystretch}{1.25}
\setlength{\tabcolsep}{10pt}
\begin{tabular}{lrr}
\hline\hline
Fit & Live points & Evaluation limit \\
\hline
MGE & 750 & 200\,000 \\
Pixelised & 300 & 100\,000 \\
S\'ersic & 100 & 100\,000 \\
\hline
\end{tabular}
}
\end{table}
\section{Preliminary round of visual inspection}\label{app:preliminary_visual_inspection}

\subsection{Description}

A preliminary round of visual inspection was undertaken to determine which diagnostic images are most informative for assessing
lens modelling results and to refine the criteria used to judge model success. A random sample of 569 of the 3000 highest
machine-learning-ranked (grade A) strong lenses was modelled, and 200 lenses were selected for inspection by 12 classifiers, all
of whom identify as lens modelling experts. Each expert reviewed 100 common lenses (to enable consistency checks between reviewers)
and 100 unique lenses (to maximise diversity in the identification of failure modes).

In the preliminary round, reviewers were given four images: (i) a $2\times2$ panel showing RGB images of the lens under different scalings;
(ii) a $4\times3$ panel showing the model fit assuming a Delaunay mesh source; (iii) the same $4\times3$ panel for a fit assuming
an MGE source; and (iv) a $2\times1$ panel showing the segmentation information of the lens. These images are not shown
in this paper for brevity. The number of images and panels given
to reviewers was intentionally larger than would be practical for evaluating \currentlenssample{} strong lenses. The motivation was
to determine what information reviewers needed to judge lenses.

Reviewers first assessed all 200 lenses with \reviewchange{three classification options} corresponding to SC, FL, and MGL,
indicating whether the lens model was considered successful. Reviewers were provided with an evaluation panel, which gave the following criteria for a model
to be considered successful:

\begin{enumerate}
    \item You have high confidence the data set is of a genuine strong lens.
    \item The pixelated lens model accurately fits all multiple images of the lensed source, and does not fit light of other nearby galaxies or foreground lens light residuals.
    \item The pixelated lens model provides an accurate and publication-worthy estimate of the Einstein radius.
    \item There are no issues with the segmentation map.
    \item The lens does not have multiple galaxies inside its Einstein radius and thus does not require multiple lens light and mass components.
\end{enumerate}

A failed model was defined as one or more of the above criteria not being met, and the MGL option was used when criterion 5 was not met.
\reviewchange{After classifying all 200 objects}, reviewers were subsequently shown all results \reviewchange{again} for lenses they
classified as failures. Next to each image shown to reviewers were buttons they could click to identify the reason
for the failure. For example, next to the segmentation image was \reviewchange{a button labelled} ``Missing Nearby Artefact/Galaxy'', meaning the segmentation
had not correctly enabled masking of the nearby galaxy. \reviewchange{Another button next to the fit panel \stylechange{was} labelled}
``Bad Source Reconstruction''.

The third and final round of inspection asked reviewers to click on the image panels that allowed them to
determine whether a lens model was SC or FL. This information was analysed to determine the final images\reviewchange{ (\cref{fig:segmentation-example,fig:vis-pix-fit})}
that were actually shown to reviewers for inspection of the full sample of \currentlenssample{} lenses.

\subsection{Discussion and lessons learned}

Reviewers agreed that judging whether a lens model is successful is inherently subjective, and that it would be
impossible to define assessment criteria interpreted identically by all inspectors. However, it was also agreed that
the criteria used during the preliminary visual inspection were unnecessarily complicated, particularly because they
mixed assessment of the lens model itself with assessment of the segmentation map, including criteria
such as ``does not fit light from nearby galaxies'' or ``there are no issues with the segmentation map''. It was
therefore decided that segmentation-map validation -- specifically checking whether genuine source emission had
been removed or nearby contaminating galaxies had not been identified correctly -- would be performed separately
from the primary visual inspection and not included directly in the model-success criteria.

The inspection criteria were subsequently simplified to those presented in \cref{sec:visual}, with the revised structure
more directly guiding inspectors towards the appropriate classification, including identifying MGL systems before
assessing model quality. Reviewers agreed that the refined criteria still retained some subjectivity, but
substantially less than in the preliminary inspection exercise.

During the preliminary inspection, reviewers first classified all 200 lenses and only afterwards returned to failed
systems to record the reason for failure. Inspectors noted that this often led to them forgetting why a particular
model had been classified as unsuccessful. The final inspection workflow therefore required reviewers to \reviewchange{select the failure category} immediately after selecting FL. The available failure categories were also simplified to
the set described in \cref{sec:visual}.

Finally, the diagnostic images used during inspection were analysed and refined into the layouts shown in
\cref{fig:segmentation-example,fig:vis-pix-fit}.
Discussion amongst reviewers motivated the option to alternate between pixelised and parametric
MGE source reconstructions, which \stylechange{provides} valuable additional diagnostic information for difficult or ambiguous systems.

\section{Additional SC examples}\label{app:additional-success-examples}

\inspectionchange{\Cref{fig:success-consensus-quality} compares three grade A models with current
SC-vote fractions of $100\%$, $75\%$, and $50\%$. Figure~\ref{fig:success-consensus-additional-high}
adds 14 grade A systems drawn randomly from the
current inspection results. Every selected system \stylechange{has} exactly four recorded votes: three \stylechange{have}
$4/4$ SC votes, four \stylechange{have} $3/4$, and seven \stylechange{have} $2/4$. The draws \stylechange{excluded} the
Figure~\ref{fig:success-consensus-quality} candidate set and \stylechange{contained} no repeated objects.
They \stylechange{were} restricted to locally available completed \stylechange{pixelised} fits and RGB images registered to
the VIS cut-outs. The four-vote restriction \stylechange{applies} to these illustrative examples, not to the entire
inspection export.}

\begin{figure*}
    \centering
    \includegraphics[width=0.96\textwidth]{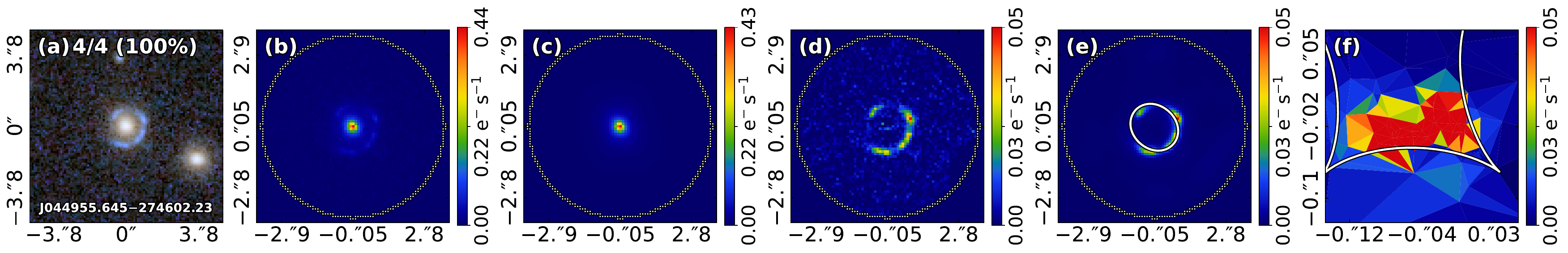}
    \includegraphics[width=0.96\textwidth]{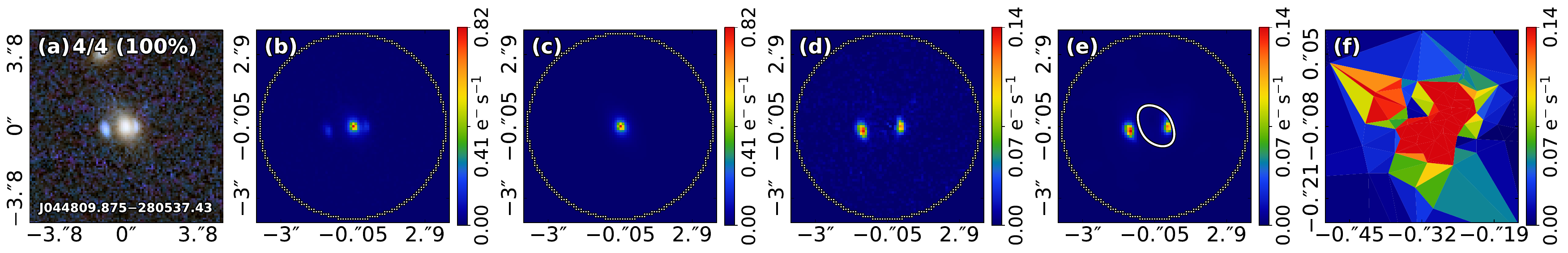}
    \includegraphics[width=0.96\textwidth]{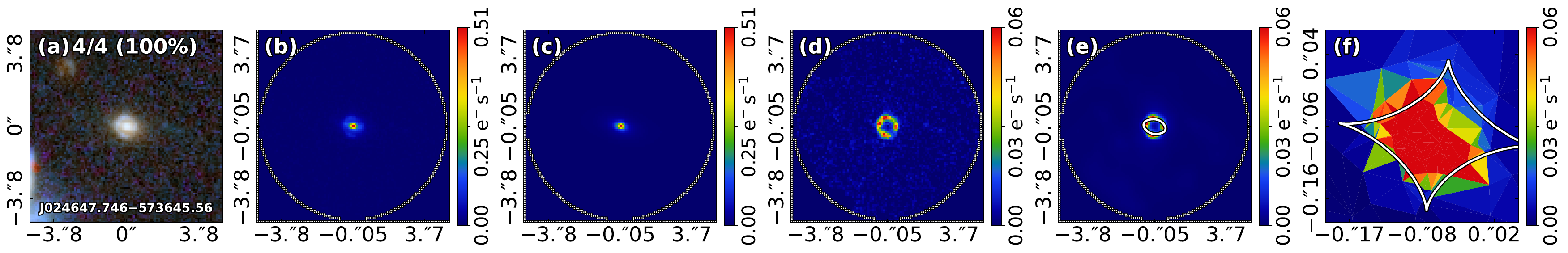}
    \includegraphics[width=0.96\textwidth]{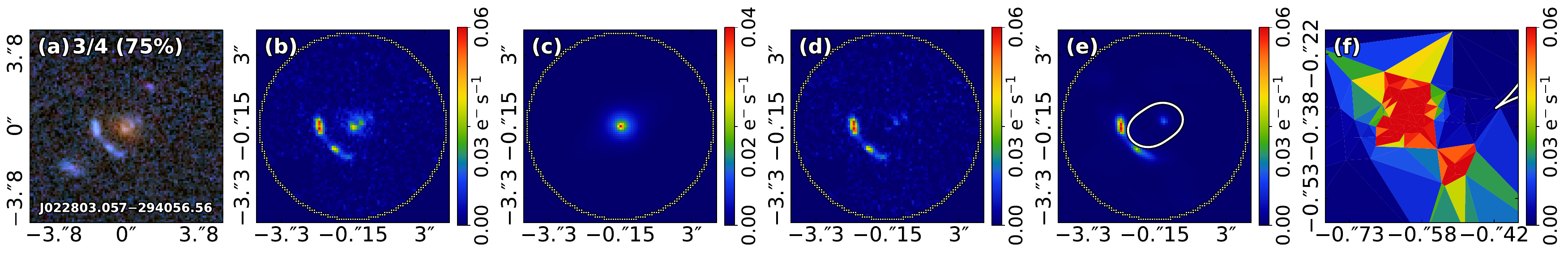}
    \includegraphics[width=0.96\textwidth]{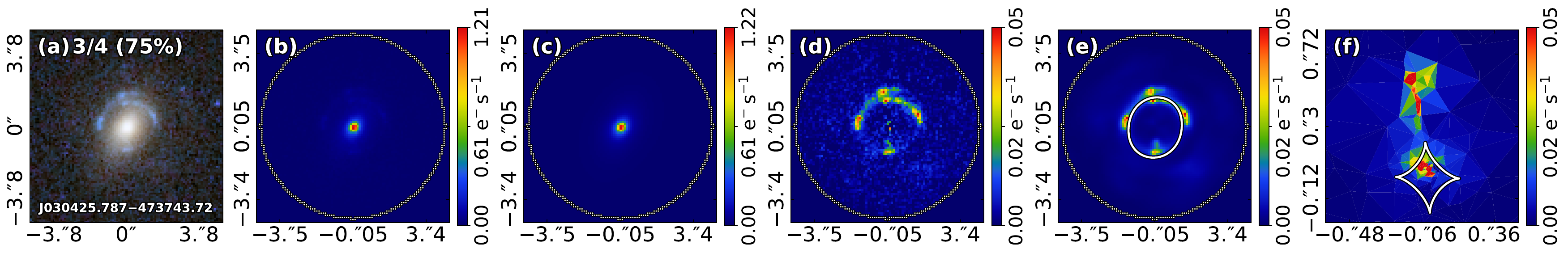}
    \includegraphics[width=0.96\textwidth]{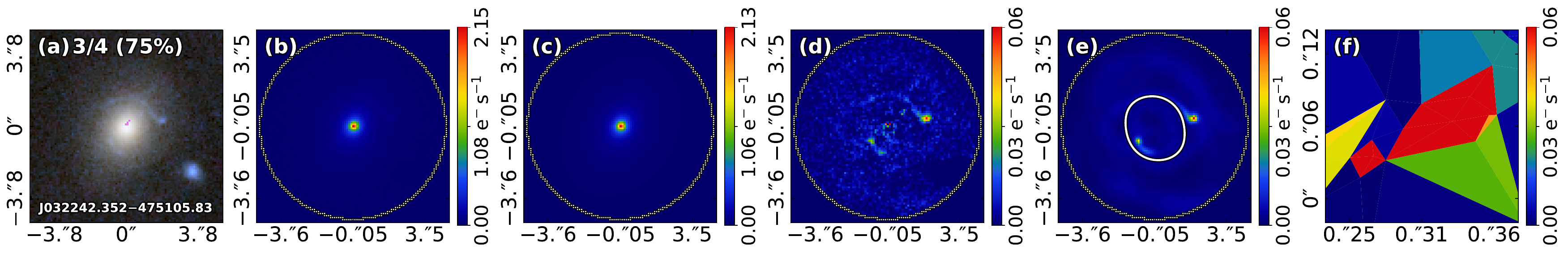}
    \includegraphics[width=0.96\textwidth]{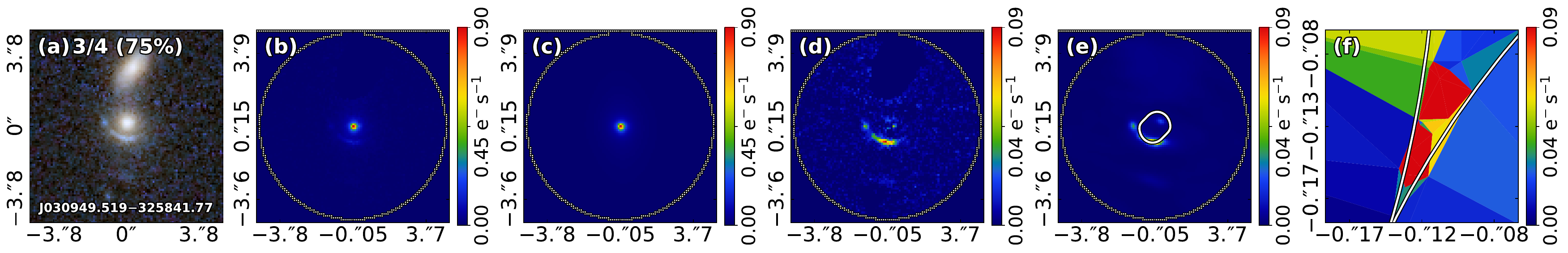}
    \caption{\inspectionchange{
        Additional randomly selected grade A lens models from the current inspection campaign.
        The first three rows have $4/4$ ($100\%$) SC votes and the remaining four have $3/4$
        ($75\%$). All seven meet the adopted $75\%$ SC threshold. Each RGB panel gives the
        SC-vote tally and percentage. All panels are shown north up, east left.
        From left to right, the panels show the RGB image (a), VIS data (b), foreground lens-light
        model (c), lens-subtracted image (d), lensed source-model image (e), and source-plane
        reconstruction (f), following \cref{fig:success-consensus-quality}. These examples illustrate
        model appearance and are not a statistical calibration of model reliability.
    }}
    \label{fig:success-consensus-additional-high}
\end{figure*}

\begin{figure*}
    \addtocounter{figure}{-1}
    \renewcommand{\theHfigure}{appendix-C-continued}
    \makeatletter
    \renewcommand{\fnum@figure}{\figurename~\thefigure\ (continued)}
    \makeatother
    \centering
    \includegraphics[width=0.96\textwidth]{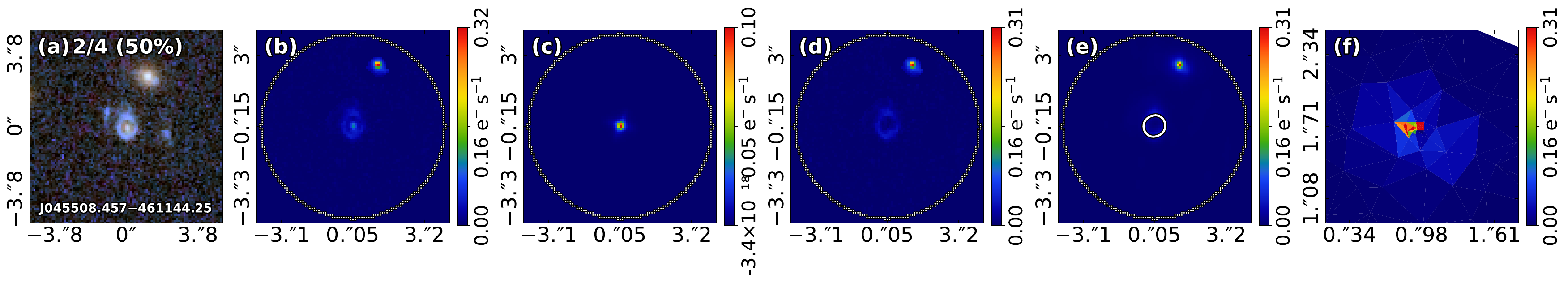}
    \includegraphics[width=0.96\textwidth]{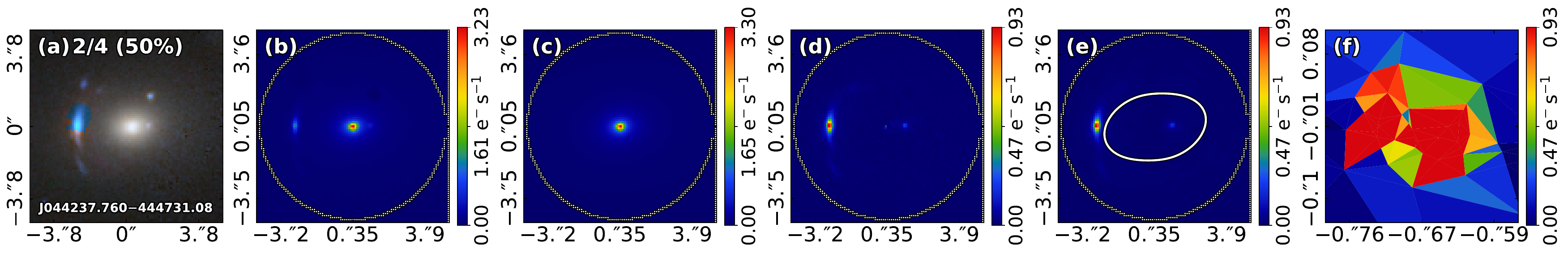}
    \includegraphics[width=0.96\textwidth]{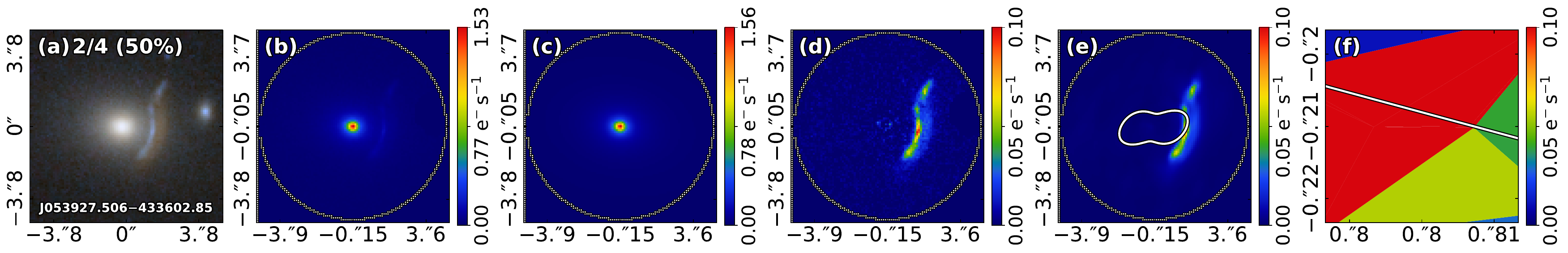}
    \includegraphics[width=0.96\textwidth]{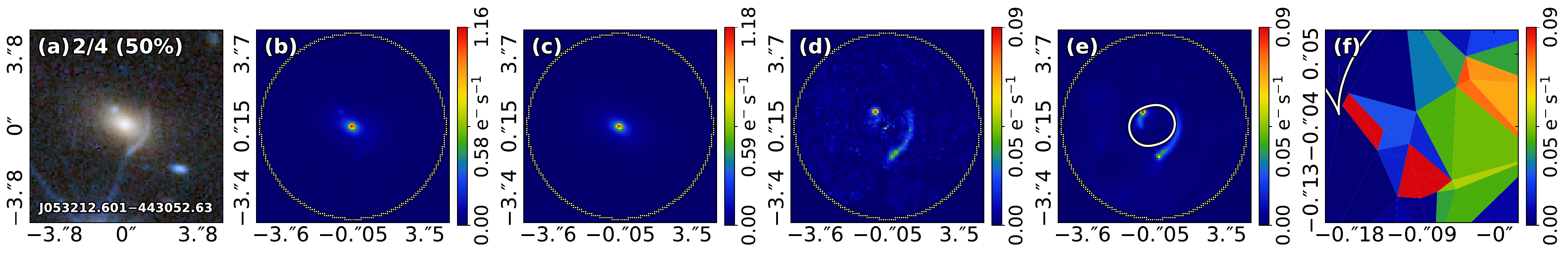}
    \includegraphics[width=0.96\textwidth]{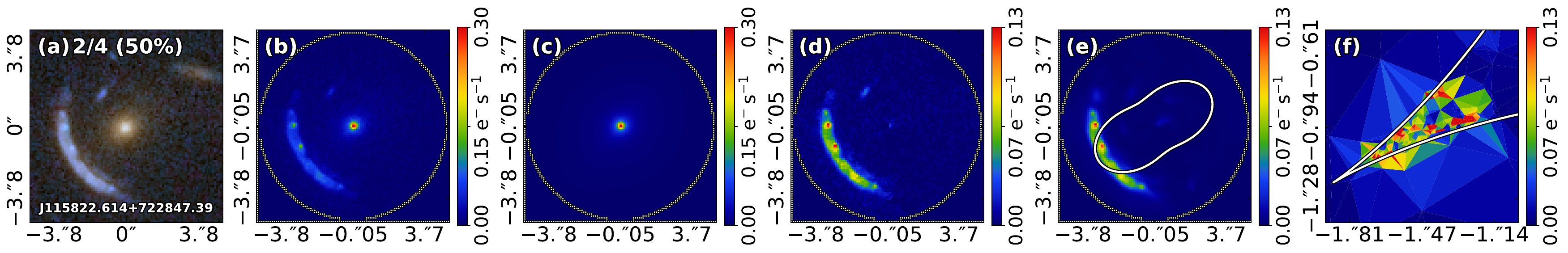}
    \includegraphics[width=0.96\textwidth]{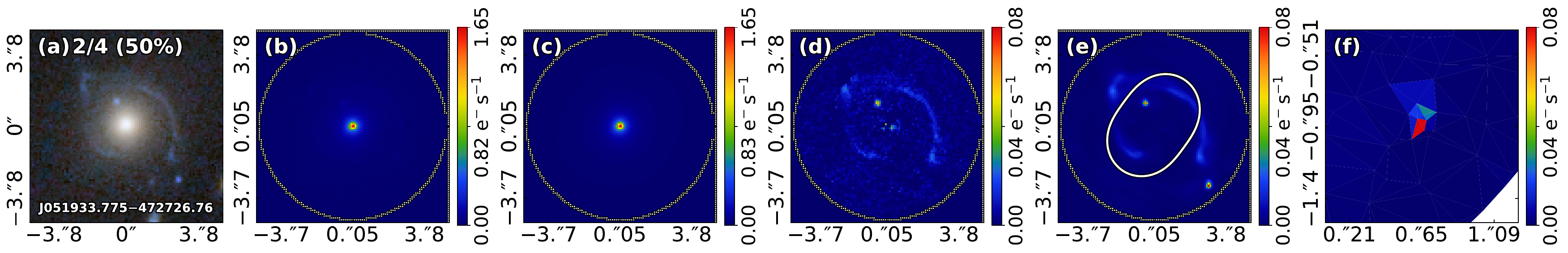}
    \includegraphics[width=0.96\textwidth]{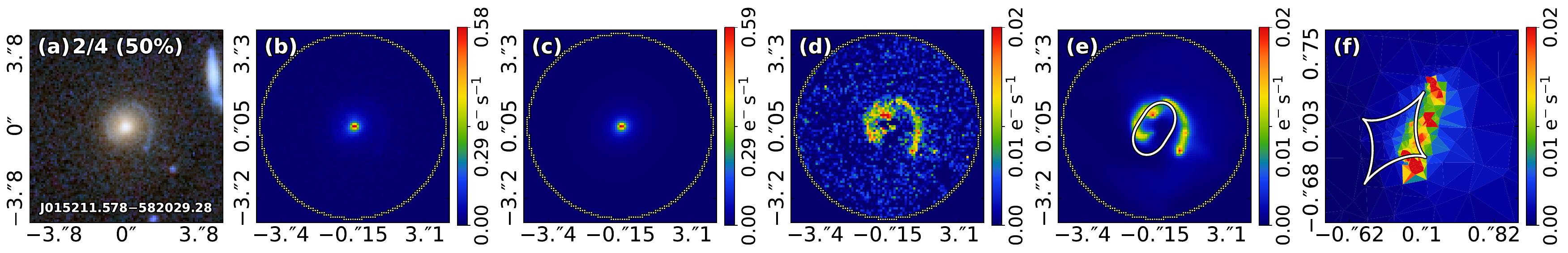}
    \caption{\inspectionchange{
        Additional randomly selected grade A lens models from the current inspection campaign,
        each with exactly $2/4$ ($50\%$) SC votes. All seven are NC at the adopted $75\%$
        threshold. Each RGB panel gives the observed SC-vote tally and percentage. All panels are
        shown north up, east left. Panels follow the layout of
        \cref{fig:success-consensus-quality}.
        These lower-agreement examples provide a qualitative comparison with the $100\%$ and
        $75\%$ examples, rather than a calibration of classification reliability.
    }}
    \label{fig:success-consensus-additional-low}
\end{figure*}

\end{document}